\documentclass[amsmath,amssymb,twocolumn, showpacs, superscriptaddress, longbibliography, floatfix, nofootinbib, aps, prb]{revtex4-2}

\usepackage{amsfonts, bm}
\usepackage{mathtools}
\usepackage[pdftex]{graphicx}
\usepackage{import}
\usepackage{ulem}
\usepackage{scalerel}
\usepackage{nccmath}
\usepackage{empheq}
\usepackage{lipsum}
\usepackage{xfrac}
\usepackage{tabularray}

\usepackage[breaklinks,hypertexnames=false]{hyperref}
\hypersetup{colorlinks,linkcolor={magenta},citecolor={blue},urlcolor={blue}} 

\usepackage{braket}
\usepackage{xcolor}

\usepackage[capitalize]{cleveref}

\usepackage{glossaries}
\glsdisablehyper

\newacronym{mbs}{MBS}{Majorana bound state}
\newacronym{mzm}{MZM}{Majorana zero mode}
\newacronym{abs}{ABS}{Andreev bound state}
\newacronym{fi}{FI}{ferromagnetic insulator}
\newacronym{jj}{JJ}{Josephson junction}
\newacronym{cpr}{CPR}{current-phase relation}

\usepackage{array}
\usepackage[margin=0.9in]{geometry}
\usepackage[utf8]{inputenc}
\usepackage[inline,shortlabels]{enumitem}
\usepackage[caption=false]{subfig}
\usepackage{multirow}

\usepackage{sidecap,tikz}
\definecolor{lime}{HTML}{A6CE39}
\DeclareRobustCommand{\orcidicon}{\hspace{-1mm}
	\begin{tikzpicture}
		\draw[lime, fill=lime] (0,0) 
		circle [radius=0.16] 
		node[white] {{\fontfamily{qag}\selectfont \tiny \,ID}};
		\draw[white, fill=white] (-0.0525,0.095) 
		circle [radius=0.007];
	\end{tikzpicture}
	\hspace{-3mm}
}
\foreach \x in {A, ..., Z}{\expandafter\xdef\csname orcid\x\endcsname{\noexpand\href{https://orcid.org/\csname orcidauthor\x\endcsname}
		{\noexpand\orcidicon}}
}

\def\be{\begin{equation}}
\def\ee{\end{equation}}
\def\bea{\begin{eqnarray}}
\def\eea{\end{eqnarray}}
\def\bmat{\begin{pmatrix}}
\def\emat{\end{pmatrix}}
\def\bs{\begin{split}}
\def\es{\end{split}}

\def\~{$\approx$}

\def\dag{\dagger}

\newcommand{\up}{\uparrow}
\newcommand{\dw}{\downarrow}

\newcommand{\quarc}{Quantum Advanced Research Center (QuARC), Consejo Superior de Investigaciones Cient{\'i}ficas (CSIC), Sor Juana In{\'e}s de la Cruz 3, 28049 Madrid, Spain}
\newcommand{\icmm}{Instituto de Ciencia de Materiales de Madrid (ICMM), Consejo Superior de Investigaciones Cient{\'i}ficas (CSIC), Sor Juana In{\'e}s de la Cruz 3, 28049 Madrid, Spain}

\begin{document}

\title{Optimal Majoranas in Mesoscopic Kitaev Chains}

\author{M. Alvarado\orcidA{}}
\email{miguel.alvarado@csic.es}
\affiliation{\quarc}\affiliation{\icmm}

\author{R. Seoane Souto\orcidB{}}
\affiliation{\quarc}\affiliation{\icmm}

\author{Mar{\'i}a Jos{\'e} Calder{\'o}n\orcidD{}}
\affiliation{\quarc}\affiliation{\icmm}

\author{Ram{\'o}n Aguado\orcidC{}}
\email{ramon.aguado@csic.es}
\affiliation{\quarc}\affiliation{\icmm}

\date{\today}

\begin{abstract}
Kitaev chains realized in quantum dots coupled via superconducting segments provide a controllable platform for engineering Majorana zero modes (MZMs). In these systems, subgap states in the hybrid region mediate the effective coupling between quantum dots and determine the emergence of sweet-spots where MZMs are strongly localized. However, existing minimal treatments often oversimplify the mesoscopic hybrid region. We perform a full microscopic treatment of this hybrid segment, capturing the quasiparticle continuum and spin-split Andreev bound states (ABSs), and show that it fundamentally alters the minimal picture. We derive analytical expressions for the renormalized couplings and sweet-spot conditions, establishing a direct link between microscopic chain parameters and Majorana optimization and identifying experimentally relevant regimes for improved device performance. Critically, we find that parity-crossings of the ABS, marking the onset of an odd-parity spin-polarized regime in the segment, identify the optimal operating windows where MZMs are simultaneously well localized with a large gap to excited states.
\end{abstract}

\maketitle

\section{Introduction}\label{SecI}
The Kitaev chain~\cite{Kitaev2001} (KC) provides a paradigmatic minimal model for one-dimensional topological superconductors hosting Majorana zero modes (MZMs)~\cite{Alicea2012, Leijnse2012_b, RevModPhys.87.137,Aguado2017, Beenakker2020, Tanaka2024, Pita2026}. Arguably, one of the field's central motivations has been the prospect of using these MZMs to realize topologically protected qubits~\cite{KITAEV2003,RevModPhys.80.1083,Sarma-Freedman-Nayak,Wilczek2009,Aguado2020}. Such qubits would encode quantum information non-locally in the parity, either even or odd, of the fermionic state shared by a pair of MZMs.

The recent realization of a Kitaev chain in quantum dots (QDs) coupled via superconductors has opened an exciting new frontier~\cite{Wang2022, Wang2023, Bordin2022, Liu2022, Dvir2023, Bordin2023, Haaf2023, Bordin2025, Bordin2025b, vanloo2025,Bas-Haaf2025,Kulesh2025}. This experimental progress follows from early theoretical proposals~\cite{Leijnse2012,Sau2012,Fulga2013} demonstrating that even the minimal version of such chains, with only two QDs coupled by a central mesoscopic superconducting (SC) segment, hosts a pair of MZMs at a fine-tuned sweet-spot -- often referred to as ``poor man's Majoranas" (PMMs). Despite being spatially separated, these PMM states remain non-topological and therefore offer limited protection compared with what longer chains could potentially provide.

Realizing PMMs requires site-resolved control over the chemical potentials, tunnel couplings, and magnetic fields in each QD~\cite{Wang2022, Wang2023, Bordin2022, Liu2022, Dvir2023, Bordin2023, Haaf2023, Bordin2025, Bordin2025b, vanloo2025}. These fine-tuned sweet-spots occur when the elastic cotunneling (ECT) and crossed Andreev reflection (CAR) amplitudes become equal by tuning the properties of the Andreev bound states (ABSs) residing in the proximitized hybrid regions of the chain~\cite{Liu2022}. 

A key consequence of this condition is a competition between two figures of merit~\cite{Wimmer2025}: the excitation gap, which protects against higher excited states, and the degree of Majorana localization, which ensures the spatial separation of the MZMs. In practice, however, the parameter values that maximize the gap do not generally coincide with those that yield the best localization, introducing a trade-off between both quantities. As a result, identifying optimal sweet-spots~\cite{Souto2022, Seoane2023, Tsintzis2024} becomes a multiobjective optimization problem, where different physical quantities of interest cannot be optimized simultaneously~\cite{Wimmer2025}. 

\begin{figure*}[t!]
\centering
\includegraphics[width=0.8\textwidth]{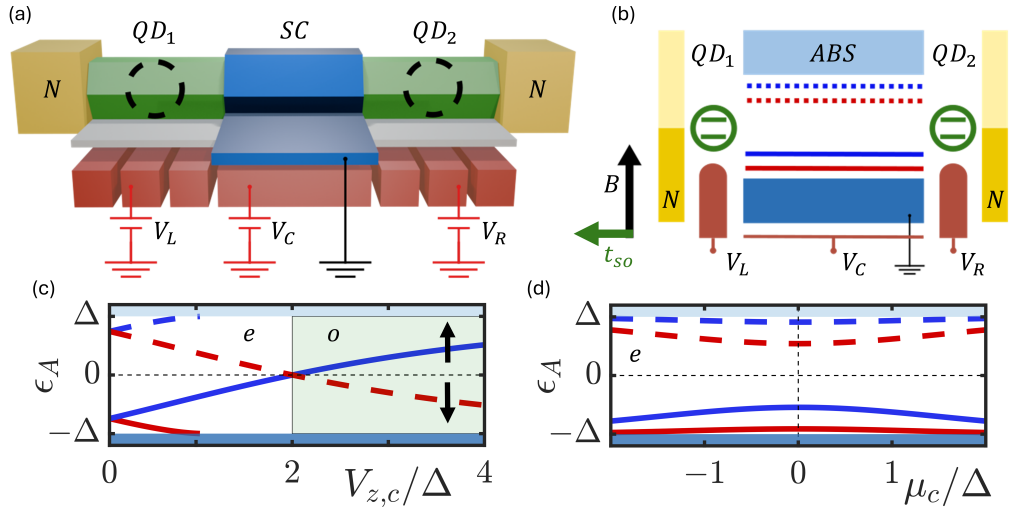}
\caption{(a) Sketch of a minimal KC consisting of two QDs coupled via a SC segment. The gate voltages $V_i$ (shown in red) tune the chemical potentials $\mu_i$ of the QDs (shown in green), while the mesoscopic SC hybrid section (shown in blue) is grounded and tuned by $V_c$. (b) Energy diagram of the system, where the red/blue subgap lines denote the spin-split ABSs. In the sketch, the directions of the magnetic field ($B$) and the spin-orbit coupling amplitude ($t_{so}$) are perpendicular. Note that the Zeeman energy in the SC hybrid section is not fully screened. (c-d) Behavior of the ABS energy ($\epsilon_A$) in the isolated proximitized region of the chain as a function of the finite Zeeman energy ($V_{z,c}$), and the chemical potential in the SC hybrid section ($\mu_c$), respectively. Unless otherwise stated, the parameters used are $\mu_c = 0$, $V_{z,c} = 0.5$, and $\Gamma_s = 2$, in units of $\Delta$. The green region in panel (c) indicates the odd-parity spin-polarized regime ($o$) of the isolated ABS occurring for $\Gamma_s\leq|V_{z,c}|$ for the particle-hole symmetric point in Eq.~\eqref{eq_sd_QPT_main}.
}
\label{fig1}
\end{figure*}

To date, this competition has been studied primarily within ideal, oversimplified Kitaev models that neglect the complex microscopic structure associated to subgap states present in the proximitized hybrid regions of the chain~\cite{Pita2026}. Low-energy effective models typically predict that increasing the coupling to the central mesoscopic SC region and enhancing spin polarization improves both the localization of the MZMs and the gap to excited states~\cite{Souto2022, Miles2023, Liu2023, Leijnse2024, Zatelli2024, Dourado2025, GomezLeon2025, Wimmer2025}. However, such models do not fully capture key microscopic degrees of freedom of the SC hybrid region, such as the quasiparticle continuum or finite spin-splitting of the ABSs, see Fig.~\ref{fig1}~(a,b). Previous works have incorporated some mesoscopic effects going beyond the idealized minimal Kitaev chain giving rise to new features~\cite{Gorm2022, SeoaneAguado2024, Alvarado2024,  Alvarado2025, Bordin2025c, Zhang2025, PhysRevB.111.L140503, Aguado2025, Wimmer2026, Zhang2026, Buonemani2026}. In particular, including both the ABS and the quasiparticle continuum above the SC gap strongly renormalizes the gap to excited states and the low-energy properties of the MZMs, leading to complex subgap dynamics in realistic chains.  

In this work, we present a unified full-microscopic treatment of the middle hybrid region, capturing the phenomena arising from its finite gap and the presence of spin-split subgap states in the proximitized segment of the chain, see blue regions in Figs.~\ref{fig1}~(a,b). Considering such microscopic degrees of freedom of the hybrid SC segment leads to qualitatively new features, including the possibility of zero-energy level crossings of spin-split ABSs~\cite{LeeNN:14} resulting in a quantum phase transition (QPT), which implies a change in the fermionic parity, {\it e.g.}, evolving from an even-parity ground state to an odd-parity spin-polarized regime~\cite{Yu1965, Shiba1968, Rusinov1969, Sakurai1970, Byers1997, Glazman1989, Vecino2003, Martin-Rodero2011, Aguado2015, Pita2026}, see Fig.~\ref{fig1}~(c,d).

Within such a framework, increasing the coupling strength or Zeeman energy does not monotonically enhance the localization of MZMs and the excitation gap. We show how these microscopic details translate into the competition between gap and degree of localization and how they modify the conditions for achieving sweet-spots. Remarkably,
the QPT of the spin-split ABSs defines a special region of parameter space in which well-localized MZMs appear in the chain, featuring the largest gap to the excited states, providing experimentally favorable conditions for the optimization of MZMs in KCs. 

While these findings focus on minimal KCs, we have 
implemented an optimization framework also valid for longer chains. This approach is based on surrogate models~\cite{Baran2023, Paaske2023} for hybrid segments, and Covariance Matrix Adaptation - Evolution Strategy (CMA-ES) algorithms~\cite{Deb2000, Hansen2016} to explore the parameter space, enabling the identification of sweet-spots.

The manuscript is organized as follows. In Sec.~\ref{SecII} we recapitulate the sweet-spot structure associated to the idealized Kitaev chain. In Sec.~\ref{SecIII} we introduce the effective mesoscopic model used to describe the system, providing analytical insights into the renormalization of the sweet-spots and gap to excited states. 

In Sec.~\ref{SecIV} we discuss the renormalization of sweet-spots within different limiting descriptions of the proximitized SC segment. First, we consider the fully-screened spin-degenerate ABS case, often discussed in the literature, in which no Zeeman splitting is considered in the hybrid segment. Second, we analyze the full microscopic limit, where screening effects emerge due to the hybridization with the SC lead, but a finite Zeeman splitting generally remains giving rise to spin-split ABSs. In this regime, we identify a relevant scenario, largely overlooked in the literature, in which the isolated ABS lies in the spin-polarized regime.
 
Finally, in Sec.~\ref{SecVI} we present our conclusions, summarizing the main results and their implications for optimizing MZMs in realistic devices. 
In the Appendices, we outline the technical steps based on the Green’s function (GF) formalism~\cite{Cuevas1996, Zazunov2016, Alvarado2020} underlying the analytical expressions used in the main text.

To clarify the terminology used throughout this work, we distinguish between two complementary descriptions. The hybrid section refers to the proximitized SC segment treated microscopically ({\it i.e.}, including subgap and above gap states, local Zeeman energy, and the coupling to the adjacent QDs), while the microscopic Kitaev chain denotes the effective model obtained after integrating out this region, where the QDs are coupled via renormalized ECT and CAR processes. This distinction provides a direct link between microscopic properties and low-energy behavior. Table~\ref{tab1} summarizes the main analytical results.

\begin{table}[t]
\caption{\label{tab1} Overview of analytical results across the different limiting regimes used to capture mesoscopic effects.}
\renewcommand{\arraystretch}{1.2}
\begin{ruledtabular}
\begin{tabular}{lll}

\multicolumn{3}{c}{\textbf{Hybrid Properties}} \\[2pt] \hline
parity-crossing condition & Sec.~\ref{SecIIIa} & Eq.~\eqref{eq_sd_QPT_main} \\
Renormalization effects   & Sec.~\ref{SecIIIa} & Eq.~\eqref{induced_Zeeman_main} \\
KC zero mode condition    & Sec.~\ref{SecIIIa} & Eq.~\eqref{Eq_muc_main} \\[2pt]

\hline
\multicolumn{3}{c}{\textbf{Microscopic Kitaev Chain}} \\[2pt] \hline
ECT and CAR amplitudes & Sec.~\ref{SecIIIb} & Eq.~\eqref{sweet_spot_Vz_main} \\
Sweet-spot condition   & Sec.~\ref{SecIIIb} & Eqs.~\eqref{mu_c_vzneq0_main} \& \eqref{renorm_mu_Vz_main} \\[2pt]

\hline
\multicolumn{3}{c}{\textbf{Spin-degenerate ABS | Weak-coupling}} \\[2pt] \hline
ECT and CAR amplitudes & Sec.~\ref{SecIIIc} & Eq.~\eqref{sweet_spot_no-Vz_main} \\
Gap to excited states  & Sec.~\ref{SecIIIc} & Eq.~\eqref{eq_gap_main} \\[2pt]

\hline
\multicolumn{3}{c}{\textbf{Spin-degenerate ABS | Strong-coupling}} \\[2pt] \hline
Gap to excited states & Sec.~\ref{SecIIId} & Eq.~\eqref{approx_dE_main} \\
Sweet-spot condition & Sec.~\ref{SecIIId} & Eq.~\eqref{renorm_muc_vz0_main} \\[2pt]

\hline
\multicolumn{3}{c}{\textbf{Spin-split ABS | Weak-coupling}} \\[2pt] \hline
Gap to excited states & Sec.~\ref{SecIIIe} & Eq.~\eqref{dE_Vzneq0_main} \\[2pt]

\hline
\multicolumn{3}{c}{\textbf{Spin-split ABS | Strong-coupling}} \\[2pt] \hline
Gap to excited states & Sec.~\ref{SecIIIf} & Eq.~\eqref{dE_Vzneq0_sc_main} \\
Sweet-spot condition & Sec.~\ref{SecIIIf} & Eqs.~\eqref{mu_c_vzneq0_t_main} \& \eqref{renorm_mu_Vz_t_main} \\

\end{tabular}
\end{ruledtabular}
\end{table}

\section{Ideal Kitaev Chain Model}\label{SecII}
 
We begin with an idealized KC description, which captures the essential ingredients for the emergence of MZMs, and allows the definition of sweet-spots in a transparent manner. This simplified perspective serves as a benchmarking guide to address more realistic situations, where the explicit hybridization with mesoscopic SC hybrids and renormalization effects become relevant. To this end, we reformulate the problem at the GF level, which naturally incorporates these effects and allows us to identify both, the conditions for the appearance of zero-energy modes, and the resulting renormalized sweet-spot configurations. 
Importantly, this framework allows us to progressively relax overly-simplified assumptions and incorporate different microscopic degrees of freedom such as the quasiparticle continuum and the finite spin-splitting of subgap states. In this manner, the results of the ideal KC model emerge as limiting cases of a more general description that will be developed in the following sections.

In the idealized limit of a KC, a strong magnetic field fully polarizes the QD spins. These QDs are coupled via an intermediate central SC region, which mediates ECT and CAR, with effective amplitudes $t$ and $\Delta$ respectively, realizing a one-dimensional spinless p-wave SC that can host MZMs~\cite{Leijnse2012}. The total Hamiltonian $H_{kc}=H_{qd}+H_T$ of the ideal KC model takes the form
\be \label{minimalmodel}
H_{qd} = \sum_i^N \mu_{i} \, n_i \ , \; H_T = \sum_i^{N-1} t \, c^\dag_{i+1}c_i + \Delta \, c^\dag_{i+1}c^\dag_i + {\rm H.c.} \, ;
\ee 
where $i$ runs over the number $N$ of QDs of the KC and $\mu_{i}$ is the on-site energy, thereby controlling the degree of localization of the MZMs. We use  $n_{i}=c_{i}^\dag c_{i}$ as the number operator, with $c_{i}^\dag$ ($c_{i}$) representing local particle creation (annihilation) operators.

From the Hamiltonian for the ideal minimal chain ($N=2$), it is possible to compute the condition for the appearance of zero-energy modes, cf. Ref.~\cite{Alvarado2024} for details, given by
\be \label{toymodel_mu}
\mu_L \, \mu_R = t^2-\Delta^2 \, .
\ee 
When representing Eq.~\eqref{toymodel_mu} in the plane defined by the chemical potentials ($\mu_R, \, \mu_L$), the zero-energy condition describes two distinct solution branches,  one for each of the two possible sign configurations of the chemical potentials along the chain as in
\be \label{toymodel_B_main}
\mu_L^\pm  = \pm \frac{t^2-\Delta^2}{|\mu^\pm_R|} \, ,
\ee 
where the super-index in $\mu^\pm_j$ indicates the sign of the chemical potential. We label the branches $B_1$ and $B_2$ for the negative and positive solutions respectively, a classification that will prove especially useful when we introduce the mesoscopic description of the chain.

It should be reminded that for having Majoranas in the chain we have to simultaneously satisfy both having a zero mode, given by Eq.~\eqref{toymodel_B_main}, and that ECT and CAR amplitudes become equal $(t=\Delta)$. 
We define optimal sweet-spots as parameter configurations that maximize Majorana localization, while keeping the ground state (GS) at zero-energy. We quantify the degree of localization by means of the Majorana polarization~\cite{Bena2015, Bena2016, Glodzik2020, Aksenov2020, Souto2022, Leijnse2024, Ola2024, Samuelson2025} (MP), which describes the Majorana character of the wavefunction. We compute the excitation gap ($\delta E$) by diagonalizing the Hamiltonian in Eq.~\eqref{minimalmodel} and extract the MP (${\cal P}$) associated to the GS following~\cite{Bena2015, Bena2016, Glodzik2020, Ola2024}
\be 
{\cal P} =  \left[ \frac{2 \, u_L \, v_L}{ \big|u_L \big|^2 +  \big|v_L \big|^2} \right]  \left[ \frac{2 \, u_R \, v_R}{ \big|u_R \big|^2 +  \big|v_R \big|^2} \right]^* \, ,
\ee 
where $u_{j} \, , \, v_{j}$ are the electron and hole local components of the eigenstate associated with the lowest-energy solution for the left ($j=L$) and right ($j=R$) QDs. Here, $|{\cal P}| = 1$ signals well-separated MZMs with negligible spatial overlap.

\begin{figure}[t!]
\centering
\includegraphics[width=\columnwidth]{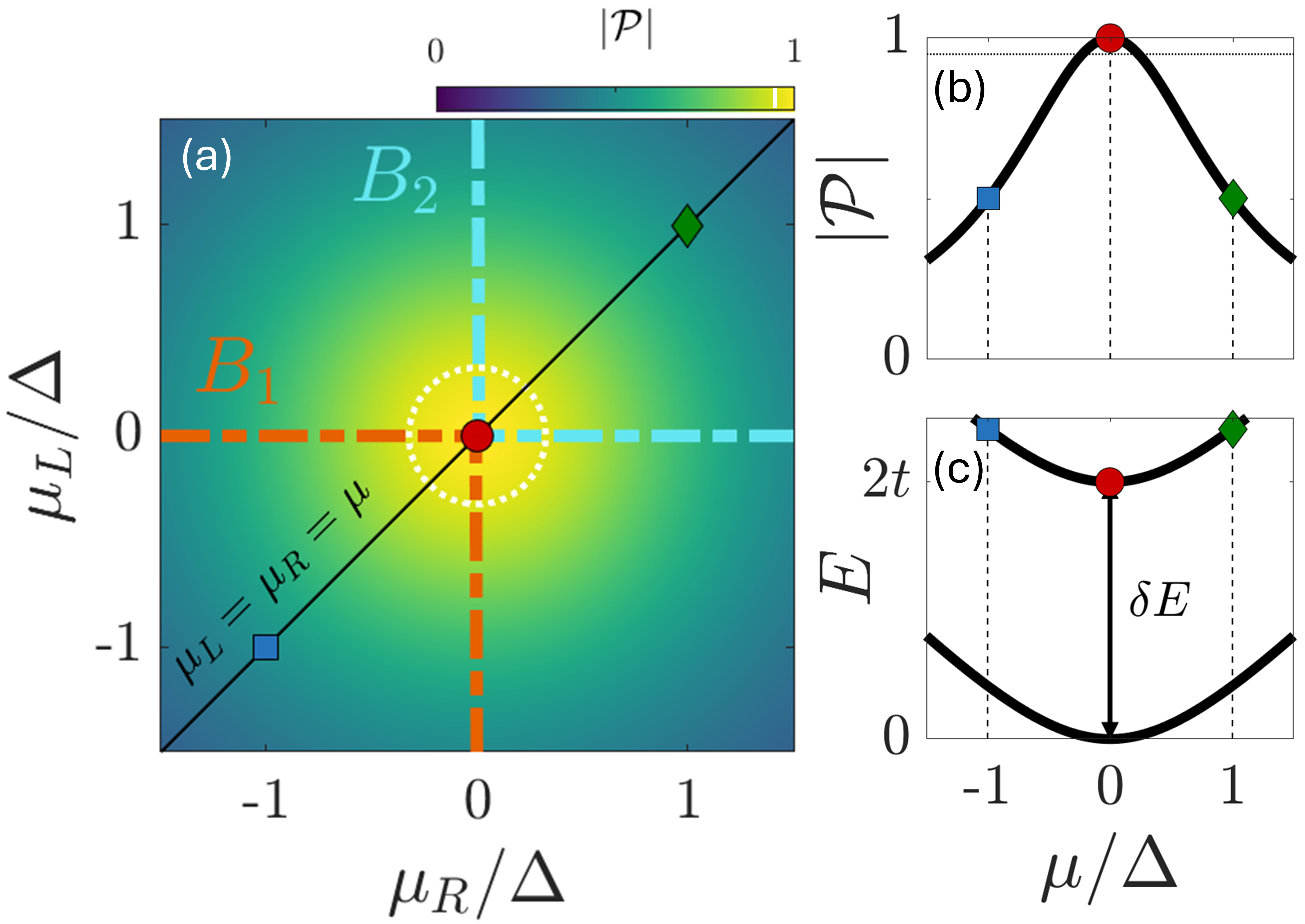}
\caption{Sweet-spot characterization of the ideal minimal KC model (N=2). (a) Majorana polarization ($|{\cal P}|$) as a function of the QD energies $\mu_L$ and $\mu_R$ for $t=\Delta$. Dot-dashed lines depict the zero-energy contours $\mu_L^\pm$ in Eq.~\eqref{toymodel_B_main} namely $B_1$ and $B_2$. The white dotted line denotes the $|{\cal P}|=0.95$ contour. Right panels: evolution of the MP (b) and the energy dispersion (c) following the black diagonal line in panel (a), corresponding to $\mu_L=\mu_R=\mu$. Markers highlight both the MP value and the associated energy gaps.
}
\label{fig2}
\end{figure}

Figure~\ref{fig2} illustrates the sweet-spot behavior of the idealized minimal KC. The main panel shows the MP for different values of the energy of the QDs, $\mu_{L/R}$, considering $t=\Delta$. The dot-dashed lines represent the two branches of zero-energy contours (blue and orange) defined by Eq.~\eqref{toymodel_B_main}. However, the MP landscape reveals that the sweet-spot occurs only at the intersection between branches at $\mu_L=\mu_R=0$, where the MP becomes maximum, indicating perfectly localized and non-overlapping MZMs, see Fig.~\ref{fig2}~(a). 

The right panels provide complementary insight of the physical behavior of the MP and the excitation gap for $\mu_L=\mu_R=\mu$. Fig.~\ref{fig2}~(b) shows that the MP peaks at the intersection between the two zero-energy contours in panel (a), and decreases smoothly away from it. 
Fig.~\ref{fig2}~(c) displays the dispersion of the energy levels when detuning the chemical potential with respect to the sweet-spot configuration at $\mu=0$. The two low energy states describe a minimum at the sweet-spot and increase parabolically.  In this ideal case, $\delta E$ is constant along this line. Together, these results demonstrate that both branches are physically equivalent and lead to identical sweet-spot properties. Importantly, Figs.~\ref{fig2}~(b,c) also highlight that the MP and the gap to excited states are not maximized for the same set of parameters. Therefore, Majorana localization and spectral protection represent competing goals, even in this idealized description.

The idealized Kitaev model provides a transparent framework for understanding the emergence of zero-energy modes and the definition of sweet-spots through the condition $t=\Delta$. However, in realistic QD-based implementations of KCs, both the effective ECT and CAR amplitudes originate from a proximitized SC segment of the chain that hosts subgap and above-gap states. As a result, the effective parameters entering the idealized minimal KC model acquire a nontrivial dependence on hybridization strength, Zeeman energy, and chemical potential.

\section{Mesoscopic Chain Models}\label{SecIII}

To describe microscopically the KC, we model a spinful chain (with $L$ sites) formed by QDs (odd sites) coupled through proximitized SC segments (even sites). The latter are described as semiconductor regions strongly coupled to a grounded SC lead where ABSs appear, see Fig.~\ref{fig1}~(a,b). The total Hamiltonian~\cite{Dominguez2016, Bordin2022, Souto2022, Liu2023, Miles2023, Souto2025} of the system is written as $H_{kc}=H_{0}+H_T+H_S$, where the local contribution in the semiconducting chain is given by 
\be 
H_{0} = \sum^L_{i=1} \sum_{\sigma} \, (\mu_{i}+s_\sigma  V_{z,i}) \; n_{i,\sigma} \, ,  
\ee 
where $i$ runs over fermionic sites. Here, $\mu_{i}$ is the on-site chemical potential, which can be tuned by independent external gates ($V_i$). The Zeeman energy is induced by a magnetic field $B$ applied along the $z$-axis, $V_{z,i}=\mu_B \, g_i \,B/2$, where $\mu_B$ is the Bohr magneton and $g_i$ is the site-dependent $g$-factor, which may be explicitly renormalized in proximitized regions of the chain. We use $s_{\up, \dw}=\pm1$, and $c_{i,\sigma}^\dag$ ($c_{i,\sigma}$) represents the particle creation (annihilation) local operators with spin $\sigma$ across the semiconducting chain.
The Coulomb repulsion is neglected in this work~\cite{Souto2022, GomezLeon2025, Buonemani2026}.

The tunneling contribution in the semiconducting chain between neighboring sites $H_T=H_t+H^{so}_t$, can be written as
\begin{subequations} 
\bea
H_{t} &=& \sum_{i=1}^{L-1}\sum_\sigma \, t \, \cos (\theta/2)\, c_{i+1,\sigma}^\dag c_{i,\sigma}  + {\rm H.c.} \, ,   \quad \\
H^{so}_{t} &=& \sum_{i=1}^{L-1}\sum_\sigma \, t \, \sin(\theta/2) \, s_\sigma \, c_{i+1,\sigma}^\dag c_{i,\bar{\sigma}} + {\rm H.c.} \, ; \quad
\eea
\end{subequations}
where $H_t$ describes spin-conserving tunneling processes, while $H^{so}_t$ accounts for the spin–orbit coupling (SOC) in the $y$-axis inducing spin-flip tunneling between sites, see Fig.~\ref{fig1}~(b). Here $t$ denotes the coupling strength, $\bar{\sigma}$ represents the spin opposite to $\sigma$, and $\theta$ is the SOC angle. We use $\theta = \pi/8$ as it gives SOC amplitudes approximately $t_{so} \sim 0.2\,t$, values commonly used in the literature that ensure that longer chains remain within the topological regime~\cite{Wimmer2025}. 

The SC contribution to the Hamiltonian $H_S$ has two terms: $H_{s,0}$ describes the grounded SC lead using an effective quasi-1D tight binding, where $\varepsilon_{k}$ is the kinetic dispersion, and $\Delta$ corresponds to the SC gap; whilst $H_{s,t}$ is the coupling between the SC and the even sites in the chain:
\begin{subequations} 
\bea
H_{s,0} &=& \sum_{{k}}\sum_\sigma \, \varepsilon_{k} \,d^\dagger_{{k}\sigma} d_{{k}\sigma} + \Delta \, d^\dagger_{{k}\up}d^\dagger_{-{k}\dw} + {\rm H.c.} \, ,\quad \\
H_{s,T} &=& \sum_{i=2j}^{L-1} \sum_{{k}} \sum_\sigma \, \lambda_s\, c_{i,\sigma}^\dag d_{k,\sigma}  + {\rm H.c.} \, ;
\eea
\end{subequations} 
where $d_{k,\sigma}^\dag$ ($d_{k,\sigma}$) represents the particle creation (annihilation) operators in the SC lead with momentum $k$. 
The coupling $\lambda_s$ gives rise to the effective hybridization parameter in the proximitized region $\Gamma_s=\lambda_s^2/t_s$, where $t_s$ sets half the bandwidth of the SC lead.
Throughout this work, we neglect the Zeeman field in the SC lead, as it is usually much smaller than the one in the semiconducting chain, where QDs and ABSs are formed. In the following, for simplicity, we focus on symmetric configurations with equal chemical potential $\mu$ at all odd sites (and equal $\mu_c$ at all proximitized even sites).

The microscopic degrees of freedom associated to the proximitized regions of the chain -- such as the quasiparticle continuum and the possible spin-splitting of the ABS -- are incorporated using the boundary Green’s function (bGF) formalism~\cite{Cuevas1996, Zazunov2016, Alvarado2018, Alvarado2020, Alvarado2022, Alvarado2025}. Thus, we obtain a minimal effective model preserving all the spinful correlations stemming from the proximitized region of the chain up to second order in perturbation theory.
To characterize the gap to excited states and MP, we diagonalize an effective Hamiltonian in which the SC lead is incorporated through a discretized tight-binding representation, cf. the Appendices for details.

It should be noted that the mesoscopic description of the semiconducting chain with proximitized regions maps onto the KC model in the Kitaev limit ({\it i.e.}, $\Gamma_s, V_z > \Delta \gg t$) for an arbitrary number of sites.

\subsection{parity-crossings of the Hybrid Segment} \label{SecIIIa}

Building on the mesoscopic description of the proximitized region of the chain discussed above, we introduce the spectral properties of an ABS in the isolated mesoscopic SC ({\it i.e.,} uncoupled from the effective QDs).
Within the GF framework, the energies of the ABSs are obtained from the poles of the bGF. This allows us to determine when the ABS approaches zero-energy and derive the condition under which it becomes resonant, cf. App.~\ref{App_A} and Ref.~\cite{Aguado2015} for details, 
\be \label{eq_sd_QPT_main}
\mu^\pm_c=\pm \sqrt{V_{z,c}^2-\Gamma_s^2} \, .
\ee
This condition defines the QPT boundary between an even-parity GS, where the ABS in the isolated hybrid section is spin-split by the presence of the Zeeman energy, and a odd-parity one, where the subgap state becomes spin-polarized by the dominant Zeeman energy, see Fig.~\ref{fig1}~(c). The transition is controlled by the competition between the Zeeman energy and the effective ABS energy $\epsilon_A\approx\sqrt{\mu_c^2 + \Gamma_s^2}$, as illustrated in Fig.~\ref{fig1}~(d). Notice that this criterion is analogous to the topological transition for MBS in nanowires~\cite{Aguado2015}, where $\Gamma_s$ plays the role of an effective gap within a low energy description~\cite{Pita2026}. 
As we demonstrate below, entering the odd-parity spin-polarized regime of the ABS in the hybrid has profound consequences for the low-energy spectrum of the KC and MZMs wavefunction properties, leading to qualitative modifications in the sweet-spot behavior and its stability. 

We now study the spectral properties of the mesoscopic hybrid in the chain when the ABSs are renormalized by the coupling with the adjacent QDs. In this way, we determine the conditions under which the chain hosts zero modes governed by the properties of the hybrid region. Throughout this work, we focus on the minimal KC ($L=3$).

Our microscopic description includes all the renormalization  effects in the hybrid section by integrating out the adjacent QDs using GFs techniques. For analytical insight, we incorporate the microscopic physics of the adjacent QDs via a perturbative expansion of the effective renormalization of the proximitized region in the weak-coupling limit ({\it i.e.}, to second order in the coupling strength $t$), see App.~\ref{App_A} for details. The leading terms appear as corrections to the Zeeman field and chemical potential of the ABS, following $V^\prime_{z,c} = V_{z,c}+V_{z,c}^{(2)}$ and $\mu_c^\prime = \mu_c+\mu_c^{(2)}$ respectively, where 
\begin{subequations} \label{induced_Zeeman_main}
\bea 
\mu_c^{(2)} &=& - \frac{2\,t^2\,\mu}{\mu^2-V_z^2} \, , \\
V_{z,c}^{(2)} &=& \frac{2\,t^2\,V_z \cos(\theta)}{\mu^2-V_z^2} \, .
\eea
\end{subequations} 
To obtain the condition to find zero energy modes in the chain we incorporate the corrections above into Eq.~\eqref{eq_sd_QPT_main}. Alternatively, it is possible to go beyond the weak coupling approximation by accounting for the renormalization effects up to infinite order via the total GF of the system projected onto the proximitized region, cf. App.~\ref{App_A} for details. From this result, which can be computed  analytically, we derive the corresponding characteristic polynomial and determine the conditions under which zero-energy modes emerge in the chain, given by
\begin{widetext} 
\be \label{Eq_muc_main}
\mu^\pm_c = \frac{2\,t^2\mu \pm\sqrt{[2 V_z \, t^2 \cos(\theta) -  (\mu^2-V_z^2)(\Gamma_s-V_{z,c})][2 V_z \, t^2 \cos(\theta) + (\mu^2-V_z^2)(\Gamma_s+V_{z,c})]}}{\mu^2-V_z^2} \, .
\ee 
\end{widetext}
In the Kitaev-chain limit ({\it i.e.}, $V_z, \, \Gamma_s > \Delta \gg t$), Eq.~\eqref{Eq_muc_main} reduces to the isolated ABS parity-crossing condition in Eq.~\eqref{eq_sd_QPT_main} describing a trivial zero mode in the chain~\cite{Zhang2026}. Moreover, similar to Eq.~\eqref{eq_sd_QPT_main} for the SC hybrid, Eq.~\eqref{Eq_muc_main} defines the zero-energy contours in the chemical potential phase diagram for a mesoscopic KC. The different signs of the square root in Eq.~\eqref{Eq_muc_main} reveal the existence of two distinct solution branches, in parallel to the ideal KC in Sec.~\ref{SecII}. In addition, when a finite Zeeman field is present at the mesoscopic hybrid $V_{z,c}\neq0$, the zero-energy contours become asymmetric with respect to $\mu_c =0$. This asymmetry plays a key role in determining the localization properties and excitation gap of the Majorana modes along each branch.

\subsection{Sweet-Spot Characterization} \label{SecIIIb}

So far, we have explored how coupling to the QDs renormalizes the properties of the hybrid region, leading to the emergence of zero modes. Here, we adopt the opposite perspective: how the SC hybrid segments effectively couple the QDs through both normal (ECT) and Andreev processes (CAR).

As discussed before, to define the sweet-spot we require the ECT and CAR amplitudes to be equal. However, it is now essential to account for renormalization effects arising from the microscopic degrees of freedom of the proximitized region. To this end, we integrate out the SC lead using the aforementioned non-equilibrium bGF techniques, capturing the effective QD–QD coupling mediated by the ABS. For analytical insight, we incorporate the microscopic physics of the SC region via a perturbative expansion of the effective coupling between QDs, see App.~\ref{App_B} for details. This perturbative treatment yields an effective Kitaev chain for the low-energy spin sector in the weak-coupling limit ({\it i.e.}, to second order in the coupling strength $t$), with the CAR and ECT amplitudes $|\Delta^{(2)}|$ and $|t^{(2)}|$ respectively, appearing as the leading-order terms:
\begin{subequations}\label{sweet_spot_Vz_main}
\bea 
|\Delta^{(2)}| &=&  \frac{t^2 \, \Gamma_s \, \sin(\theta)}{|\mu^2_c + \Gamma_s^2-V_{z,c}^2|} \, , \label{sweet_spot_Vz_main_a} \\
|t^{(2)}| &=&  \frac{t^2 \, |V_{z,c}+\mu_c \, \cos(\theta)|}{|\mu^2_c + \Gamma_s^2-V_{z,c}^2|} \, ; \label{sweet_spot_Vz_main_b}
\eea
\end{subequations}
where we have assumed $V_z, \, V_{z,c} > 0$.
It should be noted that, because the denominator in Eqs.~\eqref{sweet_spot_Vz_main_a} and \eqref{sweet_spot_Vz_main_b} reproduces the parity-crossing condition in Eq.~\eqref{eq_sd_QPT_main}, the effective low-energy description diverges when the ABS approaches zero energy.
Moreover, the gap to excited states is expected to grow following $\delta E \approx 2\,|\Delta^{(2)}|$. Therefore, the microscopic degrees of freedom of the proximitized region give rise to new behavior, exemplified by a Zeeman-driven parity-crossing, which is entirely missed in simplified treatments that neglect such microscopic details. The QPT of the ABS emerges as a point of special interest, as we discuss in what follows.

From the renormalized condition $|\Delta^{(2)}|=|t^{(2)}|$, we obtain an approximation for the chemical potential in the SC hybrid section at the sweet-spot (bar notation), fulfilling 
\be \label{mu_c_vzneq0_main}
\bar{\mu}^\pm_c = \pm \Gamma_s \tan(\theta) - V_{z,c}\sec(\theta) \, .
\ee 
By substituting this renormalized condition into Eq.~\eqref{Eq_muc_main} we can determine the local chemical potential of the QDs that simultaneously satisfies both, the conditions for a zero-energy GS and for localized Majoranas,
for a given parameter set $x_i = \{t, V_z, \, \Gamma_s\}$. Since the sweet-spot occurs when the QDs are on resonance~\cite{Wimmer2025, Zhang2026}, we linearize the chemical potential around $\bar{\mu}^\pm = V_z \pm \delta\mu^\pm$ to gain some analytical insight. From this, we obtain the sweet-spot condition in the weak-coupling limit including asymmetric contributions associated to the Zeeman field in the hybrid section. After some algebra, we get
\be \label{renorm_mu_Vz_main}
\delta \mu^\pm = \frac{t^2 \sin (2\theta)}{2 [\Gamma_s \mp V_{z,c}\sin(\theta)]} \, ,
\ee 
where we associate the solution with negative (positive) super-index to branch $B_1$ ($B_2$). This contribution originates from the mesoscopic corrections to the low-energy physics of the resonant QDs induced by the SC hybrid, namely, the renormalization of the Zeeman field and chemical potential, $V^\prime_{z} = V_{z}+V_{z}^{(2)}$ and $\mu^\prime = \mu+\mu^{(2)}$ respectively; thus $\delta\mu=V_z^{(2)}-\mu^{(2)}$, cf. App.~\ref{App_B} for details.

\subsection{Spin-Degenerate ABS: weak-coupling} \label{SecIIIc}

To establish a reference regime and provide a direct comparison with previous works, we first analyze the limit of strong screening of magnetic fields by the SC in the hybrid segment. This regime, frequently considered in the literature, corresponds to the case where the $g$-factor in the proximitized section of the chain is effectively suppressed, leading to $V_{z,c} = 0$. 
For the weak-coupling limit ($t^2\rightarrow0$), we recover the effective ECT and CAR amplitudes reported in Ref.~\cite{Wimmer2025}
\begin{subequations}\label{sweet_spot_no-Vz_main}
\bea 
|\Delta^{(2)}| &=&  \frac{t^2 \, \Gamma_s \, \sin(\theta)}{\mu^2_c + \Gamma_s^2} \, , \label{sweet_spot_noVz_main_a} \\
|t^{(2)}| &=&  \frac{t^2 \, \mu_c \, \cos(\theta)}{\mu^2_c + \Gamma_s^2} \, ; \label{sweet_spot_noVz_main_b}
\eea
\end{subequations}
which, crucially, do not capture the effects discussed above due to parity-crossings and overall gap enhancement, since they neglect Zeeman effects in the hybrid.

Having determined the condition for sweet-spots, which ensures the existence of highly localized MZMs, we now seek physical insight into the gap to excited states. Along with the MP, the gap constitutes a key target property to optimize in optimal sweet-spot configurations. To this end, we obtain a simplified expression for the excitation gap in the Kitaev limit by approximating $\delta E \approx 2\,|\Delta^{(2)}|$ in Eq.~\eqref{sweet_spot_Vz_main}, together with Eq.~\eqref{mu_c_vzneq0_main}, and assuming $V_{z,c}=0$; namely, $|\bar{\mu}_c| = \Gamma_s \tan(\theta)$, which gives
\be \label{eq_gap_main}
\delta E \approx \frac{t^2}{\Gamma_s} \sin (2\theta) \, \cos (\theta) \,  \, .
\ee
Thus, in the Kitaev limit ({\it i.e.}, $V_z, \, \Gamma_s >\Delta\gg t$), the two branches become equivalent and the excitation gap remains small, scaling quadratically with the coupling~\cite{Liu2023, Miles2023, Zatelli2024} as $\delta E \propto t^2/\Gamma_s$. 

\subsection{Spin-Degenerate ABS: strong-coupling} \label{SecIIId}

As the coupling strength increases ({\it i.e.}, $V_z,, \Gamma_s > \Delta \gtrsim t$), the microscopic degrees of freedom of the SC hybrid become increasingly relevant. To account for them, we recall the total  GF of the system. In particular, the characteristic polynomial, which was used to derive the zero-energy condition in Eq.~\eqref{Eq_muc_main}, can also be employed to extract the poles of the system and determine the excitation gap. 

However, since the degree of the characteristic polynomial is typically too high to admit an analytical solution, we resort to an approximation by performing a second-order Taylor expansion. From this, we obtain a zero mode together with two low-energy poles associated with the first excited states. This simplification, while approximate, is sufficient to accurately capture the excitation gap in this limit.
Moreover, these different poles are associated to the two solution branches $B_1$ and $B_2$, cf. App.~\ref{App_B} for details. Therefore, the excitation gap can be approximated as
\be \label{approx_dE_main}
\delta E^\pm = \frac{2 \, t^2 \, V_z \, \Gamma_s \, \sin(\theta)}{ \Gamma_s^2 \, V_z \, \sec^2(\theta) + t^2\, [2\,V_z\mp \, \Gamma_s \tan^2(\theta)]} \, \, ,
\ee 
where again we associate the solution with negative (positive) super-index to branch $B_1$ ($B_2$). 
Thus, as the coupling strength increases, the two branches show different values for the excitation gap; in particular, branch $B_2$ displays a large enhancement of the gap due to the minus sign in the denominator of Eq.~\eqref{approx_dE_main}. 

This phenomenology, completely absent in the Kitaev limit in Eq.~\eqref{eq_gap_main}, has been associated to an induced Zeeman energy in the ABS~\cite{Liu2022, Liu2023}. To check this physical interpretation, we consider all the renormalization effects described by Eq.~\eqref{induced_Zeeman_main} when assuming the sweet-spot condition in Eq.~\eqref{renorm_mu_Vz_main}, obtaining
\begin{subequations} 
\bea 
\bar{\mu}^{(2)}_{c}&\approx&-\Gamma_s \csc(\theta)\sec(\theta) - \frac{t^2}{V_z} \, , \label{induced_Zeeman_sweetspot_main_A} \\
\bar{V}^{(2)}_{z,c}&\approx&\Gamma_s \csc(\theta) \, . \label{induced_Zeeman_sweetspot_main_B}
\eea
\end{subequations}
Notably, it is possible to map this renormalized spin-degenerate ABS in the strong-coupling limit, into the weak-coupling limit of an ABS with non-zero Zeeman energy, cf. App.~\ref{App_B}. Using the substitution $\mu_c^\prime = \mu_c+\mu^{(2)}_{c}$, and $V^\prime_{z,c} = V^{(2)}_{z,c}$, into Eq.~\eqref{mu_c_vzneq0_main}, we get the corrected sweet-spot condition for the hybrid chemical potential as 
\be \label{renorm_muc_vz0_main}
\bar{\mu}^\pm_c \approx \frac{t^2}{V_z} \pm \Gamma_s \tan(\theta)\, .
\ee 
As demonstrated, the induced Zeeman field alone~\cite{Liu2022, Liu2023, Wimmer2026} is insufficient to correctly describe the strong-coupling. Only by considering all renormalization effects in the hybrid, namely, the correction to the chemical potential in Eq.~\eqref{induced_Zeeman_sweetspot_main_A} and the induced Zeeman field in  Eq.~\eqref{induced_Zeeman_sweetspot_main_B}, 
we reproduce the shift of $\bar{\mu}_c$ towards positive values.

\subsection{Spin-Split ABS: weak-coupling} \label{SecIIIe}

We now discuss in detail a mesoscopic chain including explicitly spin-splitting effects given by a finite $V_{z,c}$ in the SC hybrid section. For simplicity, we assume a homogeneous $g$-factor across the chain. Thus, the renormalization of the Zeeman energy in the ABS is mediated solely by the hybridization with the SC lead. 
Particularizing to $V_{z,c} = V_z > 0$, the sweet-spot condition satisfies Eq.~\eqref{mu_c_vzneq0_main} and Eq.~\eqref{renorm_mu_Vz_main}. For the latter case, we note that increasing either $\Gamma_s$ or $V_{z,c}$ drives $\delta\mu^-$ toward zero for branch $B_1$. However, for branch $B_2$, the correction $\delta\mu^+$ can become negative and even diverge. 

To further elucidate this behavior, we substitute the sweet-spot condition over $\bar{\mu}_c$ in Eq.~\eqref{mu_c_vzneq0_main}, into the parity-crossing of the isolated ABS point in Eq.~\eqref{eq_sd_QPT_main}, to see the effect associated to the QPT, which yields
\be \label{divergent_point_main}
\Gamma_s \mp V_{z,c}\sin(\theta) =  0 \, .
\ee
This expression appears in the denominator of the sweet-spot condition for the QD chemical potential in Eq.~\eqref{renorm_mu_Vz_main}, implying that the linear approximation breaks down as the system approaches the parity crossing.
Moreover, it should be pointed out that the QD-ABS-QD mesoscopic chain behaves as an effective 3-site chain when the ABS energy is close to resonance in the weak-coupling regime~\cite{Miles2023}. Precisely, the condition obtained in  Eq.~\eqref{divergent_point_main} for the parity-crossing coincides with the condition for equal ECT and CAR amplitudes for the effective 3-site chain configuration of the QD-ABS-QD system when the ABS is on resonance~\cite{Miles2023} ({\it i.e.}, by imposing in Eq.~\eqref{mu_c_vzneq0_main} that $\epsilon_A=\sqrt{\bar{\mu}_c^2+\Gamma_s^2}-V_{z,c}=0$). 

By considering the approximation $\delta E \approx 2\,|\Delta^{(2)}|$  under the sweet-spot condition in Eq.~\eqref{sweet_spot_Vz_main}, we obtain an expression for the gap to excited states valid in the weak-coupling and even-parity regime
\be \label{dE_Vzneq0_main}
\delta E^\pm \approx \frac{t^2 \, \Gamma_s  \sin (2\theta) \, \cos(\theta)}{[\Gamma_s \mp V_{z,c}\sin(\theta)]^2}  \,  ,
\ee 
showing two different dependencies with $V_{z,c}$ and $\Gamma_s$ for the different branches. Thus, an optimal value for the gap is expected when reaching the odd-parity spin-polarized regime due to the possible divergence of the denominator at the transition for $B_2$, while the gap associated to $B_1$ is expected to monotonically decrease. These results illustrate that the QPT of the isolated ABS emerges as a special point in parameter space, with deep implications in the properties of the sweet-spots.

\subsection{Spin-Split ABS: strong-coupling} \label{SecIIIf}

The analysis of the weak-coupling regime reveals that the effective low-energy theory develops divergences as the system approaches the parity-crossing point of the isolated ABS. These divergences originate from the resonant character of the ABS, signaling the need for a more complete description that incorporates the renormalization effects induced in the hybrid section. We address this regime by extending the treatment beyond the weak-coupling approximation, allowing us to capture both the strong-coupling limit and the spin-polarized odd-parity regime on equal footing.

We map the renormalized spin-split ABSs in this strong-coupling regime by introducing the renormalization of both the chemical potential and the Zeeman energy in Eq.~\eqref{induced_Zeeman_main}, using the renormalization substitution $\mu_c^\prime = \mu_c+\mu^{(2)}_{c}$, and $V^\prime_{z,c} = V_{z,c}+V^{(2)}_{z,c}$. This renormalized description regularizes the divergences of the ECT and CAR amplitudes $|t^{(2)}|$ and $|\Delta^{(2)}|$ in Eq.~\eqref{sweet_spot_Vz_main} at the QPT, cf. App.~\ref{App_B} for details. 

By considering the approximation $\delta E \approx 2\,|\Delta^{(2)}|$ including the renormalization substitution under the sweet-spot condition $\{\bar{\mu}_c,\, \bar{\mu}\}$ in Eqs.~\eqref{mu_c_vzneq0_main} and \eqref{renorm_mu_Vz_main}, we obtain an expression for the gap to excited states valid beyond the weak-coupling and even-parity regime 
\be \label{dE_Vzneq0_sc_main}
\delta E \approx \frac{t^2 \, \Gamma_s \, \sin(\theta)}{\left| \bar{\mu}_c^2 + \Gamma_s^2-V_{z,c}^2-\frac{4t^2\,V_z[\bar{\mu}_c+V_{z,c}\cos(\theta)]}{\bar{\mu}^2-V_z^2}\right|} \, . 
\ee 
Moreover, by considering the aforementioned renormalization substitution into Eq.~\eqref{mu_c_vzneq0_main}, we obtain an approximation for the hybrid's chemical potential up to the leading order following
\be \label{mu_c_vzneq0_t_main}
\bar{\mu}^\pm_c \approx \frac{t^2}{V_z} \pm \Gamma_s \tan(\theta) - V_{z,c}\sec(\theta) \, .
\ee 
By substituting this renormalized condition into Eq.~\eqref{Eq_muc_main}, and linearizing the chemical potential around the Zeeman energy, we obtain
\be \label{renorm_mu_Vz_t_main}
\delta \mu^\pm \approx  \frac{t^2 \sin (2\theta)}{2 [\Gamma_s \mp V_{z,c}\sin(\theta)]} - \frac{t^3 \cos^2 (\theta) \sqrt{V_z \Gamma_s \sin(2\theta)}}{ [\Gamma_s \mp V_{z,c}\sin(\theta)]^2} \, .
\ee 
This higher-order correction becomes essential for branch $B_2$, where the leading order expression for $\delta\mu^+$ in Eq.~\eqref{renorm_mu_Vz_main} diverges upon approaching the spin-polarized odd-parity regime of the ABS. In contrast, the renormalized expression provides a controlled description across the parity-crossing point and into the doublet regime.

\begin{figure*}[t!]
\centering
\includegraphics[width=\textwidth]{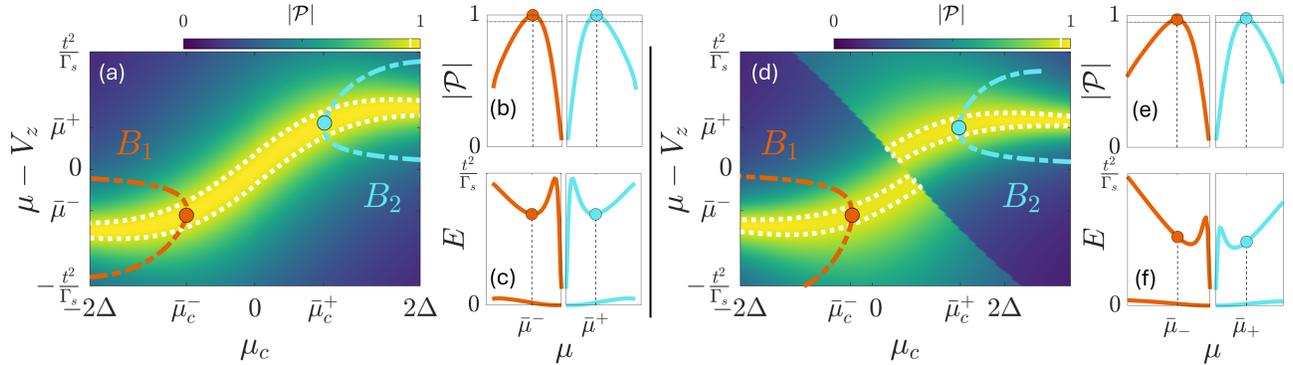}
\caption{Sweet-spot characterization routine of the mesoscopic minimal KC assuming $V_{z,c} = 0$. Left (Right) panels represent the weak (strong) coupling limit assuming $t=0.1$ ($t=1$) in $\Delta$ units. Main plots (a,d): MP $|{\cal P}(x_i)|$ as a function of $\mu$ and $\mu_c$ for the parameter set $x_i=\{t, \, V_{z}, \, \Gamma_{s} \}$ in the weak (a) and strong (d) coupling respectively. White dotted lines denote the $|{\cal P}|=0.95$ contour. Dot-dashed lines, $B_1$ and $B_2$, follow the zero-energy contours given in Eq.~\eqref{Eq_muc_main}. Orange and cyan markers indicate the sweet-spot for each of the branches. The marks $\bar{\mu}^\pm_c$ in the $y$-axis are given by Eqs.~\eqref{mu_c_vzneq0_main} and \eqref{renorm_muc_vz0_main} for the weak and strong-coupling respectively, while marks $\bar{\mu}^\pm$ in the $x$-axis are given by Eq.~\eqref{renorm_mu_Vz_main}.
Side panels: evolution of the MP and the gap dispersion for the different zero-energy branches in the weak (b,c) and strong (e,f) coupling regimes. Markers highlight both the maximum-polarization points and the associated energy gaps. We use $V_{z}=2$, $\Gamma_{s}=2$ in $\Delta$ units. 
}
\label{fig3}
\end{figure*}

\section{Numerical Results}\label{SecIV}

Having established the analytical limits of the hybrid for spin-degenerate and spin-split ABS configurations, in both the weak- and strong-coupling regimes, we now present full numerical results following the same subsection structure. Specifically, we explore the rich phenomenology of the Majorana polarization and the gap to excited states across parameter space along distinct branch trajectories, cf. App.~\ref{App_C} for details. In doing so, our results demonstrate the important role that considering microscopic effects on the hybrid segment has on the physics of the chain, well beyond the ideal Kitaev description in Fig.~\ref{fig2}.

\subsection{Spin-Degenerate ABS}\label{SecIVa}

We first study the sweet-spot properties in the limit where the Zeeman field in the central SC region is fully screened ($V_{z,c}=0$). Using the analytical expressions derived in Secs.~\ref{SecIIIc} and \ref{SecIIId}, we analyze how the renormalized sweet-spot conditions and excitation gap evolve from the weak- to the strong-coupling regime. This allows us to assess the validity of the perturbative treatment and to identify deviations from the ideal Kitaev limit that arise from microscopic degrees of freedom of the SC hybrid as the coupling between QDs is increased.

To characterize the sweet-spot properties, Fig.~\ref{fig3} shows the MP as a function of $\mu$ and $\mu_c$ for representative weak- ($t=0.1 \,\Delta$) and strong-coupling ($t=1\, \Delta$) regimes. The sweet-spots correspond to the MP maxima along the zero-energy contours. In the weak-coupling limit, Fig.~\ref{fig3}~(a), these maxima appear symmetrically in $\mu_c$ following Eq.~\eqref{mu_c_vzneq0_main}, and around the resonant QD level $(\mu - V_z)$, with amplitudes bound by $\delta\mu < t^2/\Gamma_s$, consistent with the linear approximation in Eq.~\eqref{renorm_mu_Vz_main}.  The evolution of the MP and the gap to excited states along the zero-energy contour in Eq.~\eqref{Eq_muc_main} for the two branches are shown in Fig.~\ref{fig3}~(b,c). These sweet-spots showing maximum MP along the zero-energy contour, see panel (b), coincide with a local minima of the energy gap in panel (c). For small deviations around the sweet-spot, the gap to excited states exhibits the expected parabolic dispersion. Consequently, in the small-coupling regime, the two branches remain equivalent as for the ideal KC model. 

As the coupling of the outermost QDs to the SC hybrid segment increases, however, the two branches become progressively nonequivalent. For $t \lesssim \Delta$, a pronounced asymmetry in $\mu_c$ emerges due to the combined effects of the induced Zeeman field and the renormalization of the chemical potential in the ABS, as shown in Fig.~\ref{fig3}~(d) and captured by Eq.~\eqref{renorm_muc_vz0_main}. Furthermore, Figs.~\ref{fig3}~(e,f) reveal that both the MP and the excitation gap between the two branches become non-equivalent.

\begin{figure*}[t!]
\centering
\includegraphics[width=0.8\textwidth]{FIG4.png}
\caption{Evolution of MP $|\cal P|$ and the gap to excited states, $\delta E$, at the sweet-spot  with various parameters for the two distinct branches, assuming $V_{z,c}=0$. 
Blue (red) lines correspond to $|\cal P|$ ($\delta E$).  Solid (dashed) lines indicate branch $B_1$ ($B_2$). 
(a) Dependence on the QD coupling strength $t$. 
(b) Dependence on the hybridization with the SC ($\Gamma_s$). 
(c) Dependence on the Zeeman energy ($V_z$). Black dot-dashed line represents the perturbative approximation for the gap in Eq.~\eqref{approx_dE_main}.
Unless stated otherwise, parameters are set to $\{ t = 0.5,\; V_z = 2,\; \Gamma_s = 2 \}$ in $\Delta$ units.
}
\label{fig4}
\end{figure*}

For each parameter set $x_i=\{t, \, V_{z}, \, \Gamma_{s} \}$ we can evaluate the zero-energy degeneracy lines as in Fig.~\ref{fig3} (dashed curves), select the points where the MP reaches its maximum value (marked by the dots in Fig.~\ref{fig3}) and define the sweet-spot characteristics for each branch, namely the MP and the excitation gap. Figure~\ref{fig4}~(a) shows the evolution of these quantities as a function of the inter-dot coupling strength $t$. In the weak-coupling regime, $t \ll \Delta$, the system is within the Kitaev limit, where the sweet-spots have similar MP and gap to excited states, reflecting the symmetry between the two branches. Besides, the excitation gap follows the parabolic dispersion predicted by Eq.~\eqref{eq_gap_main}, confirming the validity of the perturbative approximation.

As the system moves away from the ideal Kitaev limit and approaches the strong-coupling regime, $t \sim \Delta$, an increasing asymmetry between the sweet-spots in the two branches emerges: both feature different MP and $\delta E$, consistent with Eq.~\eqref{approx_dE_main} for $t \lesssim \Delta$. In particular, the MP decreases rapidly~\cite{Liu2023, Wimmer2025, Sanches2025, Sanches2026} and asymmetrically with increasing $t$. As a result, the sweet-spot on branch $B_2$, despite exhibiting smaller values of $\delta E$, maintains significantly higher MP, and therefore, becomes the most favorable configuration. 

Figure~\ref{fig4}~(b) shows the evolution as a function of the coupling to the SC lead. We observe a clear threshold at $\Gamma_s \gtrsim t$, above which, sufficiently large MP values are obtained to support well-defined MZMs. Moreover, as described by Eq.~\eqref{eq_gap_main}, the excitation gap is inversely proportional to $\Gamma_s$. 
For strong-coupling to the SC lead, the system approaches the Kitaev limit, in which the two branches become equivalent again, see dot-dashed line. Finally, upon increasing the Zeeman energy, Fig.~\ref{fig4}~(c), the gap $\delta E$ saturates for large spin polarizations in the outer QDs. Thus, increasing $V_z$ and $\Gamma_s$ generally improves the MP but does not enhance $\delta E$ beyond a certain threshold, while increasing the inter-dot coupling $t$ increases $\delta E$ at the expense of reducing the MP. This highlights the complicated interplay between the different parameters on the sweet-spot properties.

These trends can be understood from simple physical considerations. First, in the ideal Kitaev model the excitation gap at the sweet-spot follows $\delta E\approx2\,|t^{(2)}|$, so increasing the coupling $t$ enhances the gap to the excited states quadratically. However, larger $t$ also strengthens the hybridization between the outer QDs and the ABS, de-localizing the MZM and moving the system away from the Kitaev limit. Second, increasing the hybridization with the SC lead $\Gamma_s$ initially improves both $\delta E$ and MP by enhancing the superconducting character of the proximitized region of the chain, see Fig.~\ref{fig4}~(b). In the opposite limit $\Gamma_s \gg t$, however, the ABS becomes confined within the SC lead and weakly coupled to the outer QDs, recovering the perturbative regime and reducing the gap. Finally, increasing the Zeeman field $V_z$ enlarges the separation between opposite-spin states, improving MP and $\delta E$ until the outer QDs become fully spin polarized and the gap saturates.

\begin{figure}[b!]
\centering
\includegraphics[width=\columnwidth]{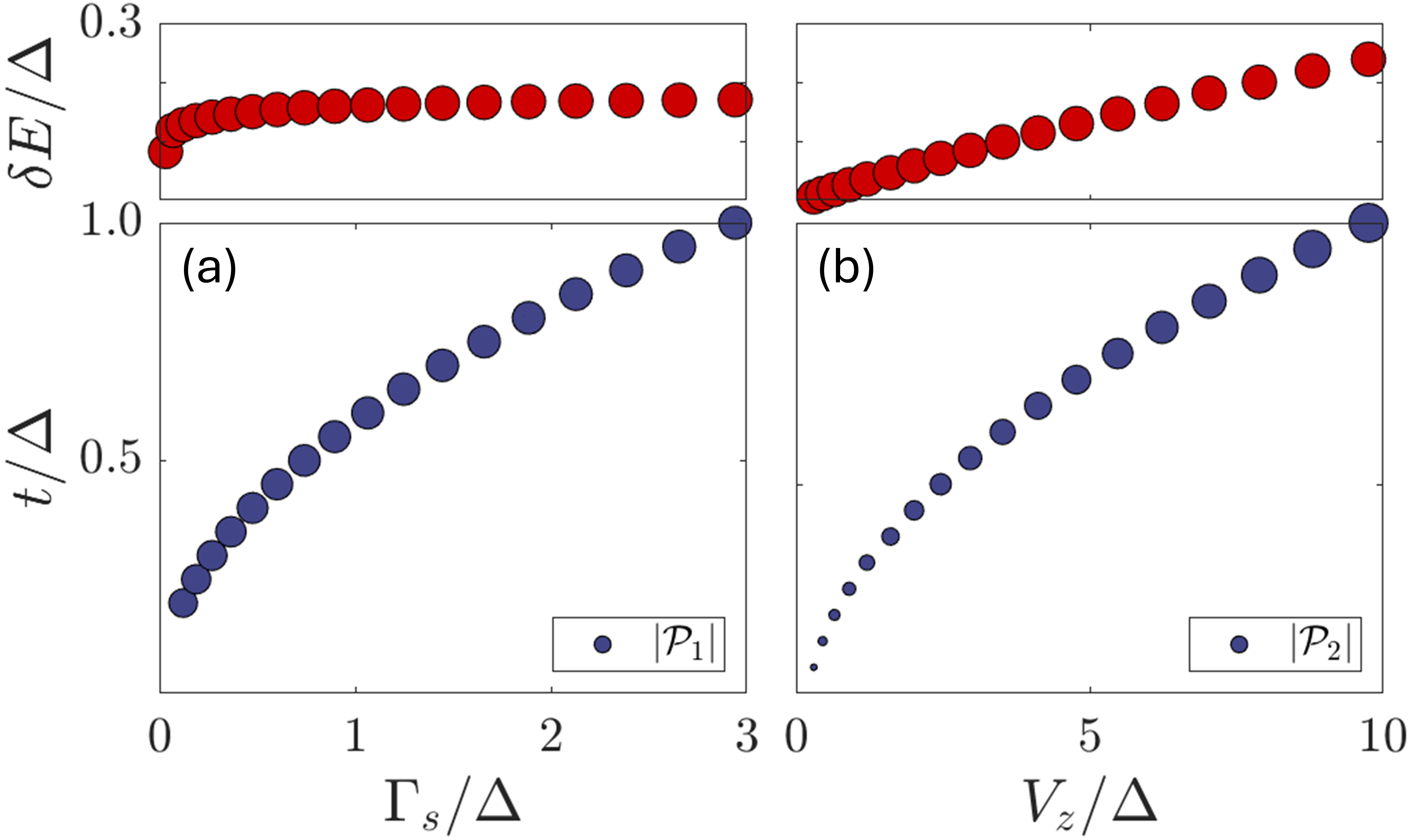}
\caption{Evolution of constant-value MP (iso-MP) in the $(\Gamma_s, \,t)$ and $(V_z,\,t)$ parameter planes respectively for $V_{z,c}=0$. 
The main panels show the evolution of the iso-MP contours. For illustration purposes, we use $|\mathcal{P}_1| = 0.99$ in panel (a), and 
$|\mathcal{P}_2| = 0.999$ in panel (b), although similar qualitative behaviors are observed for different choices of MP. The size of the markers reflects the magnitude of $\delta E$, shown explicitly in the upper panels. We use $V_z = 2$ (left panels) and $\Gamma_s= 2$ (right panels) in $\Delta$ units.
}
\label{fig5}
\end{figure}

A systematic analysis of the aforementioned competition between parameters can be used to achieve sweet-spot optimization ({i.e.}, an enhancement of the excitation gap while maintaining a constant MP). To this end, in Fig.~\ref{fig5} we analyze the evolution of constant value MP contours (iso-MP) for increasing values of the coupling strength, including corrections due to hybridization with the SC lead or Zeeman field for the left/right set of panels respectively. Figure~\ref{fig5}~(a) shows the evolution of the iso-MP, where the size of the markers reflects the magnitude of $\delta E$, as explicitly shown in the upper panel. The interplay between $t$ and $\Gamma_s$ leads to a saturation of the gap along these iso-MP contours. Consequently, increasing the coupling strength to the SC beyond a certain threshold does not yield any improvement of the sweet-spot. This behavior follows from the approximate relation $t \propto \sqrt{\Gamma_s}$ for the iso-MP lines, which, when combined with Eq.~\eqref{eq_gap_main}, results in an approximately constant excitation gap following the iso-MP contour.

\begin{figure*}[t!]
\centering
\includegraphics[width=\textwidth]{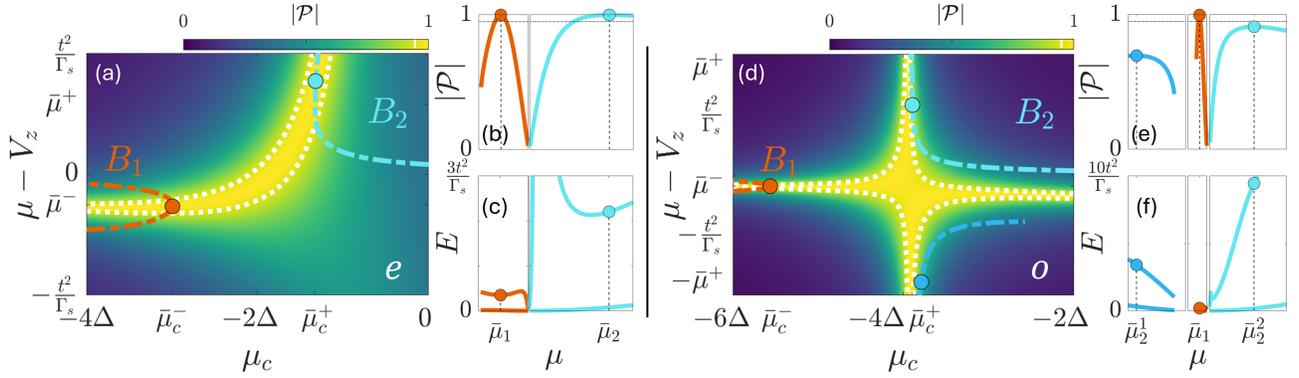}
\caption{Sweet-spot characterization of the mesoscopic minimal KC assuming $V_{z,c} \neq 0$ in the weak-coupling limit. Left (Right) side correspond to the even- (odd-) parity state of the isolated ABS in the hybrid section. We  assume $V_z=V_{z,c}=2\,\Delta$ (left) and $V_z=V_{z,c}=\sqrt{\mu_c^2+\Gamma_s^2} \approx 4.4\,\Delta$ being the QPT point for a sweet-spot at $\mu_c\approx-4\Delta $ (right). Main panels (a,d): MP $|{\cal P}(x_i)|$ as a function of $\mu$ and $\mu_c$ for the parameter set $x_i=\{t, \, V_{z}, \, \Gamma_{s} \}$. Dot-dashed lines, $B_1$ and $B_2$, depict the zero-energy contours defined in Eq.~\eqref{Eq_muc_main}. Orange and cyan markers indicate the optimal sweet-spot for each of the branches. White dotted lines denote the $|{\cal P}|=0.95$ contour. 
The marks $\bar{\mu}^\pm_c$ in the $y$-axis are given by Eqs.~\eqref{mu_c_vzneq0_main} and \eqref{mu_c_vzneq0_t_main} for the even- and odd-parity regimes, while marks $\bar{\mu}^\pm$ in the $x$-axis are given by  Eq.~\eqref{renorm_mu_Vz_main} and \eqref{renorm_mu_Vz_t_main} respectively. Side panels: evolution of the MP (b,e) and the energy dispersion for the different zero-energy branches (c,f) in the even- (b,c) and odd- (e,f) parity states of the isolated ABS. Notice that the latter case shows 3 different sweet-spots. Markers highlight both the maximum-polarization points and the associated energy gaps. We use $t=0.01, \, \Gamma_{s}=2$ in $\Delta$ units.
}
\label{fig6}
\end{figure*}

In contrast, the excitation gap in Eq.~\eqref{eq_gap_main} is largely independent of the Zeeman energy, which allows one to compensate for the loss of MP induced by increasing $t$. Figure~\ref{fig5}~(b) shows that when $t$ and $V_z$ are increased simultaneously, the sweet-spot exhibits a linear growth of the gap to excited states. This behavior follows from the iso-MP following $t \propto \sqrt{V_z}$, which, when substituted into Eq.~\eqref{eq_gap_main}, results in a linear dependence of the gap with $V_z$. Therefore, in this regime it is advantageous to increase the coupling strength provided that the Zeeman field is also increased and compensates for the loss of MP. It should be pointed out that for large enough coupling strength and Zeeman energy ({\it i.e.}, $t\approx2.5$ and $V_z\approx20$ in $\Delta$ units), the gap associated to the iso-MP deviates from the ideal linear behavior and becomes constant $\delta E \approx \Delta$ (not shown). 

\subsection{Spin-Split ABS} \label{SecIVb}

We now explore the more general and experimentally realistic situation in which the SC hybrid section exhibits a finite Zeeman splitting. Unlike low-energy effective descriptions, we explicitly resolve the emergence of spin-split ABSs in the hybrid region, enabling it to undergo a parity-crossing as system parameters are varied. Building on the analytical results of Secs.~\ref{SecIIIe} and \ref{SecIIIf}, we analyze how this transition reshapes the MP, the excitation gap, and the branch selection.

Figure~\ref{fig6} shows the MP and excitation gap across the chemical potential parameter space in the weak-coupling regime ($t=0.01\,\Delta$), revealing a richer sweet-spot structure than in the spin-degenerate case. In particular, multiple solutions can emerge depending on the GS configuration of the ABS.
In the even-parity regime of the ABS, shown in Figs.~\ref{fig6}~(a-c), we observe two different sweet-spots that are still properly described by the linear approximation in Eqs.~\eqref{mu_c_vzneq0_main} and \eqref{renorm_mu_Vz_main}.
However, in Fig.~\ref{fig6}~(a) we see the clear asymmetry in $\delta\mu^\pm$ induced by the finite Zeeman energy in the SC hybrid section, even for small values of $t$. While both sweet-spots exhibit large MP values, the gap to the excited states is highly asymmetric, favoring the sweet-spots in branch $B_2$, see Figs.~\ref{fig6}~(b,c).

As the ABS energy becomes resonant at the QPT, two distinct sweet-spots emerge for the same branch $B_2$. Figures~\ref{fig6}~(d–f) characterize the $B_2$ sweet-spots in which the ABS becomes resonant ({\it i.e.,} $\Gamma_s=2$, $\mu_c\approx-4$ and $V_z=V_{z,c}=\sqrt{\mu_c^2+\Gamma_s^2}$ in $\Delta$ units). The two distinct $B_2$ sweet-spots $\bar{\mu}^1_2$ and $\bar{\mu}^2_2$ are characterized by opposite-sign $\delta \mu$~\cite{Wimmer2025} satisfying $\delta\mu^\pm > t^2 / \Gamma_s$, see Fig.~\ref{fig6}~(d). This behavior signals the breakdown of the weak-coupling approximation and requires higher-order corrections, as captured by Eqs.~\eqref{mu_c_vzneq0_t_main} and \eqref{renorm_mu_Vz_t_main}. Although the sweet-spot associated with $B_1$ exhibits larger MP values, the corresponding extremely small excitation gap $\delta E$ renders the $B_2$ sweet-spots more favorable, see Figs.~\ref{fig6}~(e, f). This sweet-spot configuration is consistent with recent experiments on 3-site KCs~\cite{Bordin2025, Wimmer2026} showing different numbers of identifiable sweet-spots depending on the SC hybrid chemical potential. This is explained as the QD-ABS-QD mesoscopic chain behaving as an effective 3-site chain when the ABS energy is close to resonance~\cite{Miles2023}.

\begin{figure*}[t!]
\centering
\includegraphics[width=0.8\textwidth]{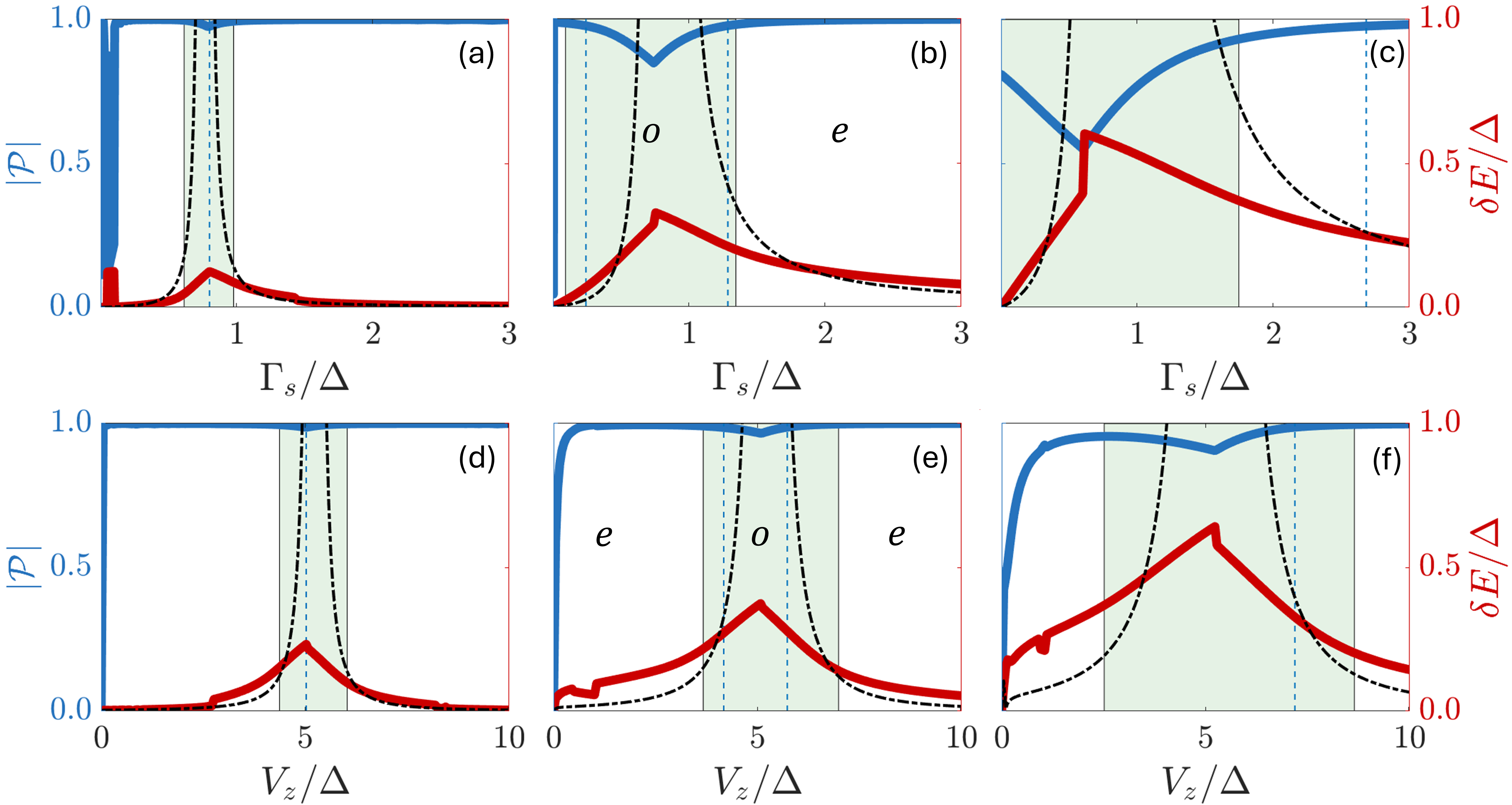}
\caption{Evolution of the optimal sweet-spot properties with the SC hybridization $\Gamma_s$ (upper panels) and Zeeman field $V_z$ (lower panels) for different coupling strengths ($t = \{0.1,\, 0.5,\, 1.0\}$ for the different columns). We assume $V_{z,c} = V_z$. 
Blue (red) lines correspond to the MP (gap to excited states). 
The odd-parity spin-polarized region ($o$) in parameter space given by Eq.~\eqref{eq_sd_QPT_main} is highlighted in green. Black dot--dashed line represents the perturbative approximation for the gap in Eqs.~\eqref{dE_Vzneq0_main} and \eqref{dE_Vzneq0_sc_main} for the weak- and strong-coupling regimes, corresponding to the first and remaining columns, respectively.
Dashed vertical blue lines indicate MP at the $\delta E$ peak shown in panel (a). We use $V_z = 2$ (upper panels) and $\Gamma_s= 2$ (lower panels) in $\Delta$ units. Note that the peak position is insensitive to the coupling strength $t$.
}
\label{fig7}
\end{figure*}

Now we can trace the evolution of the optimal sweet-spot in $B_2$ as we vary the hybridization $\Gamma_s$ or the Zeeman energy for different values of the coupling $t$, see Fig.~\ref{fig7}. Because each point of the plot requires to re-evaluate the chemical potential values $\mu$ and $\mu_c$ to find the sweet-spot, it is possible that the isolated ABS undergoes parity-crossings as parameters are varied. Fig.~\ref{fig7}~(a) shows the evolution of MP and excitation gap with the hybridization, reproducing a clear $\delta E$ peak in the middle of the odd-parity spin-polarized region. This behavior signals a change in the dominant solution along the $B_2$ branch~\cite{Wimmer2025}, corresponding to a crossover between the two possible sweet-spot configurations $\bar{\mu}^1_2$ and $\bar{\mu}^2_2$. However, as the coupling strength increases, the MP becomes progressively suppressed, as shown in Fig.~\ref{fig7}~(b,c) and similar to the spin-degenerate limit of the ABS in the previous section.

Similarly, the evolution of the sweet-spots with the Zeeman field also displays a $\delta E$ peak in the spin-polarized region, see Figs.~\ref{fig7}~(d-f). Therefore, increasing the Zeeman field does not always improve the sweet-spot, contrary to commonly assumed and in stark contrast to the spin-degenerate limit of the ABS in the previous section. Nevertheless, increasing the Zeeman field is generally favorable for typically small values of SOC, since the kink-like feature in the excitation gap occurs for $V_z =\Gamma_s \csc(\theta) \approx \Gamma_s/\theta$, given by the divergent point in Eq.~\eqref{dE_Vzneq0_main}, see dot--dashed line in Fig.~\ref{fig7}. In practice, this implies that very large Zeeman fields are required to reach the tipping point at the middle of the spin-polarized region when the SC lead is strongly hybridized with the chain and small SOC angles are considered. Notably, the transition between the even- and odd-parity regimes still exhibits near-optimal values of both MP and $\delta E$, making it a particularly accessible regime experimentally. 

\begin{figure}[b!]
\centering
\includegraphics[width=\columnwidth]{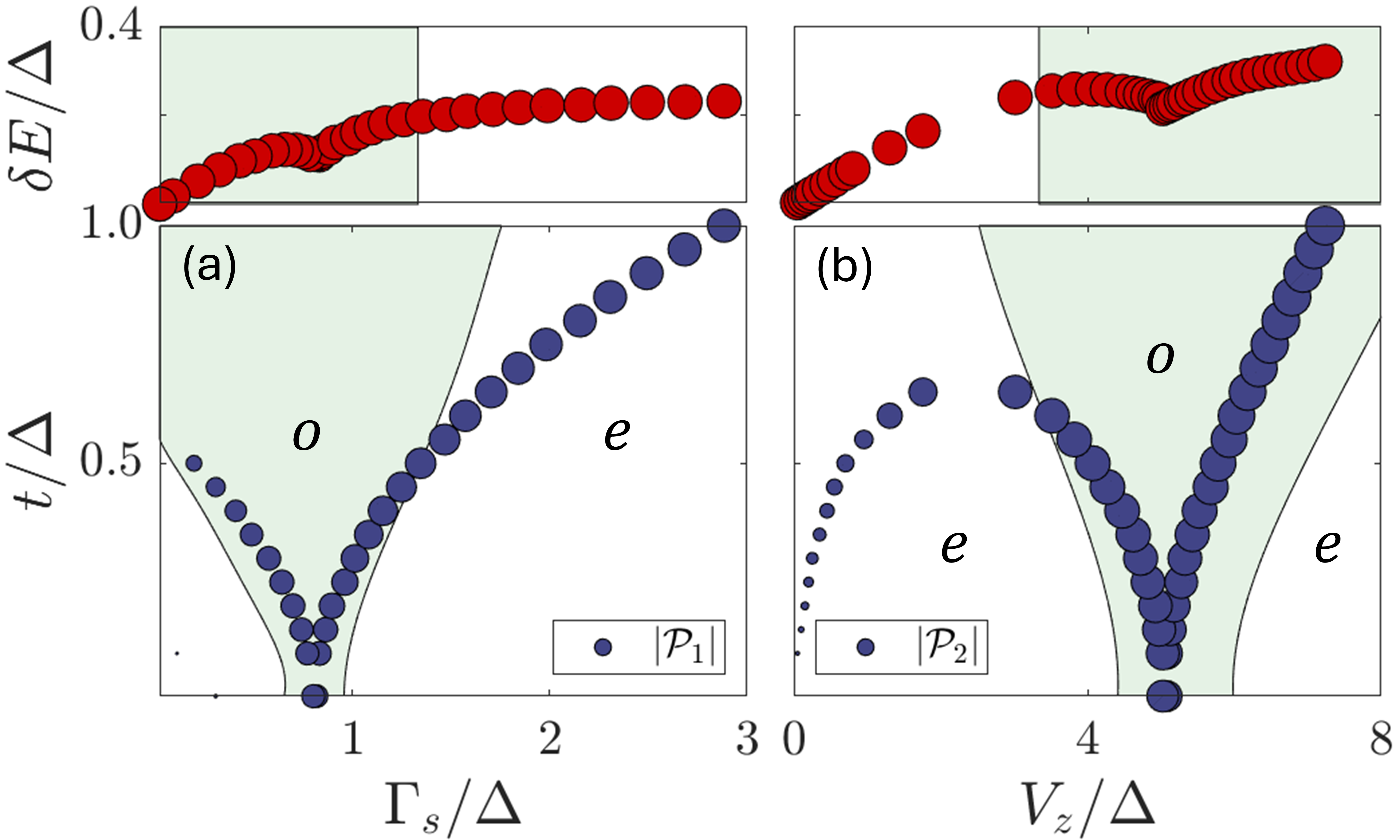}
\caption{Evolution of constant-value MP (iso-MP) in the $(\Gamma_s,\,t)$ and $(V_z,\,t)$ parameter planes respectively for $V_{z,c}\neq0$. The main panels show the evolution of the iso-MP contours evaluated at the peak values obtained at $t = 0.01\, \Delta$, associated with the vertical blue lines in Fig.~\ref{fig7}, namely 
$|\mathcal{P}_1| = 0.9810$ in panel (a), and 
$|\mathcal{P}_2| = 0.9876$ in panel (d). 
Green regions indicate the odd-parity spin-polarized state ($o$) region of the isolated ABS. The size of the markers reflects the magnitude of $\delta E$, shown explicitly in the upper panels. We use $V_z = 2$ (left panels) and $\Gamma_s= 2$ (right panels) in $\Delta$ units.
}
\label{fig8}
\end{figure}

We associate the enhanced gap in the odd-parity spin-polarized regime to the system behaving like an effective 3-site Kitaev chain where effective long-range couplings between the outer QDs mediated by the ABS~\cite{Miles2023} would explain the lower MP values. Equivalently, the reduction of the MP could be explained due to the increased spin fluctuations in the odd-parity spin-polarized regime of the ABS that delocalize the Majorana wave function ({\it i.e.}, the so called spin-exchange induced spillover effect~\cite{Sanches2025, Sanches2026}). It is worth noting that non-Hermitian effects arising from asymmetric couplings to normal leads could suppress these fluctuations through dissipative stabilization, potentially further enhancing the MP~\cite{Aguado2015, Aguado2025}.

A systematic analysis of the sweet-spot optimization when reaching the parity-crossing is made by following the evolution of a representative iso-MP, chosen as the MP value at the $\delta E$ peak in the odd-parity spin-polarized region for small couplings ($t = 0.01\,\Delta$). Fig.~\ref{fig8}~(a) displays the evolution of this iso-MP, where the size of the dots reflects the magnitude of $\delta E$, as explicitly shown in the upper panel. As a general trend, increasing $\Gamma_s$ monotonously increases the gap to the excited states, saturating for $\Gamma_s/\Delta\sim 1$. However, the corresponding value of $t$ does not follow a monotonous trend. In particular, following the iso-MP, the $t$ values decrease toward the center of the spin-polarized region, highlighting a regime where a weak SC-QD coupling can produce a large excitation gap due to resonant coupling between the different elements of the chain. Outside the odd-parity region, we observe negligible enhancements of the gap to excited states, consistent with the strongly screened regime. 

Similarly, Fig.~\ref{fig8}~(b)  shows a clearly suboptimal contour of sweet-spots in the even-parity regime. In this region, increases of $\delta E$ with the coupling strength become linear. However, this growth approaches saturation upon entering the odd-parity spin-polarized regime. Besides, as the global Zeeman field is increased, the coupling strengths required to remain on the same iso-MP contour grow rapidly. This suggests that, in more realistic experimental platforms, increasing the Zeeman field may require impractically large coupling strengths to achieve any significant improvement of the sweet-spot properties. 

Overall, the presence of spin-split ABSs qualitatively reshapes the sweet-spot landscape in such a way that mesoscopic KCs feature optimal combinations of parameters $x_i = \{t, V_z, \, \Gamma_s\}$ that optimize the gap to the excited states while keeping a good MP. These optimal regions correspond to the odd-parity regime, where the ABS hosts a single spin-polarized quasiparticle. In this regime, near-optimal sweet-spots can be achieved even for relatively small coupling strengths, owing to the resonant nature of the ABS. This is in stark contrast to approximations that only consider the subgap degrees of freedom related to the ABS in the SC, that predicts that maximizing individual parameters such  as coupling strength, spin polarization, or hybridization with the SC lead always improves the sweet-spot properties. Instead, mesoscopic effects define distinct optimal regimes, with the parity-crossing QPT providing a particularly advantageous operating point for realistic Kitaev-chain devices.

\section{Conclusions and Outlook}\label{SecVI}

In this work, we have presented a comprehensive study of sweet-spots in mesoscopic Kitaev chains formed by quantum dots coupled via mesoscopic superconducting leads. In contrast to previous approaches based on low-energy or effective models, our methodology relies on a full microscopic treatment of the hybrid section within the Green’s function formalism, allowing us to systematically incorporate both the quasiparticle continuum and spin-split Andreev bound states. This framework enables us to derive analytically the renormalized ECT and CAR amplitudes and, consequently, the modified sweet-spot conditions beyond the weak-coupling approximation. Importantly, it also captures the emergence of parity-crossings in the hybrid section, which mark the breakdown of perturbative descriptions and define key regimes where optimal sweet-spots can be achieved.

We first revisit previous results concerning the fundamental competition between the gap to excited states and Majorana localization in the weak-coupling regime and in the limit of fully screened Andreev bound states, confirming that the optimal conditions for these two quantities do not generally coincide. This competition defines an intrinsically multiobjective optimization problem, requiring a trade-off between spectral protection and spatial localization.

Going beyond this idealized limit, we show that mesoscopic superconducting hybrids hosting spin-split ABSs can undergo parity-crossings, which fundamentally reshape the sweet-spot structure. In particular, the vicinity of the parity-crossing transition -- where the ABS becomes resonant -- emerges as a privileged regime supporting near-optimal sweet-spots. This regime provides experimentally favorable conditions for realizing Majorana modes in finite chains and highlights the central role of mesoscopic renormalization in determining the low-energy properties of the system. Furthermore, we demonstrate the existence of optimal finite parameter combinations $\{t, V_z, \Gamma_s\}$, challenging the conventional expectation that increasing coupling, hybridization, or spin polarization monotonically improves sweet-spot performance.

To extend our analysis to longer chains, we have implemented an efficient approach using surrogate models~\cite{Baran2023, Paaske2023} combined with CMA-ES optimization algorithms~\cite{Deb2000,Hansen2016}, allowing us to explore high-dimensional parameter spaces that are otherwise inaccessible. This framework provides a practical route to identify optimal sweet-spots in realistic longer architectures by jointly optimizing the excitation gap and Majorana polarization. 

Finally, we envision the application of our results to more complex nanostructures, such as Josephson-junction arrays \cite{Bas-Haaf2025,Kulesh2025,PhysRevB.111.L140503} and other qubit configurations \cite{Pino-Kitmon24,PhysRevB.111.075416,Palacios_2026}. More broadly, the interplay between mesoscopic effects, Andreev physics, and optimization strategies identified here establishes a general framework for engineering robust topological modes in superconducting quantum devices.

\acknowledgments
We thank William Samuelson for valuable insights. Work supported by the Horizon Europe Framework Program of the European Commission through the European Innovation Council Pathfinder Grant No. 101115315 (QuKiT), the Spanish Comunidad de Madrid (CM) ``Talento Program'' (Project No. 2022-T1/IND-24070), the Spanish Ministry of Science, Innovation, and Universities through Grants CEX2024-001445-S (Severo Ochoa Centres of Excellence program), PID2022-140552NA-I00, and PID2024-161156NB-I00.
Support from the CSIC Interdisciplinary Thematic Platform (PTI+) on Quantum Technologies (PTI-QTEP+) is also acknowledged.

\appendix

\section{Green's Function Derivation}\label{App_A}

In this Appendix we summarize the technical details underlying the results presented in Sec.~\ref{SecIII}. While the full model definition and physical discussion are given there, here we focus on the technical aspects required for the analytical derivations. In particular, we introduce the notation and conventions used in the Green’s function (GF) formalism and outline the main steps leading to the expressions employed in the main text. 

To compute the GF of a general chain, we adopt the recursive approach of Ref.~\cite{Alvarado2022, Alvarado2025}. This formulation provides an efficient recursive construction of the bGF for finite mesoscopic chains. The spinful Hamiltonian for the semiconducting chain $H_{kc}=H_0+H_T+H_S$ written in the $4\times4$ Nambu basis (hat notation) $\hat{\Psi}_j = (\psi_{j,\up}, \psi_{j,\up}^\dagger, \psi_{j,\dw}, \psi_{j,\dw}^\dagger)^T$, where $\psi_{j,\sigma}$ are local fermion fields, reads for odd and even sites as,
\begin{subequations} 
\bea 
&\hat{H}_{0}(2k-1) =  \mu \, \sigma_0\tau_z  + V_z \,\sigma_z\tau_z \, , \qquad& \\
&\hat{H}_{0}(2k) =  \mu_c \, \sigma_0 \tau_z + V_{z,c} \,\sigma_z\tau_z  \, ; 
\eea
\end{subequations}
with $k = {1, \dots, n}$ for an $n$‑site chain. Consequently, the total length of the chain, $L = 2n - 1$, which ensures well-defined boundaries. Here, $\tau_\mu$ ($\sigma_\mu$) define Pauli matrices acting in spin (electron/hole) space. The coupling term between the sites of the chain is written as
\be 
\hat{H}_T = t \cos \left(\theta/2 \right) \, \sigma_0 \tau_z - i \, t \sin \left(\theta/2 \right) \, \sigma_y \tau_z \, .
\ee 
Within this framework, the coupling between the proximitized regions of the chain and the superconducting (SC) leads is non-zero only for even sites and is described by 
\be
\hat{H}_{s,T} = \lambda_s \, \sigma_z\tau_0 \, ,
\ee
which defines the effective hybridization parameter $\Gamma_s=\lambda_s^2/t_s$. Mesoscopic effects associated with SC hybrid regions are incorporated using advanced boundary Green’s functions (bGF) techniques~\cite{Cuevas1996, Zazunov2016, Alvarado2018, Alvarado2020, Alvarado2022}. The bGF for the semi-infinite quasi-1D SC lead reads
\be
\hat{{\cal G}}^a_s = -\frac{\omega \, \sigma_0\tau_0 - \Delta \, \sigma_y\tau_y}{t_s \sqrt{\Delta^2-\omega^2}} \, .
\ee 
The SC lead is treated within the wide-band approximation ({\it i.e.}, $t_s \gg \omega, \,\Delta$). In the strong-hybridization regime $(\Gamma_s, V_z >\Delta\gg t)$, the model reduces to an effective Kitaev chain (KC) description and can be straightforwardly extended to chains of arbitrary length.

In the advanced GF we implicitly assume the standard analytic continuation $\omega \rightarrow \omega - i\eta$ with $\eta \rightarrow 0^+$~\cite{Dynes1978, Bauer2007, Zazunov2016}, and we omit the super‑index $a$ for convenience. Following Refs.~\cite{Bauer2007, Zazunov2016}, we define the branch cut in the square root along the negative axis $\xi = \sqrt{\Delta^2-(\omega-i\eta)^2}$, which yields
\be
\xi = \left\lbrace
\begin{array}{cl}
\sqrt{\Delta^2-\omega^2}  & |\omega|\leq \Delta \, , \\
i \, \textrm{sgn}(\omega)\sqrt{\omega^2-\Delta^2}  & |\omega| > \Delta \, .
\end{array}
\right.
\ee
The bGF associated with the Andreev bound state (ABS) in the SC hybrid section admits a compact analytical form after performing a suitable unitary transformation to the basis $\hat{\Psi}^\prime = (\psi_\up, \psi_\dw^\dagger, \psi_\dw,  \psi_\up^\dagger)^T$, being block diagonal as
\begin{widetext} 
\be \label{G_ABS}
\mathcal{G}_{c,A/B} = \left [\omega \, \hat{\mathbb{I}} - \hat{H}_c - \hat{H}_{s,T}^\dag \,  \hat{{\cal G}}_s \, \hat{H}_{s,T} \right]_{A/B}^{-1} =  \frac{\big [\omega \, \Gamma_s +\xi\, (\omega \mp V_{z,c}) \, \big]  \, \nu_0 + \mu_c \, \xi \, \nu_z \pm \Delta \Gamma_s \, \nu_x}{2  \Gamma_s \, (\omega\mp V_{z,c})\,\omega + \xi \, \big [(\omega\mp V_{z,c})^2-\mu_c^2-\Gamma_s^2 \big]} \, ,
\ee   
\end{widetext}
where the labels $A/B$ denote the two diagonal blocks in the transformed basis, while $\nu_i$ acts in the effective sublattice space of the hybrid section. Here, $\hat{H}_c=\hat{H}_{0}(2k)$ corresponds to the local Hamiltonian of an $n$‑th even site. We do not include electron-electron interactions in the ABS, since charge fluctuations are efficiently screened by the grounded SC lead. From now on, we work in the $\hat{\Psi}$ basis, being straightforward to transform Eq.~\eqref{G_ABS} back to the original one.

The poles of the isolated ABS bGF are determined by the characteristic polynomial $P(\omega, \pm V_{z,c})$ appearing as the denominator in Eq.~\eqref{G_ABS}, from which the isolated Andreev energies ($\epsilon_A^\pm$) can be obtained~\cite{Bauer2007}. From the characteristic polynomial, one can extract the condition under which the ABS undergoes a parity-crossing, given by
\be \label{singlet-doublet}
V_{z,c}^2=\mu^2_c+\Gamma_s^2 \, .
\ee
This condition defines the phase boundary between two distinct ground-state configurations of the SC hybrid section. 

Within this scheme, the bGF at the $n$-th site is expressed as

\begin{subequations}
\begin{gather} \label{recursive_GF}
 \hat{\mathcal{G}}(n)  = \big [\omega \, \hat{\mathbb{I}} - \hat{H}_{0}(n)  - \hat{\Sigma}(n) - \hat{\Sigma}_s(n) \big]^{-1}\, , \\
\hat{\Sigma}(n) = \hat{H}_T \, \hat{\mathcal{G}}(n-1) \hat{H}_T^\dag  \, , \\
\hat{\Sigma}_s(n) = \hat{H}_{s,T} \, \hat{\mathcal{G}}_s \,\hat{H}_{s,T}^\dag \, ; 
\end{gather} 
\end{subequations}
where $\hat{H}_{0}(n)$ denotes the local Hamiltonian of the $n$‑th site. The self-energy $\hat{\Sigma}(n)$ accounts for the coupling between sites $n$ and $(n-1)$ via the hopping term $\hat{H}_T$, and $\hat{\Sigma}_s(n)$ accounts for the coupling to a SC lead, being non-zero only for even sites, without considering any phase difference between grounded SC leads.

From now one we study the simplified case of an isolated minimal KC ($L=3$). The renormalization effects induced in the SC hybrid section can be accounted for recalling the self-energy generated by the coupling to the outer quantum dots (QDs) as $\hat{\Sigma}^{(2)}_{0,c}=[\hat{H}_T \, \hat{G}_0^{(0)} \,\hat{H}_T^\dag + \hat{H}^\dag_T \, \hat{G}_0^{(0)} \,\hat{H}_T]$, where $\hat{G}^{(0)}_0=[\omega \, \hat{\mathbb{I}} - \hat{H}_{0}(1)]^{-1}$ is the GF of the isolated QDs. In the static limit, this self-energy is given by
\be 
\hat{\Sigma}^{(2)}_{0,c} = - \frac{2t^2\,\mu}{\mu^2-V_z^2} \sigma_z \,\tau_0+ \frac{2t^2\,V_z \cos(\theta)}{\mu^2-V_z^2} \sigma_z \, \tau_z \, ,
\ee 
where we identify two contributions, the correction to the chemical potential and the induced Zeeman field respectively. 

In order to go to infinite order in the renormalization effects from the GF projected onto the proximitized region of the chain, we follow
\be \label{G_22_eq}
\hat{G}_{c} = \big[[\hat{\mathcal{G}}_c]^{-1} - \hat{H}_T \, \hat{G}_0^{(0)} \, \hat{H}_T^\dag - \hat{H}^\dag_T \, \hat{G}_0^{(0)} \, \hat{H}_T \big]^{-1} \, .
\ee 
The full analytical expression is lengthy and not required for the following discussion, and is therefore omitted for brevity. From the resulting GF, we derive the characteristic polynomial and determine the conditions under which zero-energy modes emerge at the boundaries of the chain, given by
\begin{widetext} 
\be \label{Eq_muc}
\mu^\pm_c = \frac{2\mu \, t^2\pm\sqrt{[2 V_z \, t^2 \cos(\theta) -  (\mu^2-V_z^2)(\Gamma_s^2-V_{z,c}^2)][2 V_z \, t^2 \cos(\theta) + (\mu^2-V_z^2)(\Gamma_s^2+V_{z,c}^2)]}}{\mu^2-V_z^2} \, .
\ee 
\end{widetext} 
For completeness, in the limit $V_{z,c} = 0$ we recover the expression introduced in Ref.~\cite{Alvarado2025} as
\be  \label{Eq_muc_Vc0}
\mu^\pm_c=\frac{2\mu \, t^2\pm\sqrt{4 V_z^2 \, t^4 \cos^2(\theta) - \Gamma_s^2 (\mu^2-V_z^2)^2}}{\mu^2-V_z^2} \, ,
\ee
where the SC gap entering the zero-bandwidth approximation is effectively replaced by the hybridization with the SC lead.

\section{Sweet-spot characterization} \label{App_B}

In this Appendix, we summarize the technical details underlying the results presented in Sec.~\ref{SecIII}, focusing on the renormalization of the QDs properties induced by the microscopic degrees of freedom of the SC hybrid section. In particular, we outline how the effective coupling between the outer QDs is renormalized through the hybrid section and derive the corresponding corrections to the electron co-tunneling (ECT) and crossed Andreev reflection (CAR) amplitudes.  This framework allows us to systematically obtain the modified sweet-spot conditions and to identify the role of mesoscopic effects in shaping the low-energy properties of the chain.

First, we compute the renormalized sweet-spot condition ({\it i.e.,} ECT equals CAR) through a perturbative expansion of the effective coupling between the outer QDs mediated by the ABS. This coupling is described by the second-order self-energy $\hat{\Sigma}^{(2)}_{LR}(\omega) = \hat{H}_T \, \hat{\mathcal{G}}_c \, \hat{H}_T$, and the reciprocal process $\hat{\Sigma}^{(2)}_{RL}(\omega) = \hat{H}^\dag_T \, \hat{\mathcal{G}}_c \, \hat{H}^\dag_T$ in the static limit ($\omega \rightarrow 0$).

Second, we compute the corresponding mesoscopic correction to the QDs low energy physics through a perturbative expansion of the local renormalization in the QDs induced by the ABS.
This correction is described by the second-order self-energy for the left QD $\hat{\Sigma}_{0,L}^{(2)}(\omega)=\hat{H}_T \, \hat{\mathcal{G}}_c \, \hat{H}^\dag_T$, and for the right QD $\hat{\Sigma}_{0,R}^{(2)}(\omega)=\hat{H}_T^\dag \, \hat{\mathcal{G}}_c \, \hat{H}_T$, in the static limit ($\omega \rightarrow 0$). 

Written in the $4\times4$ Nambu basis (hat notation) $\hat{\Psi}_j = (\psi_{j,\up}, \psi_{j,\up}^\dagger, \psi_{j,\dw}, \psi_{j,\dw}^\dagger)^T$, where $\psi_{j,\sigma}$ are local fermion fields, the aforementioned self-energies take the form
\begin{widetext} 
\begin{subequations}
\bea
  \hat{\Sigma}^{(2)}_{LR/RL} &=& \bmat
  \frac{t^2 (V_{z,c} - \mu_c \cos\theta)}{(
   \mu_c^2 + \Gamma_s^2 - V_{z,c}^2)} && 
   \pm\frac{\Gamma_s t^2 \sin\theta}{(
    \mu_c^2 + \Gamma_s^2 - V_{z,c}^2)} && 
    \pm\frac{\mu_c t^2 \sin\theta}{(
   \mu_c^2 + \Gamma_s^2 - V_{z,c}^2)} && 
   \frac{\Gamma_s t^2 \cos\theta}{(
    \mu_c^2 + \Gamma_s^2 - V_{z,c}^2)} \\
  \mp\frac{\Gamma_s t^2 \sin\theta}{(\mu_c^2 
  + \Gamma_s^2 - V_{z,c}^2)} && 
  -\frac{t^2 (V_{z,c} - \mu_c \cos\theta)}{(\mu_c^2 + \Gamma_s^2 - V_{z,c}^2)} && 
  -\frac{\Gamma_s t^2 \cos\theta}{(
   \mu_c^2 + \Gamma_s^2 - V_{z,c}^2)} && 
   \mp\frac{\mu_c t^2 \sin\theta}{(
    \mu_c^2 + \Gamma_s^2 - V_{z,c}^2)} \\
  \mp\frac{\mu_c t^2 \sin\theta}{(\mu_c^2 + \Gamma_s^2 - V_{z,c}^2)} && 
  -\frac{\Gamma_s t^2 \cos\theta}{(
   \mu_c^2 + \Gamma_s^2 - V_{z,c}^2)} && 
   -\frac{t^2 (V_{z,c} + \mu_c \cos\theta)}{(
    \mu_c^2 + \Gamma_s^2 - V_{z,c}^2)} && 
    \pm\frac{\Gamma_s t^2 \sin\theta}{(
    \mu_c^2 + \Gamma_s^2 - V_{z,c}^2)} \\
  \frac{\Gamma_s t^2 \cos\theta}{(\mu_c^2 + \Gamma_s^2 - V_{z,c}^2)} && 
  \pm\frac{\mu_c t^2 \sin\theta}{(\mu_c^2 + \Gamma_s^2 - V_{z,c}^2)} && 
  \mp\frac{\Gamma_s t^2 \sin\theta}{(\mu_c^2 + \Gamma_s^2 - V_{z,c}^2)} && 
  \frac{t^2 (V_{z,c} + \mu_c \cos\theta)}{(\mu_c^2 + \Gamma_s^2 - V_{z,c}^2)} 
  \emat \, , \label{self_enrgies_A} \\ \nonumber \\
\hat{\Sigma}^{(2)}_{0,L/R} &=& \bmat
  -\frac{t^2 (\mu_c-V_{z,c}\cos\theta)}{(
   \mu_c^2 + \Gamma_s^2 - V_{z,c}^2)} && 
   0 && 
    \pm\frac{V_{z,c} t^2 \sin\theta}{(
   \mu_c^2 + \Gamma_s^2 - V_{z,c}^2)} && 
   \frac{\Gamma_s t^2}{(
    \mu_c^2 + \Gamma_s^2 - V_{z,c}^2)} \\
  0 && 
  \frac{t^2 (\mu_c-V_{z,c}\cos\theta)}{(
   \mu_c^2 + \Gamma_s^2 - V_{z,c}^2)} && 
  -\frac{\Gamma_s t^2 }{(
   \mu_c^2 + \Gamma_s^2 - V_{z,c}^2)} && 
   \mp\frac{V_{z,c} t^2 \sin\theta}{(
    \mu_c^2 + \Gamma_s^2 - V_{z,c}^2)} \\
  \pm\frac{V_{z,c} t^2 \sin\theta}{(\mu_c^2 + \Gamma_s^2 - V_{z,c}^2)} && 
  -\frac{\Gamma_s t^2}{(
   \mu_c^2 + \Gamma_s^2 - V_{z,c}^2)} && 
   -\frac{t^2 (\mu_c + V_{z,c}\cos\theta)}{(
    \mu_c^2 + \Gamma_s^2 - V_{z,c}^2)} && 
    0 \\
  \frac{\Gamma_s t^2}{(\mu_c^2 + \Gamma_s^2 - V_{z,c}^2)} && 
  \mp\frac{V_{z,c} t^2 \sin\theta}{(\mu_c^2 + \Gamma_s^2 - V_{z,c}^2)} && 
  0 && 
  \frac{t^2 (\mu_c + V_{z,c}\cos\theta)}{(\mu_c^2 + \Gamma_s^2 - V_{z,c}^2)} 
  \emat \, . \label{self_enrgies_B}
\eea
\end{subequations}
\end{widetext}
From Eq.~\eqref{self_enrgies_A} we obtain the general condition for the sweet-spot as
\begin{subequations}\label{sweet_spot_Vz}
\bea 
|\Delta^{(2)}| &=&  \frac{t^2 \, \Gamma_s \, \sin(\theta)}{|\mu^2_c + \Gamma_s^2-V_{z,c}^2|} \, , \\
|t_{\sigma}^{(2)}| &=&  \frac{t^2 \, |V_{z,c} -s_\sigma \mu_c \, \cos(\theta)|}{|\mu^2_c + \Gamma_s^2-V_{z,c}^2|} \, , 
\eea
\end{subequations}
where $\sigma = \up,\dw$ refers to the spin sector and $s_{\up,\dw}=\pm1$. It should be noted that the low-energy spin sector depends on the sign of the Zeeman field. Here, we assume $V_z>0$ and focus on spin-down sector correlations; however, it is straightforward to compute the equivalent expressions for $V_z<0$ and the spin-up sector.

From Eq.~\eqref{self_enrgies_B} we obtain the corresponding mesoscopic correction to the chemical potentials and Zeeman energy of the QDs, following $V^\prime_{z} = V_{z}+V_{z}^{(2)}$ and $\mu^\prime = \mu+\mu^{(2)}$ respectively, where 
\begin{subequations} \label{induced_Zeeman_QD}
\bea 
\mu^{(2)} &=& - \frac{2\,t^2\,\mu_c}{\mu_c^2+\Gamma_s^2-V_{z,c
}^2} \, , \\
V_{z}^{(2)} &=& \frac{2\,t^2\,V_{z,c} \cos(\theta)}{\mu_c^2+\Gamma_s^2-V_{z,c
}^2} \, .
\eea
\end{subequations} 

Alternatively, the sweet-spot condition for $\{\bar{\mu},\,\bar{\mu}_c\}$ can be formulated in terms of the mesoscopic correction to the local parameters by imposing that the renormalized QD is tuned to resonance, $\mu+\mu^{(2)}=V_z+V_z^{(2)}$, together with $|t^{(2)}|=|\Delta^{(2)}|$ as given by Eqs.~\eqref{sweet_spot_Vz} and \eqref{induced_Zeeman_QD}, cf. Ref.~\cite{Zhang2026}. Solving these two constraints leads to the same expressions obtained in the weak-coupling analysis of the main text.

Within this framework, the correction to the QD chemical potential is naturally identified as $\delta\mu=V_z^{(2)}-\mu^{(2)}$, which quantifies the detuning from resonance. Importantly, the presence of two solution branches in the main text, $B_1$ and $B_2$, emerges from the spin-dependent nature of these renormalization effects, as each spin sector acquires distinct corrections, thereby reproducing the characteristic branch structure of the sweet-spot solutions.

\subsection{Spin-Degenerate ABS}

Here, we provide an analytical characterization of the excitation gap and the renormalization effects in the spin-degenerate ABS case assuming $V_{z,c}=0$. Our goal is to explicitly account for the asymmetries observed when going beyond low energy descriptions.

Building on the GF formalism introduced previously, we approximate the gap to excited states taking advantage of the poles of the system. From the analytical expression for the total GF in Eq.~\eqref{G_22_eq}, we perform a Taylor expansion of the characteristic polynomial of the system around zero energy. For simplicity, the bGF associated to the ABS GF evaluated within the zero-bandwidth approximation~\cite{Pita2026} 
\be 
\hat{\mathcal{G}}_c =  [\omega \, \hat{\mathbb{I}} - ( \mu_c \, \sigma_0 \tau_z - \Delta \, \sigma_y\tau_y \,) \big]^{-1} \, ,
\ee 
where $\tau_\mu$ ($\sigma_\mu$) define Pauli matrices acting in spin (electron/hole) space.
The corresponding characteristic polynomial is written as
\bea
&&P^{(2)}(\omega) = (\mu^2-V_z^2)\big[4 \mu \, \mu_c \, t^2 - (\Gamma_s^2+\mu_c^2)(\mu^2-V_z^2)\big] \nonumber \\
&& \quad + 4 V_z \big[V_z^2-\mu(\mu+2\mu_c) \big] \, t^2 \cos(\theta) \, \omega \nonumber \\
&& \quad + \big\{ \mu[\mu^3- 4\mu_c \, t^2 + 2\mu(\mu_c^2+2t^2)] \nonumber \\
&& \quad +2\Gamma_s^2(\mu^2+V_z^2) + 2V_z^2(2t^2+\mu_c^2-\mu^2) + V_z^4 \big\} \, \omega^2 \, . \nonumber \\
\eea 
By substituting the simplified sweet-spot conditions $\{\bar{\mu}^+ = V_z,\,\bar{\mu}_c^\pm = \pm \Gamma_s \tan(\theta)\}$, the polynomial yields a zero-energy root $z = 0$, together with the subgap pole
\be
z^{(2)}_{1,\pm} = \frac{\pm 2\, t^2 \, V_z \Gamma_s \, \sin(\theta)}{ \Gamma_s^2 \, V_z \, \sec^2(\theta) + t^2\, [2\,V_z \mp \, \Gamma_s \tan^2(\theta)]} \, \, . 
\ee
At the complementary sweet-spot at $\bar{\mu}^- = -V_z$, the spectrum is symmetrized through the pole
\be
z^{(2)}_{2,\pm} = \frac{\pm 2\, t^2 \, V_z \Gamma_s \,  \sin(\theta)}{ \Gamma_s^2 \, V_z \, \sec^2(\theta) + t^2\, [2\,V_z\pm \, \Gamma_s \tan^2(\theta)]} \, \, . 
\ee
These results reveal two distinct subgap poles associated with opposite spin-polarized branches.
The selected branch, and consequently the spin alignment, is therefore controlled by SC hybrid chemical potential $\mu_c$. In connection with our results, Ref.~\cite{Wimmer2026} highlights that the coupling mechanism between QDs depends on the spin configuration, since spin-orbit coupling selectively enables CAR or ECT processes. Thus, the observed reduction of the MP when increasing the coupling strength $t$ can be understood as a consequence of enhanced fluctuations: the system is no longer fully spin polarized due to SOC, induced Zeeman in the ABS, and the increased coupling between sites; leading to a delocalization of the MZMs wave function, spreading into the SC hybrid section~\cite{Liu2023, Wimmer2025}.

\subsection{Spin-Split ABS}

We turn to the case of an explicitly finite $V_{z,c}>0$ in the SC hybrid section, arising from a spatially homogeneous $g$-factor along the chain. Our goal is to explicitly derive the effective coupling between the outer QDs mediated by the hybrid section when including renormalization effects going beyond the weak-coupling description. As discussed in the main text, these effects become particularly relevant near the parity-crossing point, where the perturbative expansion breaks down and a renormalized treatment is required.

To account for these effects, we introduce the renormalized parameters that capture the influence of the coupling to the QDs into the low-energy degrees of freedom of the hybrid section. In particular, we map the renormalized spin-split ABSs in this strong-coupling regime by the substitution $\mu_c^\prime = \mu_c+\mu^{(2)}_{c}$, and $V^\prime_{z,c} = V_{z,c}+V^{(2)}_{z,c}$; into Eq.~\eqref{sweet_spot_Vz}.
\begin{subequations}
\bea 
|\Delta^{(2)}| &\approx&  \frac{t^2 \, \Gamma_s \, \sin(\theta)}{\left| \mu^2_c + \Gamma_s^2-V_{z,c}^2-\frac{4t^2[\mu\mu_c+V_zV_{z,c}\cos(\theta)]}{\mu^2-V_z^2}\right|} \, , \qquad \quad \\
|t^{(2)}| &\approx&  \frac{t^2 \, \left |V_{z,c}+\mu_c \, \cos(\theta)+\frac{2t^2(\mu+V_z)\cos(\theta)}{(\mu^2-V_z^2)} \right|}{\left|\mu^2_c + \Gamma_s^2-V_{z,c}^2-\frac{4t^2[\mu\mu_c+V_zV_{z,c}\cos(\theta)]}{\mu^2-V_z^2}\right|} \, . \qquad \quad
\eea
\end{subequations}
We note that these expressions remain finite across the parity-crossing point, in contrast to the weak-coupling approximation.

\section{Majorana Polarization calculation}\label{App_C}

To compute the Majorana polarization (MP) and the gap to excited states ($\delta E$), we diagonalize the effective Hamiltonian introduced in the main text, where the SC lead is described by a discretized tight-binding model. In the $4\times4$ Nambu basis (hat notation) $\hat{\Psi}_j = (\psi_{j,\up}, \psi_{j,\up}^\dagger, \psi_{j,\dw}, \psi_{j,\dw}^\dagger)^T$, where $\psi_{j,\sigma}$ are local fermion fields, the SC lead Hamiltonian, $\hat{H}_{S}=\hat{H}_{s,0}+\hat{H}_{s,t}$, with local and tunneling terms, is given by

\be 
\hat{H}_{s,0}(k) =  - \Delta \,\sigma_y\tau_y \, , \quad \hat{H}_{s,t} = t_s \, \sigma_z\tau_0 \, ;
\ee
where $k = {1, \dots, 100}$ and $\tau_\mu$ ($\sigma_\mu$) define Pauli matrices acting in spin (electron/hole) space. We do not consider any phase difference between grounded SC leads. 

Finally, we compute the non-local MP following the generalized formulation~\cite{Bena2015, Bena2016, Glodzik2020, Ola2024, Samuelson2025}
\be 
{\cal P} = \frac{\sum_{j=1}^{L} \sum_\sigma -2 \, s_\sigma \, u_{j \sigma} v_{j \sigma}}{ \sum_{j=1}^{L} \sum_\sigma \big|u_{j \sigma} \big|^2 +  \big|v_{j \sigma} \big|^2} \, ,
\ee 
where $u_{j \sigma} \, , \, v_{j \sigma}$ are the electron and hole components with spin $\sigma$ of the eigenstate $\Phi_j$ associated with the lowest-energy solution. Here, $|{\cal P}| = 1$ signals well-separated Majorana zero modes with negligible spatial overlap.


\begin{thebibliography}{83}%
\makeatletter
\providecommand \@ifxundefined [1]{%
 \@ifx{#1\undefined}
}%
\providecommand \@ifnum [1]{%
 \ifnum #1\expandafter \@firstoftwo
 \else \expandafter \@secondoftwo
 \fi
}%
\providecommand \@ifx [1]{%
 \ifx #1\expandafter \@firstoftwo
 \else \expandafter \@secondoftwo
 \fi
}%
\providecommand \natexlab [1]{#1}%
\providecommand \enquote  [1]{``#1''}%
\providecommand \bibnamefont  [1]{#1}%
\providecommand \bibfnamefont [1]{#1}%
\providecommand \citenamefont [1]{#1}%
\providecommand \href@noop [0]{\@secondoftwo}%
\providecommand \href [0]{\begingroup \@sanitize@url \@href}%
\providecommand \@href[1]{\@@startlink{#1}\@@href}%
\providecommand \@@href[1]{\endgroup#1\@@endlink}%
\providecommand \@sanitize@url [0]{\catcode `\\12\catcode `\$12\catcode `\&12\catcode `\#12\catcode `\^12\catcode `\_12\catcode `\%12\relax}%
\providecommand \@@startlink[1]{}%
\providecommand \@@endlink[0]{}%
\providecommand \url  [0]{\begingroup\@sanitize@url \@url }%
\providecommand \@url [1]{\endgroup\@href {#1}{\urlprefix }}%
\providecommand \urlprefix  [0]{URL }%
\providecommand \Eprint [0]{\href }%
\providecommand \doibase [0]{https://doi.org/}%
\providecommand \selectlanguage [0]{\@gobble}%
\providecommand \bibinfo  [0]{\@secondoftwo}%
\providecommand \bibfield  [0]{\@secondoftwo}%
\providecommand \translation [1]{[#1]}%
\providecommand \BibitemOpen [0]{}%
\providecommand \bibitemStop [0]{}%
\providecommand \bibitemNoStop [0]{.\EOS\space}%
\providecommand \EOS [0]{\spacefactor3000\relax}%
\providecommand \BibitemShut  [1]{\csname bibitem#1\endcsname}%
\let\auto@bib@innerbib\@empty
\bibitem [{\citenamefont {Kitaev}(2001)}]{Kitaev2001}%
  \BibitemOpen
  \bibfield  {author} {\bibinfo {author} {\bibfnamefont {A.~Y.}\ \bibnamefont {Kitaev}},\ }\bibfield  {title} {\bibinfo {title} {Unpaired majorana fermions in quantum wires},\ }\href {https://doi.org/10.1070/1063-7869/44/10S/S29} {\bibfield  {journal} {\bibinfo  {journal} {Physics-Uspekhi}\ }\textbf {\bibinfo {volume} {44}},\ \bibinfo {pages} {131} (\bibinfo {year} {2001})}\BibitemShut {NoStop}%
\bibitem [{\citenamefont {Alicea}(2012)}]{Alicea2012}%
  \BibitemOpen
  \bibfield  {author} {\bibinfo {author} {\bibfnamefont {J.}~\bibnamefont {Alicea}},\ }\bibfield  {title} {\bibinfo {title} {New directions in the pursuit of majorana fermions in solid state systems},\ }\href {https://doi.org/10.1088/0034-4885/75/7/076501} {\bibfield  {journal} {\bibinfo  {journal} {Reports on progress in physics}\ }\textbf {\bibinfo {volume} {75}},\ \bibinfo {pages} {076501} (\bibinfo {year} {2012})}\BibitemShut {NoStop}%
\bibitem [{\citenamefont {Leijnse}\ and\ \citenamefont {Flensberg}(2012{\natexlab{a}})}]{Leijnse2012_b}%
  \BibitemOpen
  \bibfield  {author} {\bibinfo {author} {\bibfnamefont {M.}~\bibnamefont {Leijnse}}\ and\ \bibinfo {author} {\bibfnamefont {K.}~\bibnamefont {Flensberg}},\ }\bibfield  {title} {\bibinfo {title} {Introduction to topological superconductivity and majorana fermions},\ }\href {https://doi.org/10.1088/0268-1242/27/12/124003} {\bibfield  {journal} {\bibinfo  {journal} {Semiconductor Science and Technology}\ }\textbf {\bibinfo {volume} {27}},\ \bibinfo {pages} {124003} (\bibinfo {year} {2012}{\natexlab{a}})}\BibitemShut {NoStop}%
\bibitem [{\citenamefont {Elliott}\ and\ \citenamefont {Franz}(2015)}]{RevModPhys.87.137}%
  \BibitemOpen
  \bibfield  {author} {\bibinfo {author} {\bibfnamefont {S.~R.}\ \bibnamefont {Elliott}}\ and\ \bibinfo {author} {\bibfnamefont {M.}~\bibnamefont {Franz}},\ }\bibfield  {title} {\bibinfo {title} {Colloquium: Majorana fermions in nuclear, particle, and solid-state physics},\ }\href {https://doi.org/10.1103/RevModPhys.87.137} {\bibfield  {journal} {\bibinfo  {journal} {Rev. Mod. Phys.}\ }\textbf {\bibinfo {volume} {87}},\ \bibinfo {pages} {137} (\bibinfo {year} {2015})}\BibitemShut {NoStop}%
\bibitem [{\citenamefont {Aguado}(2017)}]{Aguado2017}%
  \BibitemOpen
  \bibfield  {author} {\bibinfo {author} {\bibfnamefont {R.}~\bibnamefont {Aguado}},\ }\bibfield  {title} {\bibinfo {title} {Majorana quasiparticles in condensed matter},\ }\href {https://doi.org/10.1393/ncr/i2017-10141-9} {\bibfield  {journal} {\bibinfo  {journal} {La Rivista del Nuovo Cimento}\ }\textbf {\bibinfo {volume} {40}},\ \bibinfo {pages} {523} (\bibinfo {year} {2017})}\BibitemShut {NoStop}%
\bibitem [{\citenamefont {Beenakker}(2020)}]{Beenakker2020}%
  \BibitemOpen
  \bibfield  {author} {\bibinfo {author} {\bibfnamefont {C.~W.~J.}\ \bibnamefont {Beenakker}},\ }\bibfield  {title} {\bibinfo {title} {{Search for non-Abelian Majorana braiding statistics in superconductors}},\ }\href {https://doi.org/10.21468/SciPostPhysLectNotes.15} {\bibfield  {journal} {\bibinfo  {journal} {SciPost Phys. Lect. Notes}\ ,\ \bibinfo {pages} {15}} (\bibinfo {year} {2020})}\BibitemShut {NoStop}%
\bibitem [{\citenamefont {Tanaka}\ \emph {et~al.}(2024)\citenamefont {Tanaka}, \citenamefont {Tamura},\ and\ \citenamefont {Cayao}}]{Tanaka2024}%
  \BibitemOpen
  \bibfield  {author} {\bibinfo {author} {\bibfnamefont {Y.}~\bibnamefont {Tanaka}}, \bibinfo {author} {\bibfnamefont {S.}~\bibnamefont {Tamura}},\ and\ \bibinfo {author} {\bibfnamefont {J.}~\bibnamefont {Cayao}},\ }\bibfield  {title} {\bibinfo {title} {Theory of majorana zero modes in unconventional superconductors},\ }\href {https://doi.org/10.1093/ptep/ptae065} {\bibfield  {journal} {\bibinfo  {journal} {Progress of Theoretical and Experimental Physics}\ ,\ \bibinfo {pages} {ptae065}} (\bibinfo {year} {2024})}\BibitemShut {NoStop}%
\bibitem [{\citenamefont {Pita-Vidal}\ \emph {et~al.}(2025)\citenamefont {Pita-Vidal}, \citenamefont {Souto}, \citenamefont {Goswami}, \citenamefont {Andersen}, \citenamefont {Katsaros}, \citenamefont {Shabani},\ and\ \citenamefont {Aguado}}]{Pita2026}%
  \BibitemOpen
  \bibfield  {author} {\bibinfo {author} {\bibfnamefont {M.}~\bibnamefont {Pita-Vidal}}, \bibinfo {author} {\bibfnamefont {R.~S.}\ \bibnamefont {Souto}}, \bibinfo {author} {\bibfnamefont {S.}~\bibnamefont {Goswami}}, \bibinfo {author} {\bibfnamefont {C.~K.}\ \bibnamefont {Andersen}}, \bibinfo {author} {\bibfnamefont {G.}~\bibnamefont {Katsaros}}, \bibinfo {author} {\bibfnamefont {J.}~\bibnamefont {Shabani}},\ and\ \bibinfo {author} {\bibfnamefont {R.}~\bibnamefont {Aguado}},\ }\bibfield  {title} {\bibinfo {title} {Novel qubits in hybrid semiconductor-superconductor nanostructures},\ }\href {https://arxiv.org/abs/2512.23336} {\bibfield  {journal} {\bibinfo  {journal} {arXiv preprint arXiv:2512.23336}\ } (\bibinfo {year} {2025})}\BibitemShut {NoStop}%
\bibitem [{\citenamefont {Kitaev}(2003)}]{KITAEV2003}%
  \BibitemOpen
  \bibfield  {author} {\bibinfo {author} {\bibfnamefont {A.~Y.}\ \bibnamefont {Kitaev}},\ }\bibfield  {title} {\bibinfo {title} {Fault-tolerant quantum computation by anyons},\ }\href {https://doi.org/10.1016/S0003-4916(02)00018-0} {\bibfield  {journal} {\bibinfo  {journal} {Annals of physics}\ }\textbf {\bibinfo {volume} {303}},\ \bibinfo {pages} {2} (\bibinfo {year} {2003})}\BibitemShut {NoStop}%
\bibitem [{\citenamefont {Nayak}\ \emph {et~al.}(2008)\citenamefont {Nayak}, \citenamefont {Simon}, \citenamefont {Stern}, \citenamefont {Freedman},\ and\ \citenamefont {Das~Sarma}}]{RevModPhys.80.1083}%
  \BibitemOpen
  \bibfield  {author} {\bibinfo {author} {\bibfnamefont {C.}~\bibnamefont {Nayak}}, \bibinfo {author} {\bibfnamefont {S.~H.}\ \bibnamefont {Simon}}, \bibinfo {author} {\bibfnamefont {A.}~\bibnamefont {Stern}}, \bibinfo {author} {\bibfnamefont {M.}~\bibnamefont {Freedman}},\ and\ \bibinfo {author} {\bibfnamefont {S.}~\bibnamefont {Das~Sarma}},\ }\bibfield  {title} {\bibinfo {title} {Non-abelian anyons and topological quantum computation},\ }\href {https://doi.org/10.1103/RevModPhys.80.1083} {\bibfield  {journal} {\bibinfo  {journal} {Rev. Mod. Phys.}\ }\textbf {\bibinfo {volume} {80}},\ \bibinfo {pages} {1083} (\bibinfo {year} {2008})}\BibitemShut {NoStop}%
\bibitem [{\citenamefont {Sarma}\ \emph {et~al.}(2015)\citenamefont {Sarma}, \citenamefont {Freedman},\ and\ \citenamefont {Nayak}}]{Sarma-Freedman-Nayak}%
  \BibitemOpen
  \bibfield  {author} {\bibinfo {author} {\bibfnamefont {S.~D.}\ \bibnamefont {Sarma}}, \bibinfo {author} {\bibfnamefont {M.}~\bibnamefont {Freedman}},\ and\ \bibinfo {author} {\bibfnamefont {C.}~\bibnamefont {Nayak}},\ }\bibfield  {title} {\bibinfo {title} {Majorana zero modes and topological quantum computation},\ }\href {https://doi.org/10.1038/npjqi.2015.1} {\bibfield  {journal} {\bibinfo  {journal} {npj Quantum Information}\ }\textbf {\bibinfo {volume} {1}},\ \bibinfo {pages} {15001} (\bibinfo {year} {2015})}\BibitemShut {NoStop}%
\bibitem [{\citenamefont {Wilczek}(2009)}]{Wilczek2009}%
  \BibitemOpen
  \bibfield  {author} {\bibinfo {author} {\bibfnamefont {F.}~\bibnamefont {Wilczek}},\ }\bibfield  {title} {\bibinfo {title} {Majorana returns},\ }\href {https://doi.org/10.1038/nphys1380} {\bibfield  {journal} {\bibinfo  {journal} {Nature Physics}\ }\textbf {\bibinfo {volume} {5}},\ \bibinfo {pages} {614} (\bibinfo {year} {2009})}\BibitemShut {NoStop}%
\bibitem [{\citenamefont {Aguado}\ and\ \citenamefont {Kouwenhoven}(2020)}]{Aguado2020}%
  \BibitemOpen
  \bibfield  {author} {\bibinfo {author} {\bibfnamefont {R.}~\bibnamefont {Aguado}}\ and\ \bibinfo {author} {\bibfnamefont {L.~P.}\ \bibnamefont {Kouwenhoven}},\ }\bibfield  {title} {\bibinfo {title} {Majorana qubits for topological quantum computing},\ }\href {https://doi.org/10.1063/PT.3.4499} {\bibfield  {journal} {\bibinfo  {journal} {Physics Today}\ }\textbf {\bibinfo {volume} {73}},\ \bibinfo {pages} {44} (\bibinfo {year} {2020})}\BibitemShut {NoStop}%
\bibitem [{\citenamefont {Wang}\ \emph {et~al.}(2022)\citenamefont {Wang}, \citenamefont {Dvir}, \citenamefont {Mazur}, \citenamefont {Liu}, \citenamefont {van Loo}, \citenamefont {Ten~Haaf}, \citenamefont {Bordin}, \citenamefont {Gazibegovic}, \citenamefont {Badawy}, \citenamefont {Bakkers} \emph {et~al.}}]{Wang2022}%
  \BibitemOpen
  \bibfield  {author} {\bibinfo {author} {\bibfnamefont {G.}~\bibnamefont {Wang}}, \bibinfo {author} {\bibfnamefont {T.}~\bibnamefont {Dvir}}, \bibinfo {author} {\bibfnamefont {G.~P.}\ \bibnamefont {Mazur}}, \bibinfo {author} {\bibfnamefont {C.-X.}\ \bibnamefont {Liu}}, \bibinfo {author} {\bibfnamefont {N.}~\bibnamefont {van Loo}}, \bibinfo {author} {\bibfnamefont {S.~L.}\ \bibnamefont {Ten~Haaf}}, \bibinfo {author} {\bibfnamefont {A.}~\bibnamefont {Bordin}}, \bibinfo {author} {\bibfnamefont {S.}~\bibnamefont {Gazibegovic}}, \bibinfo {author} {\bibfnamefont {G.}~\bibnamefont {Badawy}}, \bibinfo {author} {\bibfnamefont {E.~P.}\ \bibnamefont {Bakkers}}, \emph {et~al.},\ }\bibfield  {title} {\bibinfo {title} {Singlet and triplet cooper pair splitting in hybrid superconducting nanowires},\ }\href {https://doi.org/10.1038/s41586-022-05352-2} {\bibfield  {journal} {\bibinfo  {journal} {Nature}\ }\textbf {\bibinfo {volume} {612}},\ \bibinfo {pages} {448} (\bibinfo {year} {2022})}\BibitemShut {NoStop}%
\bibitem [{\citenamefont {Wang}\ \emph {et~al.}(2023)\citenamefont {Wang}, \citenamefont {Ten~Haaf}, \citenamefont {Kulesh}, \citenamefont {Xiao}, \citenamefont {Thomas}, \citenamefont {Manfra},\ and\ \citenamefont {Goswami}}]{Wang2023}%
  \BibitemOpen
  \bibfield  {author} {\bibinfo {author} {\bibfnamefont {Q.}~\bibnamefont {Wang}}, \bibinfo {author} {\bibfnamefont {S.~L.}\ \bibnamefont {Ten~Haaf}}, \bibinfo {author} {\bibfnamefont {I.}~\bibnamefont {Kulesh}}, \bibinfo {author} {\bibfnamefont {D.}~\bibnamefont {Xiao}}, \bibinfo {author} {\bibfnamefont {C.}~\bibnamefont {Thomas}}, \bibinfo {author} {\bibfnamefont {M.~J.}\ \bibnamefont {Manfra}},\ and\ \bibinfo {author} {\bibfnamefont {S.}~\bibnamefont {Goswami}},\ }\bibfield  {title} {\bibinfo {title} {Triplet correlations in cooper pair splitters realized in a two-dimensional electron gas},\ }\href {https://doi.org/10.1038/s41467-023-40551-z} {\bibfield  {journal} {\bibinfo  {journal} {Nature Communications}\ }\textbf {\bibinfo {volume} {14}},\ \bibinfo {pages} {4876} (\bibinfo {year} {2023})}\BibitemShut {NoStop}%
\bibitem [{\citenamefont {Bordin}\ \emph {et~al.}(2023)\citenamefont {Bordin}, \citenamefont {Wang}, \citenamefont {Liu}, \citenamefont {Ten~Haaf}, \citenamefont {Van~Loo}, \citenamefont {Mazur}, \citenamefont {Xu}, \citenamefont {Van~Driel}, \citenamefont {Zatelli}, \citenamefont {Gazibegovic} \emph {et~al.}}]{Bordin2022}%
  \BibitemOpen
  \bibfield  {author} {\bibinfo {author} {\bibfnamefont {A.}~\bibnamefont {Bordin}}, \bibinfo {author} {\bibfnamefont {G.}~\bibnamefont {Wang}}, \bibinfo {author} {\bibfnamefont {C.-X.}\ \bibnamefont {Liu}}, \bibinfo {author} {\bibfnamefont {S.~L.}\ \bibnamefont {Ten~Haaf}}, \bibinfo {author} {\bibfnamefont {N.}~\bibnamefont {Van~Loo}}, \bibinfo {author} {\bibfnamefont {G.~P.}\ \bibnamefont {Mazur}}, \bibinfo {author} {\bibfnamefont {D.}~\bibnamefont {Xu}}, \bibinfo {author} {\bibfnamefont {D.}~\bibnamefont {Van~Driel}}, \bibinfo {author} {\bibfnamefont {F.}~\bibnamefont {Zatelli}}, \bibinfo {author} {\bibfnamefont {S.}~\bibnamefont {Gazibegovic}}, \emph {et~al.},\ }\bibfield  {title} {\bibinfo {title} {Tunable crossed andreev reflection and elastic cotunneling in hybrid nanowires},\ }\href {https://doi.org/10.1103/physrevx.13.031031} {\bibfield  {journal} {\bibinfo  {journal} {Physical Review X}\ }\textbf {\bibinfo {volume} {13}},\ \bibinfo {pages} {031031} (\bibinfo {year} {2023})}\BibitemShut {NoStop}%
\bibitem [{\citenamefont {Liu}\ \emph {et~al.}(2022)\citenamefont {Liu}, \citenamefont {Wang}, \citenamefont {Dvir},\ and\ \citenamefont {Wimmer}}]{Liu2022}%
  \BibitemOpen
  \bibfield  {author} {\bibinfo {author} {\bibfnamefont {C.-X.}\ \bibnamefont {Liu}}, \bibinfo {author} {\bibfnamefont {G.}~\bibnamefont {Wang}}, \bibinfo {author} {\bibfnamefont {T.}~\bibnamefont {Dvir}},\ and\ \bibinfo {author} {\bibfnamefont {M.}~\bibnamefont {Wimmer}},\ }\bibfield  {title} {\bibinfo {title} {Tunable superconducting coupling of quantum dots via andreev bound states in semiconductor-superconductor nanowires},\ }\href {https://doi.org/10.1103/PhysRevLett.129.267701} {\bibfield  {journal} {\bibinfo  {journal} {Phys. Rev. Lett.}\ }\textbf {\bibinfo {volume} {129}},\ \bibinfo {pages} {267701} (\bibinfo {year} {2022})}\BibitemShut {NoStop}%
\bibitem [{\citenamefont {Dvir}\ \emph {et~al.}(2023)\citenamefont {Dvir}, \citenamefont {Wang}, \citenamefont {van Loo}, \citenamefont {Liu}, \citenamefont {Mazur}, \citenamefont {Bordin}, \citenamefont {Ten~Haaf}, \citenamefont {Wang}, \citenamefont {van Driel}, \citenamefont {Zatelli} \emph {et~al.}}]{Dvir2023}%
  \BibitemOpen
  \bibfield  {author} {\bibinfo {author} {\bibfnamefont {T.}~\bibnamefont {Dvir}}, \bibinfo {author} {\bibfnamefont {G.}~\bibnamefont {Wang}}, \bibinfo {author} {\bibfnamefont {N.}~\bibnamefont {van Loo}}, \bibinfo {author} {\bibfnamefont {C.-X.}\ \bibnamefont {Liu}}, \bibinfo {author} {\bibfnamefont {G.~P.}\ \bibnamefont {Mazur}}, \bibinfo {author} {\bibfnamefont {A.}~\bibnamefont {Bordin}}, \bibinfo {author} {\bibfnamefont {S.~L.}\ \bibnamefont {Ten~Haaf}}, \bibinfo {author} {\bibfnamefont {J.-Y.}\ \bibnamefont {Wang}}, \bibinfo {author} {\bibfnamefont {D.}~\bibnamefont {van Driel}}, \bibinfo {author} {\bibfnamefont {F.}~\bibnamefont {Zatelli}}, \emph {et~al.},\ }\bibfield  {title} {\bibinfo {title} {Realization of a minimal kitaev chain in coupled quantum dots},\ }\href {https://doi.org/10.1038/s41586-022-05585-1} {\bibfield  {journal} {\bibinfo  {journal} {Nature}\ }\textbf {\bibinfo {volume} {614}},\ \bibinfo {pages} {445} (\bibinfo {year} {2023})}\BibitemShut {NoStop}%
\bibitem [{\citenamefont {Bordin}\ \emph {et~al.}(2024)\citenamefont {Bordin}, \citenamefont {Li}, \citenamefont {van Driel}, \citenamefont {Wolff}, \citenamefont {Wang}, \citenamefont {ten Haaf}, \citenamefont {Wang}, \citenamefont {van Loo}, \citenamefont {Kouwenhoven},\ and\ \citenamefont {Dvir}}]{Bordin2023}%
  \BibitemOpen
  \bibfield  {author} {\bibinfo {author} {\bibfnamefont {A.}~\bibnamefont {Bordin}}, \bibinfo {author} {\bibfnamefont {X.}~\bibnamefont {Li}}, \bibinfo {author} {\bibfnamefont {D.}~\bibnamefont {van Driel}}, \bibinfo {author} {\bibfnamefont {J.~C.}\ \bibnamefont {Wolff}}, \bibinfo {author} {\bibfnamefont {Q.}~\bibnamefont {Wang}}, \bibinfo {author} {\bibfnamefont {S.~L.~D.}\ \bibnamefont {ten Haaf}}, \bibinfo {author} {\bibfnamefont {G.}~\bibnamefont {Wang}}, \bibinfo {author} {\bibfnamefont {N.}~\bibnamefont {van Loo}}, \bibinfo {author} {\bibfnamefont {L.~P.}\ \bibnamefont {Kouwenhoven}},\ and\ \bibinfo {author} {\bibfnamefont {T.}~\bibnamefont {Dvir}},\ }\bibfield  {title} {\bibinfo {title} {Crossed andreev reflection and elastic cotunneling in three quantum dots coupled by superconductors},\ }\href {https://doi.org/10.1103/PhysRevLett.132.056602} {\bibfield  {journal} {\bibinfo  {journal} {Phys. Rev. Lett.}\ }\textbf {\bibinfo {volume} {132}},\ \bibinfo {pages} {056602} (\bibinfo {year}
  {2024})}\BibitemShut {NoStop}%
\bibitem [{\citenamefont {Ten~Haaf}\ \emph {et~al.}(2024)\citenamefont {Ten~Haaf}, \citenamefont {Wang}, \citenamefont {Bozkurt}, \citenamefont {Liu}, \citenamefont {Kulesh}, \citenamefont {Kim}, \citenamefont {Xiao}, \citenamefont {Thomas}, \citenamefont {Manfra}, \citenamefont {Dvir} \emph {et~al.}}]{Haaf2023}%
  \BibitemOpen
  \bibfield  {author} {\bibinfo {author} {\bibfnamefont {S.~L.}\ \bibnamefont {Ten~Haaf}}, \bibinfo {author} {\bibfnamefont {Q.}~\bibnamefont {Wang}}, \bibinfo {author} {\bibfnamefont {A.~M.}\ \bibnamefont {Bozkurt}}, \bibinfo {author} {\bibfnamefont {C.-X.}\ \bibnamefont {Liu}}, \bibinfo {author} {\bibfnamefont {I.}~\bibnamefont {Kulesh}}, \bibinfo {author} {\bibfnamefont {P.}~\bibnamefont {Kim}}, \bibinfo {author} {\bibfnamefont {D.}~\bibnamefont {Xiao}}, \bibinfo {author} {\bibfnamefont {C.}~\bibnamefont {Thomas}}, \bibinfo {author} {\bibfnamefont {M.~J.}\ \bibnamefont {Manfra}}, \bibinfo {author} {\bibfnamefont {T.}~\bibnamefont {Dvir}}, \emph {et~al.},\ }\bibfield  {title} {\bibinfo {title} {A two-site kitaev chain in a two-dimensional electron gas},\ }\href {https://doi.org/10.1038/s41586-024-07434-9} {\bibfield  {journal} {\bibinfo  {journal} {Nature}\ }\textbf {\bibinfo {volume} {630}},\ \bibinfo {pages} {329} (\bibinfo {year} {2024})}\BibitemShut {NoStop}%
\bibitem [{\citenamefont {Bordin}\ \emph {et~al.}(2025{\natexlab{a}})\citenamefont {Bordin}, \citenamefont {Liu}, \citenamefont {Dvir}, \citenamefont {Zatelli}, \citenamefont {Ten~Haaf}, \citenamefont {van Driel}, \citenamefont {Wang}, \citenamefont {Van~Loo}, \citenamefont {Zhang}, \citenamefont {Wolff} \emph {et~al.}}]{Bordin2025}%
  \BibitemOpen
  \bibfield  {author} {\bibinfo {author} {\bibfnamefont {A.}~\bibnamefont {Bordin}}, \bibinfo {author} {\bibfnamefont {C.-X.}\ \bibnamefont {Liu}}, \bibinfo {author} {\bibfnamefont {T.}~\bibnamefont {Dvir}}, \bibinfo {author} {\bibfnamefont {F.}~\bibnamefont {Zatelli}}, \bibinfo {author} {\bibfnamefont {S.~L.}\ \bibnamefont {Ten~Haaf}}, \bibinfo {author} {\bibfnamefont {D.}~\bibnamefont {van Driel}}, \bibinfo {author} {\bibfnamefont {G.}~\bibnamefont {Wang}}, \bibinfo {author} {\bibfnamefont {N.}~\bibnamefont {Van~Loo}}, \bibinfo {author} {\bibfnamefont {Y.}~\bibnamefont {Zhang}}, \bibinfo {author} {\bibfnamefont {J.~C.}\ \bibnamefont {Wolff}}, \emph {et~al.},\ }\bibfield  {title} {\bibinfo {title} {Enhanced majorana stability in a three-site kitaev chain},\ }\href {https://doi.org/https://doi.org/10.1038/s41565-025-01894-4} {\bibfield  {journal} {\bibinfo  {journal} {Nature Nanotechnology}\ ,\ \bibinfo {pages} {1}} (\bibinfo {year} {2025}{\natexlab{a}})}\BibitemShut {NoStop}%
\bibitem [{\citenamefont {Bordin}\ \emph {et~al.}(2026)\citenamefont {Bordin}, \citenamefont {Bennebroek~Evertsz’}, \citenamefont {Roovers}, \citenamefont {Torres~Luna}, \citenamefont {Huisman}, \citenamefont {Zatelli}, \citenamefont {Mazur}, \citenamefont {Ten~Haaf}, \citenamefont {Badawy}, \citenamefont {Bakkers} \emph {et~al.}}]{Bordin2025b}%
  \BibitemOpen
  \bibfield  {author} {\bibinfo {author} {\bibfnamefont {A.}~\bibnamefont {Bordin}}, \bibinfo {author} {\bibfnamefont {F.~J.}\ \bibnamefont {Bennebroek~Evertsz’}}, \bibinfo {author} {\bibfnamefont {B.}~\bibnamefont {Roovers}}, \bibinfo {author} {\bibfnamefont {J.~D.}\ \bibnamefont {Torres~Luna}}, \bibinfo {author} {\bibfnamefont {W.~D.}\ \bibnamefont {Huisman}}, \bibinfo {author} {\bibfnamefont {F.}~\bibnamefont {Zatelli}}, \bibinfo {author} {\bibfnamefont {G.~P.}\ \bibnamefont {Mazur}}, \bibinfo {author} {\bibfnamefont {S.~L.}\ \bibnamefont {Ten~Haaf}}, \bibinfo {author} {\bibfnamefont {G.}~\bibnamefont {Badawy}}, \bibinfo {author} {\bibfnamefont {E.~P.}\ \bibnamefont {Bakkers}}, \emph {et~al.},\ }\bibfield  {title} {\bibinfo {title} {Probing majorana localization of a phase-controlled three-site kitaev chain with an additional quantum dot},\ }\href {https://doi.org/10.1038/s41467-026-68897-0} {\bibfield  {journal} {\bibinfo  {journal} {Nature Communications}\ } (\bibinfo {year} {2026})}\BibitemShut
  {NoStop}%
\bibitem [{\citenamefont {van Loo}\ \emph {et~al.}(2026)\citenamefont {van Loo}, \citenamefont {Zatelli}, \citenamefont {Steffensen}, \citenamefont {Roovers}, \citenamefont {Wang}, \citenamefont {Van~Caekenberghe}, \citenamefont {Bordin}, \citenamefont {van Driel}, \citenamefont {Zhang}, \citenamefont {Huisman} \emph {et~al.}}]{vanloo2025}%
  \BibitemOpen
  \bibfield  {author} {\bibinfo {author} {\bibfnamefont {N.}~\bibnamefont {van Loo}}, \bibinfo {author} {\bibfnamefont {F.}~\bibnamefont {Zatelli}}, \bibinfo {author} {\bibfnamefont {G.~O.}\ \bibnamefont {Steffensen}}, \bibinfo {author} {\bibfnamefont {B.}~\bibnamefont {Roovers}}, \bibinfo {author} {\bibfnamefont {G.}~\bibnamefont {Wang}}, \bibinfo {author} {\bibfnamefont {T.}~\bibnamefont {Van~Caekenberghe}}, \bibinfo {author} {\bibfnamefont {A.}~\bibnamefont {Bordin}}, \bibinfo {author} {\bibfnamefont {D.}~\bibnamefont {van Driel}}, \bibinfo {author} {\bibfnamefont {Y.}~\bibnamefont {Zhang}}, \bibinfo {author} {\bibfnamefont {W.~D.}\ \bibnamefont {Huisman}}, \emph {et~al.},\ }\bibfield  {title} {\bibinfo {title} {Single-shot parity readout of a minimal kitaev chain},\ }\href {https://doi.org/10.1038/s41586-025-09927-7} {\bibfield  {journal} {\bibinfo  {journal} {Nature}\ }\textbf {\bibinfo {volume} {650}},\ \bibinfo {pages} {334} (\bibinfo {year} {2026})}\BibitemShut {NoStop}%
\bibitem [{\citenamefont {ten Haaf}\ \emph {et~al.}(2025)\citenamefont {ten Haaf}, \citenamefont {Zhang}, \citenamefont {Wang}, \citenamefont {Bordin}, \citenamefont {Liu}, \citenamefont {Kulesh}, \citenamefont {Sietses}, \citenamefont {Prosko}, \citenamefont {Xiao}, \citenamefont {Thomas}, \citenamefont {Manfra}, \citenamefont {Wimmer},\ and\ \citenamefont {Goswami}}]{Bas-Haaf2025}%
  \BibitemOpen
  \bibfield  {author} {\bibinfo {author} {\bibfnamefont {S.~L.~D.}\ \bibnamefont {ten Haaf}}, \bibinfo {author} {\bibfnamefont {Y.}~\bibnamefont {Zhang}}, \bibinfo {author} {\bibfnamefont {Q.}~\bibnamefont {Wang}}, \bibinfo {author} {\bibfnamefont {A.}~\bibnamefont {Bordin}}, \bibinfo {author} {\bibfnamefont {C.-X.}\ \bibnamefont {Liu}}, \bibinfo {author} {\bibfnamefont {I.}~\bibnamefont {Kulesh}}, \bibinfo {author} {\bibfnamefont {V.~P.~M.}\ \bibnamefont {Sietses}}, \bibinfo {author} {\bibfnamefont {C.~G.}\ \bibnamefont {Prosko}}, \bibinfo {author} {\bibfnamefont {D.}~\bibnamefont {Xiao}}, \bibinfo {author} {\bibfnamefont {C.}~\bibnamefont {Thomas}}, \bibinfo {author} {\bibfnamefont {M.~J.}\ \bibnamefont {Manfra}}, \bibinfo {author} {\bibfnamefont {M.}~\bibnamefont {Wimmer}},\ and\ \bibinfo {author} {\bibfnamefont {S.}~\bibnamefont {Goswami}},\ }\bibfield  {title} {\bibinfo {title} {Observation of edge and bulk states in a three-site kitaev chain},\ }\href {https://doi.org/10.1038/s41586-025-08892-5}
  {\bibfield  {journal} {\bibinfo  {journal} {Nature}\ }\textbf {\bibinfo {volume} {641}},\ \bibinfo {pages} {890} (\bibinfo {year} {2025})}\BibitemShut {NoStop}%
\bibitem [{\citenamefont {Kulesh}\ \emph {et~al.}(2025)\citenamefont {Kulesh}, \citenamefont {ten Haaf}, \citenamefont {Wang}, \citenamefont {Sietses}, \citenamefont {Zhang}, \citenamefont {Roelofs}, \citenamefont {Prosko}, \citenamefont {Xiao}, \citenamefont {Thomas}, \citenamefont {Manfra},\ and\ \citenamefont {Goswami}}]{Kulesh2025}%
  \BibitemOpen
  \bibfield  {author} {\bibinfo {author} {\bibfnamefont {I.}~\bibnamefont {Kulesh}}, \bibinfo {author} {\bibfnamefont {S.~L.~D.}\ \bibnamefont {ten Haaf}}, \bibinfo {author} {\bibfnamefont {Q.}~\bibnamefont {Wang}}, \bibinfo {author} {\bibfnamefont {V.~P.~M.}\ \bibnamefont {Sietses}}, \bibinfo {author} {\bibfnamefont {Y.}~\bibnamefont {Zhang}}, \bibinfo {author} {\bibfnamefont {S.~R.}\ \bibnamefont {Roelofs}}, \bibinfo {author} {\bibfnamefont {C.~G.}\ \bibnamefont {Prosko}}, \bibinfo {author} {\bibfnamefont {D.}~\bibnamefont {Xiao}}, \bibinfo {author} {\bibfnamefont {C.}~\bibnamefont {Thomas}}, \bibinfo {author} {\bibfnamefont {M.~J.}\ \bibnamefont {Manfra}},\ and\ \bibinfo {author} {\bibfnamefont {S.}~\bibnamefont {Goswami}},\ }\bibfield  {title} {\bibinfo {title} {Flux-controlled two-site kitaev chain},\ }\href {https://doi.org/10.1103/r9pv-2prs} {\bibfield  {journal} {\bibinfo  {journal} {Phys. Rev. Lett.}\ }\textbf {\bibinfo {volume} {135}},\ \bibinfo {pages} {056301} (\bibinfo {year} {2025})}\BibitemShut
  {NoStop}%
\bibitem [{\citenamefont {Leijnse}\ and\ \citenamefont {Flensberg}(2012{\natexlab{b}})}]{Leijnse2012}%
  \BibitemOpen
  \bibfield  {author} {\bibinfo {author} {\bibfnamefont {M.}~\bibnamefont {Leijnse}}\ and\ \bibinfo {author} {\bibfnamefont {K.}~\bibnamefont {Flensberg}},\ }\bibfield  {title} {\bibinfo {title} {Parity qubits and poor man's majorana bound states in double quantum dots},\ }\href {https://doi.org/10.1103/PhysRevB.86.134528} {\bibfield  {journal} {\bibinfo  {journal} {Phys. Rev. B}\ }\textbf {\bibinfo {volume} {86}},\ \bibinfo {pages} {134528} (\bibinfo {year} {2012}{\natexlab{b}})}\BibitemShut {NoStop}%
\bibitem [{\citenamefont {Sau}\ and\ \citenamefont {Sarma}(2012)}]{Sau2012}%
  \BibitemOpen
  \bibfield  {author} {\bibinfo {author} {\bibfnamefont {J.~D.}\ \bibnamefont {Sau}}\ and\ \bibinfo {author} {\bibfnamefont {S.~D.}\ \bibnamefont {Sarma}},\ }\bibfield  {title} {\bibinfo {title} {Realizing a robust practical majorana chain in a quantum-dot-superconductor linear array},\ }\href {https://doi.org/10.1038/ncomms1966} {\bibfield  {journal} {\bibinfo  {journal} {Nature communications}\ }\textbf {\bibinfo {volume} {3}},\ \bibinfo {pages} {964} (\bibinfo {year} {2012})}\BibitemShut {NoStop}%
\bibitem [{\citenamefont {Fulga}\ \emph {et~al.}(2013)\citenamefont {Fulga}, \citenamefont {Haim}, \citenamefont {Akhmerov},\ and\ \citenamefont {Oreg}}]{Fulga2013}%
  \BibitemOpen
  \bibfield  {author} {\bibinfo {author} {\bibfnamefont {I.~C.}\ \bibnamefont {Fulga}}, \bibinfo {author} {\bibfnamefont {A.}~\bibnamefont {Haim}}, \bibinfo {author} {\bibfnamefont {A.~R.}\ \bibnamefont {Akhmerov}},\ and\ \bibinfo {author} {\bibfnamefont {Y.}~\bibnamefont {Oreg}},\ }\bibfield  {title} {\bibinfo {title} {Adaptive tuning of majorana fermions in a quantum dot chain},\ }\href {https://doi.org/10.1088/1367-2630/15/4/045020} {\bibfield  {journal} {\bibinfo  {journal} {New journal of physics}\ }\textbf {\bibinfo {volume} {15}},\ \bibinfo {pages} {045020} (\bibinfo {year} {2013})}\BibitemShut {NoStop}%
\bibitem [{\citenamefont {Luna}\ \emph {et~al.}(2025)\citenamefont {Luna}, \citenamefont {Miles}, \citenamefont {Bozkurt}, \citenamefont {Liu}, \citenamefont {Manesco}, \citenamefont {Akhmerov},\ and\ \citenamefont {Wimmer}}]{Wimmer2025}%
  \BibitemOpen
  \bibfield  {author} {\bibinfo {author} {\bibfnamefont {J.~D.~T.}\ \bibnamefont {Luna}}, \bibinfo {author} {\bibfnamefont {S.}~\bibnamefont {Miles}}, \bibinfo {author} {\bibfnamefont {A.~M.}\ \bibnamefont {Bozkurt}}, \bibinfo {author} {\bibfnamefont {C.-X.}\ \bibnamefont {Liu}}, \bibinfo {author} {\bibfnamefont {A.~L.}\ \bibnamefont {Manesco}}, \bibinfo {author} {\bibfnamefont {A.~R.}\ \bibnamefont {Akhmerov}},\ and\ \bibinfo {author} {\bibfnamefont {M.}~\bibnamefont {Wimmer}},\ }\bibfield  {title} {\bibinfo {title} {Pareto-optimality of majoranas in hybrid platforms},\ }\href {https://doi.org/10.48550/arXiv.2510.07406} {\bibfield  {journal} {\bibinfo  {journal} {arXiv preprint arXiv:2510.07406}\ } (\bibinfo {year} {2025})}\BibitemShut {NoStop}%
\bibitem [{\citenamefont {Tsintzis}\ \emph {et~al.}(2022)\citenamefont {Tsintzis}, \citenamefont {Souto},\ and\ \citenamefont {Leijnse}}]{Souto2022}%
  \BibitemOpen
  \bibfield  {author} {\bibinfo {author} {\bibfnamefont {A.}~\bibnamefont {Tsintzis}}, \bibinfo {author} {\bibfnamefont {R.~S.}\ \bibnamefont {Souto}},\ and\ \bibinfo {author} {\bibfnamefont {M.}~\bibnamefont {Leijnse}},\ }\bibfield  {title} {\bibinfo {title} {Creating and detecting poor man's majorana bound states in interacting quantum dots},\ }\href {https://doi.org/10.1103/PhysRevB.106.L201404} {\bibfield  {journal} {\bibinfo  {journal} {Phys. Rev. B}\ }\textbf {\bibinfo {volume} {106}},\ \bibinfo {pages} {L201404} (\bibinfo {year} {2022})}\BibitemShut {NoStop}%
\bibitem [{\citenamefont {Souto}\ \emph {et~al.}(2023)\citenamefont {Souto}, \citenamefont {Tsintzis}, \citenamefont {Leijnse},\ and\ \citenamefont {Danon}}]{Seoane2023}%
  \BibitemOpen
  \bibfield  {author} {\bibinfo {author} {\bibfnamefont {R.~S.}\ \bibnamefont {Souto}}, \bibinfo {author} {\bibfnamefont {A.}~\bibnamefont {Tsintzis}}, \bibinfo {author} {\bibfnamefont {M.}~\bibnamefont {Leijnse}},\ and\ \bibinfo {author} {\bibfnamefont {J.}~\bibnamefont {Danon}},\ }\bibfield  {title} {\bibinfo {title} {Probing majorana localization in minimal kitaev chains through a quantum dot},\ }\href {https://doi.org/10.1103/PhysRevResearch.5.043182} {\bibfield  {journal} {\bibinfo  {journal} {Phys. Rev. Res.}\ }\textbf {\bibinfo {volume} {5}},\ \bibinfo {pages} {043182} (\bibinfo {year} {2023})}\BibitemShut {NoStop}%
\bibitem [{\citenamefont {Tsintzis}\ \emph {et~al.}(2024)\citenamefont {Tsintzis}, \citenamefont {Souto}, \citenamefont {Flensberg}, \citenamefont {Danon},\ and\ \citenamefont {Leijnse}}]{Tsintzis2024}%
  \BibitemOpen
  \bibfield  {author} {\bibinfo {author} {\bibfnamefont {A.}~\bibnamefont {Tsintzis}}, \bibinfo {author} {\bibfnamefont {R.~S.}\ \bibnamefont {Souto}}, \bibinfo {author} {\bibfnamefont {K.}~\bibnamefont {Flensberg}}, \bibinfo {author} {\bibfnamefont {J.}~\bibnamefont {Danon}},\ and\ \bibinfo {author} {\bibfnamefont {M.}~\bibnamefont {Leijnse}},\ }\bibfield  {title} {\bibinfo {title} {Majorana qubits and non-abelian physics in quantum dot--based minimal kitaev chains},\ }\href {https://doi.org/10.1103/PRXQuantum.5.010323} {\bibfield  {journal} {\bibinfo  {journal} {PRX Quantum}\ }\textbf {\bibinfo {volume} {5}},\ \bibinfo {pages} {010323} (\bibinfo {year} {2024})}\BibitemShut {NoStop}%
\bibitem [{\citenamefont {Miles}\ \emph {et~al.}(2024)\citenamefont {Miles}, \citenamefont {van Driel}, \citenamefont {Wimmer},\ and\ \citenamefont {Liu}}]{Miles2023}%
  \BibitemOpen
  \bibfield  {author} {\bibinfo {author} {\bibfnamefont {S.}~\bibnamefont {Miles}}, \bibinfo {author} {\bibfnamefont {D.}~\bibnamefont {van Driel}}, \bibinfo {author} {\bibfnamefont {M.}~\bibnamefont {Wimmer}},\ and\ \bibinfo {author} {\bibfnamefont {C.-X.}\ \bibnamefont {Liu}},\ }\bibfield  {title} {\bibinfo {title} {Kitaev chain in an alternating quantum dot-andreev bound state array},\ }\href {https://doi.org/10.1103/PhysRevB.110.024520} {\bibfield  {journal} {\bibinfo  {journal} {Phys. Rev. B}\ }\textbf {\bibinfo {volume} {110}},\ \bibinfo {pages} {024520} (\bibinfo {year} {2024})}\BibitemShut {NoStop}%
\bibitem [{\citenamefont {Liu}\ \emph {et~al.}(2024)\citenamefont {Liu}, \citenamefont {Bozkurt}, \citenamefont {Zatelli}, \citenamefont {ten Haaf}, \citenamefont {Dvir},\ and\ \citenamefont {Wimmer}}]{Liu2023}%
  \BibitemOpen
  \bibfield  {author} {\bibinfo {author} {\bibfnamefont {C.-X.}\ \bibnamefont {Liu}}, \bibinfo {author} {\bibfnamefont {A.~M.}\ \bibnamefont {Bozkurt}}, \bibinfo {author} {\bibfnamefont {F.}~\bibnamefont {Zatelli}}, \bibinfo {author} {\bibfnamefont {S.~L.}\ \bibnamefont {ten Haaf}}, \bibinfo {author} {\bibfnamefont {T.}~\bibnamefont {Dvir}},\ and\ \bibinfo {author} {\bibfnamefont {M.}~\bibnamefont {Wimmer}},\ }\bibfield  {title} {\bibinfo {title} {Enhancing the excitation gap of a quantum-dot-based kitaev chain},\ }\href {https://doi.org/10.1038/s42005-024-01715-5} {\bibfield  {journal} {\bibinfo  {journal} {Communications Physics}\ }\textbf {\bibinfo {volume} {7}},\ \bibinfo {pages} {235} (\bibinfo {year} {2024})}\BibitemShut {NoStop}%
\bibitem [{\citenamefont {Samuelson}\ \emph {et~al.}(2024)\citenamefont {Samuelson}, \citenamefont {Svensson},\ and\ \citenamefont {Leijnse}}]{Leijnse2024}%
  \BibitemOpen
  \bibfield  {author} {\bibinfo {author} {\bibfnamefont {W.}~\bibnamefont {Samuelson}}, \bibinfo {author} {\bibfnamefont {V.}~\bibnamefont {Svensson}},\ and\ \bibinfo {author} {\bibfnamefont {M.}~\bibnamefont {Leijnse}},\ }\bibfield  {title} {\bibinfo {title} {Minimal quantum dot based kitaev chain with only local superconducting proximity effect},\ }\href {https://doi.org/10.1103/PhysRevB.109.035415} {\bibfield  {journal} {\bibinfo  {journal} {Phys. Rev. B}\ }\textbf {\bibinfo {volume} {109}},\ \bibinfo {pages} {035415} (\bibinfo {year} {2024})}\BibitemShut {NoStop}%
\bibitem [{\citenamefont {Zatelli}\ \emph {et~al.}(2024)\citenamefont {Zatelli}, \citenamefont {van Driel}, \citenamefont {Xu}, \citenamefont {Wang}, \citenamefont {Liu}, \citenamefont {Bordin}, \citenamefont {Roovers}, \citenamefont {Mazur}, \citenamefont {van Loo}, \citenamefont {Wolff} \emph {et~al.}}]{Zatelli2024}%
  \BibitemOpen
  \bibfield  {author} {\bibinfo {author} {\bibfnamefont {F.}~\bibnamefont {Zatelli}}, \bibinfo {author} {\bibfnamefont {D.}~\bibnamefont {van Driel}}, \bibinfo {author} {\bibfnamefont {D.}~\bibnamefont {Xu}}, \bibinfo {author} {\bibfnamefont {G.}~\bibnamefont {Wang}}, \bibinfo {author} {\bibfnamefont {C.-X.}\ \bibnamefont {Liu}}, \bibinfo {author} {\bibfnamefont {A.}~\bibnamefont {Bordin}}, \bibinfo {author} {\bibfnamefont {B.}~\bibnamefont {Roovers}}, \bibinfo {author} {\bibfnamefont {G.~P.}\ \bibnamefont {Mazur}}, \bibinfo {author} {\bibfnamefont {N.}~\bibnamefont {van Loo}}, \bibinfo {author} {\bibfnamefont {J.~C.}\ \bibnamefont {Wolff}}, \emph {et~al.},\ }\bibfield  {title} {\bibinfo {title} {Robust poor man’s majorana zero modes using yu-shiba-rusinov states},\ }\href {https://doi.org/10.1038/s41467-024-52066-2} {\bibfield  {journal} {\bibinfo  {journal} {Nature Communications}\ }\textbf {\bibinfo {volume} {15}},\ \bibinfo {pages} {7933} (\bibinfo {year} {2024})}\BibitemShut {NoStop}%
\bibitem [{\citenamefont {Dourado}\ \emph {et~al.}(2026)\citenamefont {Dourado}, \citenamefont {Egues},\ and\ \citenamefont {Penteado}}]{Dourado2025}%
  \BibitemOpen
  \bibfield  {author} {\bibinfo {author} {\bibfnamefont {R.~A.}\ \bibnamefont {Dourado}}, \bibinfo {author} {\bibfnamefont {J.~C.}\ \bibnamefont {Egues}},\ and\ \bibinfo {author} {\bibfnamefont {P.~H.}\ \bibnamefont {Penteado}},\ }\bibfield  {title} {\bibinfo {title} {Two-site kitaev sweet spots evolving into topological islands},\ }\href {https://doi.org/10.1103/wptk-lvc5} {\bibfield  {journal} {\bibinfo  {journal} {Phys. Rev. B}\ }\textbf {\bibinfo {volume} {113}},\ \bibinfo {pages} {035432} (\bibinfo {year} {2026})}\BibitemShut {NoStop}%
\bibitem [{\citenamefont {G\'omez-Le\'on}\ \emph {et~al.}(2025)\citenamefont {G\'omez-Le\'on}, \citenamefont {Schir\`o},\ and\ \citenamefont {Dmytruk}}]{GomezLeon2025}%
  \BibitemOpen
  \bibfield  {author} {\bibinfo {author} {\bibfnamefont {A.}~\bibnamefont {G\'omez-Le\'on}}, \bibinfo {author} {\bibfnamefont {M.}~\bibnamefont {Schir\`o}},\ and\ \bibinfo {author} {\bibfnamefont {O.}~\bibnamefont {Dmytruk}},\ }\bibfield  {title} {\bibinfo {title} {Majorana bound states from cavity embedding in an interacting two-site kitaev chain},\ }\href {https://doi.org/10.1103/PhysRevB.111.155410} {\bibfield  {journal} {\bibinfo  {journal} {Phys. Rev. B}\ }\textbf {\bibinfo {volume} {111}},\ \bibinfo {pages} {155410} (\bibinfo {year} {2025})}\BibitemShut {NoStop}%
\bibitem [{\citenamefont {Steffensen}\ \emph {et~al.}(2022)\citenamefont {Steffensen}, \citenamefont {Salda{\~n}a}, \citenamefont {Vekris}, \citenamefont {Krogstrup}, \citenamefont {Grove-Rasmussen}, \citenamefont {Nyg\aa{}rd}, \citenamefont {Yeyati},\ and\ \citenamefont {Paaske}}]{Gorm2022}%
  \BibitemOpen
  \bibfield  {author} {\bibinfo {author} {\bibfnamefont {G.~O.}\ \bibnamefont {Steffensen}}, \bibinfo {author} {\bibfnamefont {J.~C.~E.}\ \bibnamefont {Salda{\~n}a}}, \bibinfo {author} {\bibfnamefont {A.}~\bibnamefont {Vekris}}, \bibinfo {author} {\bibfnamefont {P.}~\bibnamefont {Krogstrup}}, \bibinfo {author} {\bibfnamefont {K.}~\bibnamefont {Grove-Rasmussen}}, \bibinfo {author} {\bibfnamefont {J.}~\bibnamefont {Nyg\aa{}rd}}, \bibinfo {author} {\bibfnamefont {A.~L.}\ \bibnamefont {Yeyati}},\ and\ \bibinfo {author} {\bibfnamefont {J.}~\bibnamefont {Paaske}},\ }\bibfield  {title} {\bibinfo {title} {Direct transport between superconducting subgap states in a double quantum dot},\ }\href {https://doi.org/10.1103/PhysRevB.105.L161302} {\bibfield  {journal} {\bibinfo  {journal} {Phys. Rev. B}\ }\textbf {\bibinfo {volume} {105}},\ \bibinfo {pages} {L161302} (\bibinfo {year} {2022})}\BibitemShut {NoStop}%
\bibitem [{\citenamefont {Seoane~Souto}\ and\ \citenamefont {Aguado}(2024)}]{SeoaneAguado2024}%
  \BibitemOpen
  \bibfield  {author} {\bibinfo {author} {\bibfnamefont {R.}~\bibnamefont {Seoane~Souto}}\ and\ \bibinfo {author} {\bibfnamefont {R.}~\bibnamefont {Aguado}},\ }\bibinfo {title} {Subgap states in semiconductor-superconductor devices for quantum technologies: Andreev qubits and minimal majorana chains},\ in\ \href {https://doi.org/10.1007/978-3-031-55657-9_3} {\emph {\bibinfo {booktitle} {New Trends and Platforms for Quantum Technologies}}},\ \bibinfo {editor} {edited by\ \bibinfo {editor} {\bibfnamefont {R.}~\bibnamefont {Aguado}}, \bibinfo {editor} {\bibfnamefont {R.}~\bibnamefont {Citro}}, \bibinfo {editor} {\bibfnamefont {M.}~\bibnamefont {Lewenstein}},\ and\ \bibinfo {editor} {\bibfnamefont {M.}~\bibnamefont {Stern}}}\ (\bibinfo  {publisher} {Springer Nature Switzerland},\ \bibinfo {address} {Cham},\ \bibinfo {year} {2024})\ pp.\ \bibinfo {pages} {133--223}\BibitemShut {NoStop}%
\bibitem [{\citenamefont {Alvarado}\ \emph {et~al.}(2024)\citenamefont {Alvarado}, \citenamefont {Yeyati}, \citenamefont {Aguado},\ and\ \citenamefont {Souto}}]{Alvarado2024}%
  \BibitemOpen
  \bibfield  {author} {\bibinfo {author} {\bibfnamefont {M.}~\bibnamefont {Alvarado}}, \bibinfo {author} {\bibfnamefont {A.~L.}\ \bibnamefont {Yeyati}}, \bibinfo {author} {\bibfnamefont {R.}~\bibnamefont {Aguado}},\ and\ \bibinfo {author} {\bibfnamefont {R.~S.}\ \bibnamefont {Souto}},\ }\bibfield  {title} {\bibinfo {title} {Interplay between majorana and shiba states in a minimal kitaev chain coupled to a superconductor},\ }\href {https://doi.org/10.1103/PhysRevB.110.245144} {\bibfield  {journal} {\bibinfo  {journal} {Phys. Rev. B}\ }\textbf {\bibinfo {volume} {110}},\ \bibinfo {pages} {245144} (\bibinfo {year} {2024})}\BibitemShut {NoStop}%
\bibitem [{\citenamefont {Alvarado}\ \emph {et~al.}(2025)\citenamefont {Alvarado}, \citenamefont {Yeyati}, \citenamefont {Aguado},\ and\ \citenamefont {Souto}}]{Alvarado2025}%
  \BibitemOpen
  \bibfield  {author} {\bibinfo {author} {\bibfnamefont {M.}~\bibnamefont {Alvarado}}, \bibinfo {author} {\bibfnamefont {A.~L.}\ \bibnamefont {Yeyati}}, \bibinfo {author} {\bibfnamefont {R.}~\bibnamefont {Aguado}},\ and\ \bibinfo {author} {\bibfnamefont {R.~S.}\ \bibnamefont {Souto}},\ }\bibfield  {title} {\bibinfo {title} {Characterizing local majorana properties using andreev states},\ }\href {https://doi.org/10.1103/1jqt-3ht5} {\bibfield  {journal} {\bibinfo  {journal} {Phys. Rev. B}\ }\textbf {\bibinfo {volume} {112}},\ \bibinfo {pages} {L241403} (\bibinfo {year} {2025})}\BibitemShut {NoStop}%
\bibitem [{\citenamefont {Bordin}\ \emph {et~al.}(2025{\natexlab{b}})\citenamefont {Bordin}, \citenamefont {Bennebroek~Evertsz'}, \citenamefont {Steffensen}, \citenamefont {Dvir}, \citenamefont {Mazur}, \citenamefont {van Driel}, \citenamefont {van Loo}, \citenamefont {Wolff}, \citenamefont {Bakkers}, \citenamefont {Yeyati},\ and\ \citenamefont {Kouwenhoven}}]{Bordin2025c}%
  \BibitemOpen
  \bibfield  {author} {\bibinfo {author} {\bibfnamefont {A.}~\bibnamefont {Bordin}}, \bibinfo {author} {\bibfnamefont {F.~J.}\ \bibnamefont {Bennebroek~Evertsz'}}, \bibinfo {author} {\bibfnamefont {G.~O.}\ \bibnamefont {Steffensen}}, \bibinfo {author} {\bibfnamefont {T.}~\bibnamefont {Dvir}}, \bibinfo {author} {\bibfnamefont {G.~P.}\ \bibnamefont {Mazur}}, \bibinfo {author} {\bibfnamefont {D.}~\bibnamefont {van Driel}}, \bibinfo {author} {\bibfnamefont {N.}~\bibnamefont {van Loo}}, \bibinfo {author} {\bibfnamefont {J.~C.}\ \bibnamefont {Wolff}}, \bibinfo {author} {\bibfnamefont {E.~P. A.~M.}\ \bibnamefont {Bakkers}}, \bibinfo {author} {\bibfnamefont {A.~L.}\ \bibnamefont {Yeyati}},\ and\ \bibinfo {author} {\bibfnamefont {L.~P.}\ \bibnamefont {Kouwenhoven}},\ }\bibfield  {title} {\bibinfo {title} {Impact of andreev bound states within the leads of a quantum dot josephson junction},\ }\href {https://doi.org/10.1103/PhysRevX.15.011046} {\bibfield  {journal} {\bibinfo  {journal} {Phys. Rev. X}\ }\textbf {\bibinfo
  {volume} {15}},\ \bibinfo {pages} {011046} (\bibinfo {year} {2025}{\natexlab{b}})}\BibitemShut {NoStop}%
\bibitem [{\citenamefont {Zhang}\ \emph {et~al.}(2025)\citenamefont {Zhang}, \citenamefont {Qiao},\ and\ \citenamefont {Sun}}]{Zhang2025}%
  \BibitemOpen
  \bibfield  {author} {\bibinfo {author} {\bibfnamefont {Z.-L.}\ \bibnamefont {Zhang}}, \bibinfo {author} {\bibfnamefont {G.-J.}\ \bibnamefont {Qiao}},\ and\ \bibinfo {author} {\bibfnamefont {C.}~\bibnamefont {Sun}},\ }\bibfield  {title} {\bibinfo {title} {Poor man's majoranon in two quantum dots dressed by superconducting quasi-excitations},\ }\href {https://arxiv.org/abs/2506.10367} {\bibfield  {journal} {\bibinfo  {journal} {arXiv preprint arXiv:2506.10367}\ } (\bibinfo {year} {2025})}\BibitemShut {NoStop}%
\bibitem [{\citenamefont {Pino}\ \emph {et~al.}(2025)\citenamefont {Pino}, \citenamefont {Meir},\ and\ \citenamefont {Aguado}}]{PhysRevB.111.L140503}%
  \BibitemOpen
  \bibfield  {author} {\bibinfo {author} {\bibfnamefont {D.~M.}\ \bibnamefont {Pino}}, \bibinfo {author} {\bibfnamefont {Y.}~\bibnamefont {Meir}},\ and\ \bibinfo {author} {\bibfnamefont {R.}~\bibnamefont {Aguado}},\ }\bibfield  {title} {\bibinfo {title} {Thermodynamics of non-hermitian josephson junctions with exceptional points},\ }\href {https://doi.org/10.1103/PhysRevB.111.L140503} {\bibfield  {journal} {\bibinfo  {journal} {Phys. Rev. B}\ }\textbf {\bibinfo {volume} {111}},\ \bibinfo {pages} {L140503} (\bibinfo {year} {2025})}\BibitemShut {NoStop}%
\bibitem [{\citenamefont {Cayao}\ and\ \citenamefont {Aguado}(2025)}]{Aguado2025}%
  \BibitemOpen
  \bibfield  {author} {\bibinfo {author} {\bibfnamefont {J.}~\bibnamefont {Cayao}}\ and\ \bibinfo {author} {\bibfnamefont {R.}~\bibnamefont {Aguado}},\ }\bibfield  {title} {\bibinfo {title} {Non-hermitian minimal kitaev chains},\ }\href {https://doi.org/10.1103/PhysRevB.111.205432} {\bibfield  {journal} {\bibinfo  {journal} {Phys. Rev. B}\ }\textbf {\bibinfo {volume} {111}},\ \bibinfo {pages} {205432} (\bibinfo {year} {2025})}\BibitemShut {NoStop}%
\bibitem [{\citenamefont {Huisman}\ \emph {et~al.}(2026)\citenamefont {Huisman}, \citenamefont {ten Haaf}, \citenamefont {Liu}, \citenamefont {Wang}, \citenamefont {Bordin}, \citenamefont {Evertsz'}, \citenamefont {Roovers}, \citenamefont {Wimmer},\ and\ \citenamefont {Goswami}}]{Wimmer2026}%
  \BibitemOpen
  \bibfield  {author} {\bibinfo {author} {\bibfnamefont {W.~D.}\ \bibnamefont {Huisman}}, \bibinfo {author} {\bibfnamefont {S.~L.~D.}\ \bibnamefont {ten Haaf}}, \bibinfo {author} {\bibfnamefont {C.-X.}\ \bibnamefont {Liu}}, \bibinfo {author} {\bibfnamefont {Q.}~\bibnamefont {Wang}}, \bibinfo {author} {\bibfnamefont {A.}~\bibnamefont {Bordin}}, \bibinfo {author} {\bibfnamefont {F.~J.~B.}\ \bibnamefont {Evertsz'}}, \bibinfo {author} {\bibfnamefont {B.}~\bibnamefont {Roovers}}, \bibinfo {author} {\bibfnamefont {M.}~\bibnamefont {Wimmer}},\ and\ \bibinfo {author} {\bibfnamefont {S.}~\bibnamefont {Goswami}},\ }\bibfield  {title} {\bibinfo {title} {Using andreev bound states and spin to remove domain walls in a kitaev chain},\ }\href {https://arxiv.org/abs/2601.12891} {\bibfield  {journal} {\bibinfo  {journal} {arXiv preprint arXiv:2601.12891}\ } (\bibinfo {year} {2026})}\BibitemShut {NoStop}%
\bibitem [{\citenamefont {Zhang}\ \emph {et~al.}(2026)\citenamefont {Zhang}, \citenamefont {Yue}, \citenamefont {Qiao},\ and\ \citenamefont {Sun}}]{Zhang2026}%
  \BibitemOpen
  \bibfield  {author} {\bibinfo {author} {\bibfnamefont {Z.-L.}\ \bibnamefont {Zhang}}, \bibinfo {author} {\bibfnamefont {X.}~\bibnamefont {Yue}}, \bibinfo {author} {\bibfnamefont {G.-J.}\ \bibnamefont {Qiao}},\ and\ \bibinfo {author} {\bibfnamefont {C.~P.}\ \bibnamefont {Sun}},\ }\bibfield  {title} {\bibinfo {title} {Sensitive dependence of poor man's majorana modes on the length of superconductor},\ }\href {https://arxiv.org/abs/2604.12950} {\bibfield  {journal} {\bibinfo  {journal} {arXiv preprint arXiv:2604.12950}\ } (\bibinfo {year} {2026})}\BibitemShut {NoStop}%
\bibitem [{\citenamefont {Buonemani}\ \emph {et~al.}(2026)\citenamefont {Buonemani}, \citenamefont {Gómez-León}, \citenamefont {Schirò},\ and\ \citenamefont {Dmytruk}}]{Buonemani2026}%
  \BibitemOpen
  \bibfield  {author} {\bibinfo {author} {\bibfnamefont {F.}~\bibnamefont {Buonemani}}, \bibinfo {author} {\bibfnamefont {A.}~\bibnamefont {Gómez-León}}, \bibinfo {author} {\bibfnamefont {M.}~\bibnamefont {Schirò}},\ and\ \bibinfo {author} {\bibfnamefont {O.}~\bibnamefont {Dmytruk}},\ }\bibfield  {title} {\bibinfo {title} {Poor man's majorana bound states in quantum dot based kitaev chain coupled to a photonic cavity},\ }\href {https://arxiv.org/abs/2604.15036} {\bibfield  {journal} {\bibinfo  {journal} {arXiv preprint arXiv:2604.15036}\ } (\bibinfo {year} {2026})}\BibitemShut {NoStop}%
\bibitem [{\citenamefont {Lee}\ \emph {et~al.}(2014)\citenamefont {Lee}, \citenamefont {Jiang}, \citenamefont {Houzet}, \citenamefont {Aguado}, \citenamefont {Lieber},\ and\ \citenamefont {De~Franceschi}}]{LeeNN:14}%
  \BibitemOpen
  \bibfield  {author} {\bibinfo {author} {\bibfnamefont {E.~J.~H.}\ \bibnamefont {Lee}}, \bibinfo {author} {\bibfnamefont {X.}~\bibnamefont {Jiang}}, \bibinfo {author} {\bibfnamefont {M.}~\bibnamefont {Houzet}}, \bibinfo {author} {\bibfnamefont {R.}~\bibnamefont {Aguado}}, \bibinfo {author} {\bibfnamefont {C.~M.}\ \bibnamefont {Lieber}},\ and\ \bibinfo {author} {\bibfnamefont {S.}~\bibnamefont {De~Franceschi}},\ }\bibfield  {title} {\bibinfo {title} {Spin-resolved andreev levels and parity crossings in hybrid superconductor--semiconductor nanostructures},\ }\href {https://doi.org/10.1038/nnano.2013.267} {\bibfield  {journal} {\bibinfo  {journal} {Nature Nanotechnology}\ }\textbf {\bibinfo {volume} {9}},\ \bibinfo {pages} {79} (\bibinfo {year} {2014})}\BibitemShut {NoStop}%
\bibitem [{\citenamefont {Luh}(1965)}]{Yu1965}%
  \BibitemOpen
  \bibfield  {author} {\bibinfo {author} {\bibfnamefont {Y.}~\bibnamefont {Luh}},\ }\bibfield  {title} {\bibinfo {title} {Bound state in superconductors with paramagnetic impurities},\ }\href {https://doi.org/10.7498/aps.21.75} {\bibfield  {journal} {\bibinfo  {journal} {Acta Physica Sinica}\ }\textbf {\bibinfo {volume} {21}},\ \bibinfo {pages} {75} (\bibinfo {year} {1965})}\BibitemShut {NoStop}%
\bibitem [{\citenamefont {Shiba}(1968)}]{Shiba1968}%
  \BibitemOpen
  \bibfield  {author} {\bibinfo {author} {\bibfnamefont {H.}~\bibnamefont {Shiba}},\ }\bibfield  {title} {\bibinfo {title} {Classical spins in superconductors},\ }\href {https://doi.org/10.1143/PTP.40.435} {\bibfield  {journal} {\bibinfo  {journal} {Progress of theoretical Physics}\ }\textbf {\bibinfo {volume} {40}},\ \bibinfo {pages} {435} (\bibinfo {year} {1968})}\BibitemShut {NoStop}%
\bibitem [{\citenamefont {{Rusinov}}(1969)}]{Rusinov1969}%
  \BibitemOpen
  \bibfield  {author} {\bibinfo {author} {\bibfnamefont {A.~I.}\ \bibnamefont {{Rusinov}}},\ }\bibfield  {title} {\bibinfo {title} {{On the Theory of Gapless Superconductivity in Alloys Containing Paramagnetic Impurities}},\ }\href {http://jetp.ras.ru/cgi-bin/e/index/e/29/6/p1101?a=list} {\bibfield  {journal} {\bibinfo  {journal} {Soviet Journal of Experimental and Theoretical Physics}\ }\textbf {\bibinfo {volume} {29}},\ \bibinfo {pages} {1101} (\bibinfo {year} {1969})}\BibitemShut {NoStop}%
\bibitem [{\citenamefont {Sakurai}(1970)}]{Sakurai1970}%
  \BibitemOpen
  \bibfield  {author} {\bibinfo {author} {\bibfnamefont {A.}~\bibnamefont {Sakurai}},\ }\bibfield  {title} {\bibinfo {title} {{Comments on Superconductors with Magnetic Impurities}},\ }\href {https://doi.org/10.1143/PTP.44.1472} {\bibfield  {journal} {\bibinfo  {journal} {Progress of Theoretical Physics}\ }\textbf {\bibinfo {volume} {44}},\ \bibinfo {pages} {1472} (\bibinfo {year} {1970})}\BibitemShut {NoStop}%
\bibitem [{\citenamefont {Flatt\'e}\ and\ \citenamefont {Byers}(1997)}]{Byers1997}%
  \BibitemOpen
  \bibfield  {author} {\bibinfo {author} {\bibfnamefont {M.~E.}\ \bibnamefont {Flatt\'e}}\ and\ \bibinfo {author} {\bibfnamefont {J.~M.}\ \bibnamefont {Byers}},\ }\bibfield  {title} {\bibinfo {title} {Local electronic structure of a single magnetic impurity in a superconductor},\ }\href {https://doi.org/10.1103/PhysRevLett.78.3761} {\bibfield  {journal} {\bibinfo  {journal} {Phys. Rev. Lett.}\ }\textbf {\bibinfo {volume} {78}},\ \bibinfo {pages} {3761} (\bibinfo {year} {1997})}\BibitemShut {NoStop}%
\bibitem [{\citenamefont {Glazman}\ and\ \citenamefont {Matveev}(1989)}]{Glazman1989}%
  \BibitemOpen
  \bibfield  {author} {\bibinfo {author} {\bibfnamefont {L.}~\bibnamefont {Glazman}}\ and\ \bibinfo {author} {\bibfnamefont {K.}~\bibnamefont {Matveev}},\ }\bibfield  {title} {\bibinfo {title} {Resonant josephson current through kondo impurities in a tunnel barrier},\ }\href {https://doi.org/10.1007/978-3-642-77274-0_20} {\bibfield  {journal} {\bibinfo  {journal} {JETP Lett}\ }\textbf {\bibinfo {volume} {49}},\ \bibinfo {pages} {659} (\bibinfo {year} {1989})}\BibitemShut {NoStop}%
\bibitem [{\citenamefont {Vecino}\ \emph {et~al.}(2003)\citenamefont {Vecino}, \citenamefont {Mart\'{\i}n-Rodero},\ and\ \citenamefont {Yeyati}}]{Vecino2003}%
  \BibitemOpen
  \bibfield  {author} {\bibinfo {author} {\bibfnamefont {E.}~\bibnamefont {Vecino}}, \bibinfo {author} {\bibfnamefont {A.}~\bibnamefont {Mart\'{\i}n-Rodero}},\ and\ \bibinfo {author} {\bibfnamefont {A.~L.}\ \bibnamefont {Yeyati}},\ }\bibfield  {title} {\bibinfo {title} {Josephson current through a correlated quantum level: Andreev states and $\ensuremath{\pi}$ junction behavior},\ }\href {https://doi.org/10.1103/PhysRevB.68.035105} {\bibfield  {journal} {\bibinfo  {journal} {Phys. Rev. B}\ }\textbf {\bibinfo {volume} {68}},\ \bibinfo {pages} {035105} (\bibinfo {year} {2003})}\BibitemShut {NoStop}%
\bibitem [{\citenamefont {Mart{\'\i}n-Rodero}\ and\ \citenamefont {Levy~Yeyati}(2011)}]{Martin-Rodero2011}%
  \BibitemOpen
  \bibfield  {author} {\bibinfo {author} {\bibfnamefont {A.}~\bibnamefont {Mart{\'\i}n-Rodero}}\ and\ \bibinfo {author} {\bibfnamefont {A.}~\bibnamefont {Levy~Yeyati}},\ }\bibfield  {title} {\bibinfo {title} {Josephson and andreev transport through quantum dots},\ }\href {https://doi.org/10.1080/00018732.2011.624266} {\bibfield  {journal} {\bibinfo  {journal} {Advances in Physics}\ }\textbf {\bibinfo {volume} {60}},\ \bibinfo {pages} {899} (\bibinfo {year} {2011})}\BibitemShut {NoStop}%
\bibitem [{\citenamefont {\ifmmode~\check{Z}\else \v{Z}\fi{}itko}\ \emph {et~al.}(2015)\citenamefont {\ifmmode~\check{Z}\else \v{Z}\fi{}itko}, \citenamefont {Lim}, \citenamefont {L\'opez},\ and\ \citenamefont {Aguado}}]{Aguado2015}%
  \BibitemOpen
  \bibfield  {author} {\bibinfo {author} {\bibfnamefont {R.}~\bibnamefont {\ifmmode~\check{Z}\else \v{Z}\fi{}itko}}, \bibinfo {author} {\bibfnamefont {J.~S.}\ \bibnamefont {Lim}}, \bibinfo {author} {\bibfnamefont {R.}~\bibnamefont {L\'opez}},\ and\ \bibinfo {author} {\bibfnamefont {R.}~\bibnamefont {Aguado}},\ }\bibfield  {title} {\bibinfo {title} {Shiba states and zero-bias anomalies in the hybrid normal-superconductor anderson model},\ }\href {https://doi.org/10.1103/PhysRevB.91.045441} {\bibfield  {journal} {\bibinfo  {journal} {Phys. Rev. B}\ }\textbf {\bibinfo {volume} {91}},\ \bibinfo {pages} {045441} (\bibinfo {year} {2015})}\BibitemShut {NoStop}%
\bibitem [{\citenamefont {Baran}\ and\ \citenamefont {Nichita}(2023)}]{Baran2023}%
  \BibitemOpen
  \bibfield  {author} {\bibinfo {author} {\bibfnamefont {V.~V.}\ \bibnamefont {Baran}}\ and\ \bibinfo {author} {\bibfnamefont {D.~R.}\ \bibnamefont {Nichita}},\ }\bibfield  {title} {\bibinfo {title} {Reduced basis emulation of pairing in finite systems},\ }\href {https://doi.org/10.1103/PhysRevB.107.144503} {\bibfield  {journal} {\bibinfo  {journal} {Phys. Rev. B}\ }\textbf {\bibinfo {volume} {107}},\ \bibinfo {pages} {144503} (\bibinfo {year} {2023})}\BibitemShut {NoStop}%
\bibitem [{\citenamefont {Baran}\ \emph {et~al.}(2023)\citenamefont {Baran}, \citenamefont {Frost},\ and\ \citenamefont {Paaske}}]{Paaske2023}%
  \BibitemOpen
  \bibfield  {author} {\bibinfo {author} {\bibfnamefont {V.~V.}\ \bibnamefont {Baran}}, \bibinfo {author} {\bibfnamefont {E.~J.~P.}\ \bibnamefont {Frost}},\ and\ \bibinfo {author} {\bibfnamefont {J.}~\bibnamefont {Paaske}},\ }\bibfield  {title} {\bibinfo {title} {Surrogate model solver for impurity-induced superconducting subgap states},\ }\href {https://doi.org/10.1103/PhysRevB.108.L220506} {\bibfield  {journal} {\bibinfo  {journal} {Phys. Rev. B}\ }\textbf {\bibinfo {volume} {108}},\ \bibinfo {pages} {L220506} (\bibinfo {year} {2023})}\BibitemShut {NoStop}%
\bibitem [{\citenamefont {Deb}(2000)}]{Deb2000}%
  \BibitemOpen
  \bibfield  {author} {\bibinfo {author} {\bibfnamefont {K.}~\bibnamefont {Deb}},\ }\bibfield  {title} {\bibinfo {title} {An efficient constraint handling method for genetic algorithms},\ }\href {https://doi.org/10.1016/S0045-7825(99)00389-8} {\bibfield  {journal} {\bibinfo  {journal} {Computer methods in applied mechanics and engineering}\ }\textbf {\bibinfo {volume} {186}},\ \bibinfo {pages} {311} (\bibinfo {year} {2000})}\BibitemShut {NoStop}%
\bibitem [{\citenamefont {Hansen}(2016)}]{Hansen2016}%
  \BibitemOpen
  \bibfield  {author} {\bibinfo {author} {\bibfnamefont {N.}~\bibnamefont {Hansen}},\ }\bibfield  {title} {\bibinfo {title} {The cma evolution strategy: A tutorial},\ }\href {https://arxiv.org/abs/1604.00772} {\bibfield  {journal} {\bibinfo  {journal} {arXiv preprint arXiv:1604.00772}\ } (\bibinfo {year} {2016})}\BibitemShut {NoStop}%
\bibitem [{\citenamefont {Cuevas}\ \emph {et~al.}(1996)\citenamefont {Cuevas}, \citenamefont {Mart\'{\i}n-Rodero},\ and\ \citenamefont {Yeyati}}]{Cuevas1996}%
  \BibitemOpen
  \bibfield  {author} {\bibinfo {author} {\bibfnamefont {J.~C.}\ \bibnamefont {Cuevas}}, \bibinfo {author} {\bibfnamefont {A.}~\bibnamefont {Mart\'{\i}n-Rodero}},\ and\ \bibinfo {author} {\bibfnamefont {A.~L.}\ \bibnamefont {Yeyati}},\ }\bibfield  {title} {\bibinfo {title} {Hamiltonian approach to the transport properties of superconducting quantum point contacts},\ }\href {https://doi.org/10.1103/PhysRevB.54.7366} {\bibfield  {journal} {\bibinfo  {journal} {Phys. Rev. B}\ }\textbf {\bibinfo {volume} {54}},\ \bibinfo {pages} {7366} (\bibinfo {year} {1996})}\BibitemShut {NoStop}%
\bibitem [{\citenamefont {Zazunov}\ \emph {et~al.}(2016)\citenamefont {Zazunov}, \citenamefont {Egger},\ and\ \citenamefont {Levy~Yeyati}}]{Zazunov2016}%
  \BibitemOpen
  \bibfield  {author} {\bibinfo {author} {\bibfnamefont {A.}~\bibnamefont {Zazunov}}, \bibinfo {author} {\bibfnamefont {R.}~\bibnamefont {Egger}},\ and\ \bibinfo {author} {\bibfnamefont {A.}~\bibnamefont {Levy~Yeyati}},\ }\bibfield  {title} {\bibinfo {title} {Low-energy theory of transport in majorana wire junctions},\ }\href {https://doi.org/10.1103/PhysRevB.94.014502} {\bibfield  {journal} {\bibinfo  {journal} {Phys. Rev. B}\ }\textbf {\bibinfo {volume} {94}},\ \bibinfo {pages} {014502} (\bibinfo {year} {2016})}\BibitemShut {NoStop}%
\bibitem [{\citenamefont {Alvarado}\ \emph {et~al.}(2020)\citenamefont {Alvarado}, \citenamefont {Iks}, \citenamefont {Zazunov}, \citenamefont {Egger},\ and\ \citenamefont {Yeyati}}]{Alvarado2020}%
  \BibitemOpen
  \bibfield  {author} {\bibinfo {author} {\bibfnamefont {M.}~\bibnamefont {Alvarado}}, \bibinfo {author} {\bibfnamefont {A.}~\bibnamefont {Iks}}, \bibinfo {author} {\bibfnamefont {A.}~\bibnamefont {Zazunov}}, \bibinfo {author} {\bibfnamefont {R.}~\bibnamefont {Egger}},\ and\ \bibinfo {author} {\bibfnamefont {A.~L.}\ \bibnamefont {Yeyati}},\ }\bibfield  {title} {\bibinfo {title} {Boundary green's function approach for spinful single-channel and multichannel majorana nanowires},\ }\href {https://doi.org/10.1103/PhysRevB.101.094511} {\bibfield  {journal} {\bibinfo  {journal} {Physical Review B}\ }\textbf {\bibinfo {volume} {101}},\ \bibinfo {pages} {094511} (\bibinfo {year} {2020})}\BibitemShut {NoStop}%
\bibitem [{\citenamefont {Sedlmayr}\ and\ \citenamefont {Bena}(2015)}]{Bena2015}%
  \BibitemOpen
  \bibfield  {author} {\bibinfo {author} {\bibfnamefont {N.}~\bibnamefont {Sedlmayr}}\ and\ \bibinfo {author} {\bibfnamefont {C.}~\bibnamefont {Bena}},\ }\bibfield  {title} {\bibinfo {title} {Visualizing majorana bound states in one and two dimensions using the generalized majorana polarization},\ }\href {https://doi.org/10.1103/PhysRevB.92.115115} {\bibfield  {journal} {\bibinfo  {journal} {Phys. Rev. B}\ }\textbf {\bibinfo {volume} {92}},\ \bibinfo {pages} {115115} (\bibinfo {year} {2015})}\BibitemShut {NoStop}%
\bibitem [{\citenamefont {Sedlmayr}\ \emph {et~al.}(2016)\citenamefont {Sedlmayr}, \citenamefont {Aguiar-Hualde},\ and\ \citenamefont {Bena}}]{Bena2016}%
  \BibitemOpen
  \bibfield  {author} {\bibinfo {author} {\bibfnamefont {N.}~\bibnamefont {Sedlmayr}}, \bibinfo {author} {\bibfnamefont {J.~M.}\ \bibnamefont {Aguiar-Hualde}},\ and\ \bibinfo {author} {\bibfnamefont {C.}~\bibnamefont {Bena}},\ }\bibfield  {title} {\bibinfo {title} {Majorana bound states in open quasi-one-dimensional and two-dimensional systems with transverse rashba coupling},\ }\href {https://doi.org/10.1103/PhysRevB.93.155425} {\bibfield  {journal} {\bibinfo  {journal} {Phys. Rev. B}\ }\textbf {\bibinfo {volume} {93}},\ \bibinfo {pages} {155425} (\bibinfo {year} {2016})}\BibitemShut {NoStop}%
\bibitem [{\citenamefont {G\l{}odzik}\ \emph {et~al.}(2020)\citenamefont {G\l{}odzik}, \citenamefont {Sedlmayr},\ and\ \citenamefont {Doma\ifmmode~\acute{n}\else \'{n}\fi{}ski}}]{Glodzik2020}%
  \BibitemOpen
  \bibfield  {author} {\bibinfo {author} {\bibfnamefont {S.}~\bibnamefont {G\l{}odzik}}, \bibinfo {author} {\bibfnamefont {N.}~\bibnamefont {Sedlmayr}},\ and\ \bibinfo {author} {\bibfnamefont {T.}~\bibnamefont {Doma\ifmmode~\acute{n}\else \'{n}\fi{}ski}},\ }\bibfield  {title} {\bibinfo {title} {How to measure the majorana polarization of a topological planar josephson junction},\ }\href {https://doi.org/10.1103/PhysRevB.102.085411} {\bibfield  {journal} {\bibinfo  {journal} {Phys. Rev. B}\ }\textbf {\bibinfo {volume} {102}},\ \bibinfo {pages} {085411} (\bibinfo {year} {2020})}\BibitemShut {NoStop}%
\bibitem [{\citenamefont {Aksenov}\ \emph {et~al.}(2020)\citenamefont {Aksenov}, \citenamefont {Zlotnikov},\ and\ \citenamefont {Shustin}}]{Aksenov2020}%
  \BibitemOpen
  \bibfield  {author} {\bibinfo {author} {\bibfnamefont {S.~V.}\ \bibnamefont {Aksenov}}, \bibinfo {author} {\bibfnamefont {A.~O.}\ \bibnamefont {Zlotnikov}},\ and\ \bibinfo {author} {\bibfnamefont {M.~S.}\ \bibnamefont {Shustin}},\ }\bibfield  {title} {\bibinfo {title} {Strong coulomb interactions in the problem of majorana modes in a wire of the nontrivial topological class bdi},\ }\href {https://doi.org/10.1103/PhysRevB.101.125431} {\bibfield  {journal} {\bibinfo  {journal} {Phys. Rev. B}\ }\textbf {\bibinfo {volume} {101}},\ \bibinfo {pages} {125431} (\bibinfo {year} {2020})}\BibitemShut {NoStop}%
\bibitem [{\citenamefont {Awoga}\ and\ \citenamefont {Cayao}(2024)}]{Ola2024}%
  \BibitemOpen
  \bibfield  {author} {\bibinfo {author} {\bibfnamefont {O.~A.}\ \bibnamefont {Awoga}}\ and\ \bibinfo {author} {\bibfnamefont {J.}~\bibnamefont {Cayao}},\ }\bibfield  {title} {\bibinfo {title} {Identifying trivial and majorana zero-energy modes using the majorana polarization},\ }\href {https://doi.org/10.1103/PhysRevB.110.165404} {\bibfield  {journal} {\bibinfo  {journal} {Phys. Rev. B}\ }\textbf {\bibinfo {volume} {110}},\ \bibinfo {pages} {165404} (\bibinfo {year} {2024})}\BibitemShut {NoStop}%
\bibitem [{\citenamefont {Samuelson}\ \emph {et~al.}(2026)\citenamefont {Samuelson}, \citenamefont {Torres~Luna}, \citenamefont {Miles}, \citenamefont {Bozkurt}, \citenamefont {Leijnse}, \citenamefont {Wimmer},\ and\ \citenamefont {Svensson}}]{Samuelson2025}%
  \BibitemOpen
  \bibfield  {author} {\bibinfo {author} {\bibfnamefont {W.}~\bibnamefont {Samuelson}}, \bibinfo {author} {\bibfnamefont {J.~D.}\ \bibnamefont {Torres~Luna}}, \bibinfo {author} {\bibfnamefont {S.}~\bibnamefont {Miles}}, \bibinfo {author} {\bibfnamefont {A.~M.}\ \bibnamefont {Bozkurt}}, \bibinfo {author} {\bibfnamefont {M.}~\bibnamefont {Leijnse}}, \bibinfo {author} {\bibfnamefont {M.}~\bibnamefont {Wimmer}},\ and\ \bibinfo {author} {\bibfnamefont {V.}~\bibnamefont {Svensson}},\ }\bibfield  {title} {\bibinfo {title} {Quantifying robustness and locality of majorana bound states in interacting systems},\ }\href {https://doi.org/10.1103/pw51-tnsz} {\bibfield  {journal} {\bibinfo  {journal} {PRX Quantum}\ }\textbf {\bibinfo {volume} {7}},\ \bibinfo {pages} {020341} (\bibinfo {year} {2026})}\BibitemShut {NoStop}%
\bibitem [{\citenamefont {Dom{\'\i}nguez}\ and\ \citenamefont {Yeyati}(2016)}]{Dominguez2016}%
  \BibitemOpen
  \bibfield  {author} {\bibinfo {author} {\bibfnamefont {F.}~\bibnamefont {Dom{\'\i}nguez}}\ and\ \bibinfo {author} {\bibfnamefont {A.~L.}\ \bibnamefont {Yeyati}},\ }\bibfield  {title} {\bibinfo {title} {Quantum interference in a cooper pair splitter: The three sites model},\ }\href {https://doi.org/10.1016/j.physe.2015.09.040} {\bibfield  {journal} {\bibinfo  {journal} {Physica E: Low-dimensional Systems and Nanostructures}\ }\textbf {\bibinfo {volume} {75}},\ \bibinfo {pages} {322} (\bibinfo {year} {2016})}\BibitemShut {NoStop}%
\bibitem [{\citenamefont {Dourado}\ \emph {et~al.}(2025)\citenamefont {Dourado}, \citenamefont {Leijnse},\ and\ \citenamefont {Souto}}]{Souto2025}%
  \BibitemOpen
  \bibfield  {author} {\bibinfo {author} {\bibfnamefont {R.~A.}\ \bibnamefont {Dourado}}, \bibinfo {author} {\bibfnamefont {M.}~\bibnamefont {Leijnse}},\ and\ \bibinfo {author} {\bibfnamefont {R.~S.}\ \bibnamefont {Souto}},\ }\bibfield  {title} {\bibinfo {title} {Majorana sweet spots in three-site kitaev chains},\ }\href {https://doi.org/10.1103/PhysRevB.111.235409} {\bibfield  {journal} {\bibinfo  {journal} {Phys. Rev. B}\ }\textbf {\bibinfo {volume} {111}},\ \bibinfo {pages} {235409} (\bibinfo {year} {2025})}\BibitemShut {NoStop}%
\bibitem [{\citenamefont {Zazunov}\ \emph {et~al.}(2018)\citenamefont {Zazunov}, \citenamefont {Iks}, \citenamefont {Alvarado}, \citenamefont {Yeyati},\ and\ \citenamefont {Egger}}]{Alvarado2018}%
  \BibitemOpen
  \bibfield  {author} {\bibinfo {author} {\bibfnamefont {A.}~\bibnamefont {Zazunov}}, \bibinfo {author} {\bibfnamefont {A.}~\bibnamefont {Iks}}, \bibinfo {author} {\bibfnamefont {M.}~\bibnamefont {Alvarado}}, \bibinfo {author} {\bibfnamefont {A.~L.}\ \bibnamefont {Yeyati}},\ and\ \bibinfo {author} {\bibfnamefont {R.}~\bibnamefont {Egger}},\ }\bibfield  {title} {\bibinfo {title} {Josephson effect in junctions of conventional and topological superconductors},\ }\href {https://doi.org/10.3762/bjnano.9.158} {\bibfield  {journal} {\bibinfo  {journal} {Beilstein Journal of Nanotechnology}\ }\textbf {\bibinfo {volume} {9}},\ \bibinfo {pages} {1659} (\bibinfo {year} {2018})}\BibitemShut {NoStop}%
\bibitem [{\citenamefont {Alvarado}\ and\ \citenamefont {Yeyati}(2022)}]{Alvarado2022}%
  \BibitemOpen
  \bibfield  {author} {\bibinfo {author} {\bibfnamefont {M.}~\bibnamefont {Alvarado}}\ and\ \bibinfo {author} {\bibfnamefont {A.~L.}\ \bibnamefont {Yeyati}},\ }\bibfield  {title} {\bibinfo {title} {{2D topological matter from a boundary Green's functions perspective: Faddeev-LeVerrier algorithm implementation}},\ }\href {https://doi.org/10.21468/SciPostPhys.13.1.009} {\bibfield  {journal} {\bibinfo  {journal} {SciPost Phys.}\ }\textbf {\bibinfo {volume} {13}},\ \bibinfo {pages} {009} (\bibinfo {year} {2022})}\BibitemShut {NoStop}%
\bibitem [{\citenamefont {Sanches}\ \emph {et~al.}(2025)\citenamefont {Sanches}, \citenamefont {Lustosa}, \citenamefont {Ricco}, \citenamefont {Sigursson}, \citenamefont {de~Souza}, \citenamefont {Figueira}, \citenamefont {Marinho~Jr},\ and\ \citenamefont {Seridonio}}]{Sanches2025}%
  \BibitemOpen
  \bibfield  {author} {\bibinfo {author} {\bibfnamefont {J.~E.}\ \bibnamefont {Sanches}}, \bibinfo {author} {\bibfnamefont {L.~T.}\ \bibnamefont {Lustosa}}, \bibinfo {author} {\bibfnamefont {L.~S.}\ \bibnamefont {Ricco}}, \bibinfo {author} {\bibfnamefont {H.}~\bibnamefont {Sigursson}}, \bibinfo {author} {\bibfnamefont {M.}~\bibnamefont {de~Souza}}, \bibinfo {author} {\bibfnamefont {M.~S.}\ \bibnamefont {Figueira}}, \bibinfo {author} {\bibfnamefont {E.}~\bibnamefont {Marinho~Jr}},\ and\ \bibinfo {author} {\bibfnamefont {A.~C.}\ \bibnamefont {Seridonio}},\ }\bibfield  {title} {\bibinfo {title} {Spin-exchange induced spillover on poor man's majoranas in minimal kitaev chains},\ }\href {https://doi.org/10.1088/1361-648X/adce6a} {\bibfield  {journal} {\bibinfo  {journal} {Journal of Physics: Condensed Matter}\ }\textbf {\bibinfo {volume} {37}},\ \bibinfo {pages} {205601} (\bibinfo {year} {2025})}\BibitemShut {NoStop}%
\bibitem [{\citenamefont {Sanches}\ \emph {et~al.}(2026)\citenamefont {Sanches}, \citenamefont {Sobreira}, \citenamefont {Ricco}, \citenamefont {Figueira},\ and\ \citenamefont {Seridonio}}]{Sanches2026}%
  \BibitemOpen
  \bibfield  {author} {\bibinfo {author} {\bibfnamefont {J.}~\bibnamefont {Sanches}}, \bibinfo {author} {\bibfnamefont {T.}~\bibnamefont {Sobreira}}, \bibinfo {author} {\bibfnamefont {L.}~\bibnamefont {Ricco}}, \bibinfo {author} {\bibfnamefont {M.}~\bibnamefont {Figueira}},\ and\ \bibinfo {author} {\bibfnamefont {A.}~\bibnamefont {Seridonio}},\ }\bibfield  {title} {\bibinfo {title} {Revisiting the poor man’s majoranas: the spin--exchange induced spillover effect},\ }\href {https://doi.org/10.1088/1361-648X/ae2f13} {\bibfield  {journal} {\bibinfo  {journal} {Journal of Physics: Condensed Matter}\ }\textbf {\bibinfo {volume} {38}},\ \bibinfo {pages} {023001} (\bibinfo {year} {2026})}\BibitemShut {NoStop}%
\bibitem [{\citenamefont {Pino}\ \emph {et~al.}(2024)\citenamefont {Pino}, \citenamefont {Souto},\ and\ \citenamefont {Aguado}}]{Pino-Kitmon24}%
  \BibitemOpen
  \bibfield  {author} {\bibinfo {author} {\bibfnamefont {D.~M.}\ \bibnamefont {Pino}}, \bibinfo {author} {\bibfnamefont {R.~S.}\ \bibnamefont {Souto}},\ and\ \bibinfo {author} {\bibfnamefont {R.}~\bibnamefont {Aguado}},\ }\bibfield  {title} {\bibinfo {title} {Minimal kitaev-transmon qubit based on double quantum dots},\ }\href {https://doi.org/10.1103/PhysRevB.109.075101} {\bibfield  {journal} {\bibinfo  {journal} {Phys. Rev. B}\ }\textbf {\bibinfo {volume} {109}},\ \bibinfo {pages} {075101} (\bibinfo {year} {2024})}\BibitemShut {NoStop}%
\bibitem [{\citenamefont {Pan}\ \emph {et~al.}(2025)\citenamefont {Pan}, \citenamefont {Das~Sarma},\ and\ \citenamefont {Liu}}]{PhysRevB.111.075416}%
  \BibitemOpen
  \bibfield  {author} {\bibinfo {author} {\bibfnamefont {H.}~\bibnamefont {Pan}}, \bibinfo {author} {\bibfnamefont {S.}~\bibnamefont {Das~Sarma}},\ and\ \bibinfo {author} {\bibfnamefont {C.-X.}\ \bibnamefont {Liu}},\ }\bibfield  {title} {\bibinfo {title} {Rabi and ramsey oscillations of a majorana qubit in a quantum dot-superconductor array},\ }\href {https://doi.org/10.1103/PhysRevB.111.075416} {\bibfield  {journal} {\bibinfo  {journal} {Phys. Rev. B}\ }\textbf {\bibinfo {volume} {111}},\ \bibinfo {pages} {075416} (\bibinfo {year} {2025})}\BibitemShut {NoStop}%
\bibitem [{\citenamefont {Palacios}\ and\ \citenamefont {Aligia}(2026)}]{Palacios_2026}%
  \BibitemOpen
  \bibfield  {author} {\bibinfo {author} {\bibfnamefont {C.}~\bibnamefont {Palacios}}\ and\ \bibinfo {author} {\bibfnamefont {A.~A.}\ \bibnamefont {Aligia}},\ }\bibfield  {title} {\bibinfo {title} {Effect of interatomic repulsion and quasi-degenerate states on a kitaev-transmon qubit based on double quantum dots},\ }\href {https://doi.org/10.1088/1361-648X/ae3bc7} {\bibfield  {journal} {\bibinfo  {journal} {Journal of Physics: Condensed Matter}\ }\textbf {\bibinfo {volume} {38}},\ \bibinfo {pages} {045302} (\bibinfo {year} {2026})}\BibitemShut {NoStop}%
\bibitem [{\citenamefont {Dynes}\ \emph {et~al.}(1978)\citenamefont {Dynes}, \citenamefont {Narayanamurti},\ and\ \citenamefont {Garno}}]{Dynes1978}%
  \BibitemOpen
  \bibfield  {author} {\bibinfo {author} {\bibfnamefont {R.~C.}\ \bibnamefont {Dynes}}, \bibinfo {author} {\bibfnamefont {V.}~\bibnamefont {Narayanamurti}},\ and\ \bibinfo {author} {\bibfnamefont {J.~P.}\ \bibnamefont {Garno}},\ }\bibfield  {title} {\bibinfo {title} {Direct measurement of quasiparticle-lifetime broadening in a strong-coupled superconductor},\ }\href {https://doi.org/10.1103/PhysRevLett.41.1509} {\bibfield  {journal} {\bibinfo  {journal} {Phys. Rev. Lett.}\ }\textbf {\bibinfo {volume} {41}},\ \bibinfo {pages} {1509} (\bibinfo {year} {1978})}\BibitemShut {NoStop}%
\bibitem [{\citenamefont {Bauer}\ \emph {et~al.}(2007)\citenamefont {Bauer}, \citenamefont {Oguri},\ and\ \citenamefont {Hewson}}]{Bauer2007}%
  \BibitemOpen
  \bibfield  {author} {\bibinfo {author} {\bibfnamefont {J.}~\bibnamefont {Bauer}}, \bibinfo {author} {\bibfnamefont {A.}~\bibnamefont {Oguri}},\ and\ \bibinfo {author} {\bibfnamefont {A.~C.}\ \bibnamefont {Hewson}},\ }\bibfield  {title} {\bibinfo {title} {Spectral properties of locally correlated electrons in a bardeen–cooper–schrieffer superconductor},\ }\href {https://doi.org/10.1088/0953-8984/19/48/486211} {\bibfield  {journal} {\bibinfo  {journal} {Journal of Physics: Condensed Matter}\ }\textbf {\bibinfo {volume} {19}},\ \bibinfo {pages} {486211} (\bibinfo {year} {2007})}\BibitemShut {NoStop}%
\end{thebibliography}

\providecommand{\noopsort}[1]{}\providecommand{\singleletter}[1]{#1}%

\end{document}